\documentclass{pas}
\usepackage{multirow}
\usepackage{subcaption}
\usepackage[percent]{overpic} 
\usepackage{float} 
\usepackage{stackengine}
\usepackage{adjustbox} 
\usepackage{caption}
\usepackage{rotating} 
\usepackage{xcolor}

\usepackage{placeins}

\newcommand*\arcsec{\ensuremath{^{\prime\prime}}}

\begin{document}

\lefttitle{Publications of the Astronomical Society of Australia}
\righttitle{Cambridge Author}

\jnlPage{1}{4}
\jnlDoiYr{2021}
\doival{10.1017/pasa.xxxx.xx}

\articletitt{Research Paper}

\title{COPAS: Compact Objects and the Physics of Accretion Survey I. Accreting Binaries with $P<83$ Minutes} 

\author{\gn{Wendy} \sn{Mendoza}$^{1}$,
        \gn{Liliana} \sn{Rivera Sandoval}$^{1,2}$, 
        \gn{Jan} \sn{K\'{a}ra}$^{1}$, 
        \gn{Ryan J.} \sn{Oelkers}$^{1,2}$, 
        \gn{Yuri} \sn{Cavecchi}$^{3,4}$, 
        \gn{Thomas J.} \sn{Maccarone}$^{5}$, 
        \gn{Manuel} \sn{Pichardo Marcano}$^{6}$,  
        \gn{Meryem K.} \sn{Da\u{g}}$^{7}$,
        \gn{Simone} \sn{Scaringi}$^{7}$, 
        \gn{Kieran} \sn{O'Brien}$^{7}$,
        \gn{Martina} \sn{Veresvarska}$^{8}$, and
        \gn{Christian} \sn{Knigge}$^{9}$}

\affil{$^1$Department of Physics and Astronomy, University of Texas Rio Grande Valley, Brownsville, TX 78520, USA}
\affil{$^2$South Texas Space Science Institute, University of Texas Rio Grande Valley, Brownsville, TX 78520, USA}
\affil{$^3$Departamento de Astrofísica, Universidad de La Laguna, Ave. Astrofísico Francisco Sánchez, s/n, \\ \phantom{$^3$} San Crist\'obal de La Laguna 38206, S/C de Tenerife, Spain}
\affil{$^4$Instituto de Astrofísica de Canarias (IAC), Vía Láctea s/n, San Crist\'obal de La Laguna 38205, S/C de Tenerife, Spain}
\affil{$^5$Department of Physics \& Astronomy, Texas Tech University, Box 41051, Lubbock, TX, 79409-1051, USA}
\affil{$^6$Universidad Nacional Aut\'onoma de M\'exico, Instituto de Astronom\'ia, Ciudad Universitaria, 04510 Ciudad de M\'exico, Mexico}
\affil{$^7$Department of Physics, Centre for Extragalactic Astronomy, Durham University, South Road, Durham DH1 3LE, UK}
\affil{$^8$Institute of Space Sciences (ICE, CSIC), Campus UAB, Carrer de Can Magrans s/n, 08193, Barcelona, Spain}
\affil{$^9$School of Physics and Astronomy, University of Southampton, Highfield, Southampton SO17 1BJ, UK}

%\author{\sn{Cambridge} \gn{Author1}$^{1}$ and \sn{Cambridge} \gn{Author2}$^{2}$}

%\affil{$^1$School of Physics, The University of Melbourne, Parkville, VIC 3010, Australia and $^2$ARC Centre of Excellence for All-Sky Astrophysics in 3 Dimensions (ASTRO-3D)}

\corresp{L. Rivera Sandoval, Email: Liliana.riverasandoval@utrgv.edu}

%\author{Etc..}

%\citeauth{Author1 C and Author2 C, an open-source python tool for simulations of source recovery and completeness in galaxy surveys. {\it Publications of the Astronomical Society of Australia} {\bf 00}, 1--12. https://doi.org/10.1017/pasa.xxxx.xx}

\history{(Received xx xx xxxx; revised xx xx xxxx; accepted xx xx xxxx)}

\begin{abstract}
We present the first results from the Compact Objects and the Physics of Accretion Survey (COPAS), a high-cadence photometric survey focused on the detection of accreting white dwarf binaries mainly through their outbursts using TESS. From an initial sample of $2\,054$ cataclysmic variable (CV) candidates observed during TESS Cycle 6 and 7, we identify 11 systems with periods below 83 minutes. This includes 
7 AM~CVns and 4 CVs. Our Gemini spectroscopic data confirm ZTF18aaxuusk and ASASSN-21eo as AM~CVns, while Gaia21akb is confirmed as a CV.
We provide the first period measurements for 6 out of the 11 systems. Complementing our TESS data with ground-based observations, we measured superoutburst recurrence times of approximately one year for the AM~CVn systems ASASSN-21in and ZTF18aaxuusk, and 1.8-2.0 years for the short-period CVs ASASSN-18rd and ASASSN-15ev. 
Using TESS, we also estimate the mass-ratio for ASASSN-19ct ($q = 0.05 \pm 0.02$), which is consistent with a highly evolved low-mass donor star. 
Though the number of ultracompact systems detected among the total sample is low (0.54\%), and likely the result  of observational biases and intrinsic population rareness, the reclassification of  5 systems as AM CVns highlights the importance of continuous monitoring to identify more systems through their outbursts and short periods. 
\end{abstract}


\maketitle

\section{Introduction} 
The accreting compact systems known as Cataclysmic Variables (CVs) and AM Canum Venaticorum (AM~CVn) are two distinct but related classes of semi-detached white dwarf binaries, in which a white dwarf primary star accretes matter from the Roche-lobe-filling secondary star. These systems serve as important laboratories for studying the process(es) of mass transfer rate, the loss of angular momentum through gravitational radiation, and binary evolution, with implications for many other types of binaries such as low-mass X-ray binaries (LMXBs), supersoft X-ray sources (SSXSs or SSSs), and progenitors of Type Ia Supernovae (SN Ia) \citep{tauris2023physicsbinarystarevolution, Belloni_2023, 1984ApJ...277..355W}. Many of these binaries are expected to contribute to the low-frequency gravitational wave background and serve as verification sources for the Laser Interferometer Space Antenna (LISA) \citep{2023LRR....26....2A, 2023MNRAS.525L..50S}. 

Most CVs contain a hydrogen-rich main-sequence donor star, and their mass transfer typically begins at orbital periods of $6-10$ h, and their long-term evolution is instead driven by angular momentum loss (AML), dominated by magnetic braking and gravitational wave emission \citep{Belloni_2023, 2011ApJS..194...28K, 1980MNRAS.190..801W, 1967AcA....17..287P, 1966AnAp...29..331H}. \footnote{In a subset of these systems containing evolved or subgiant donors, their mass transfer is primarily driven by nuclear expansion of the donor rather than AML \citep{2003MNRAS.340.1214P}.} As a CV evolves towards $P_{\rm orb} \simeq 3\,\mathrm{h}$, the donor star becomes fully convective. This convection disrupts the magnetic braking mechanism, causing the system to temporarily detach, leading to the formation of the \textit{period gap} at $2\mathrm{h} < P_{\rm orb} < 3\mathrm{h}$ \citep{2025A&A...696A..92B, 2024A&A...682L...7S, 2011ApJS..194...28K}. Once the orbital period decreases below $\sim$2 h, mass transfer resumes as the donor refills its Roche lobe, driven by AML through gravitational radiation. The system then evolves towards a theoretical \textit{minimum period}, typically around 80 minutes \citep{2026ApJ...998..153X, Belloni_2023, 1985SvAL...11...52T}. At this evolutionary stage, the donor star becomes partially degenerate, and further mass loss increases its radius. As a result, the system evolves back toward longer orbital periods, becoming what is known as a \textit{period bouncer}.   

A fundamental discrepancy exists between standard models and the observed population of post-period minimum CVs. In particular, evolutionary models predict a minimum orbital period of $P_{orb} \simeq 65$--$70\,\mathrm{min}$ \citep[e.g.][]{kolb1998cvperiodminimum}. This theoretical prediction arises because these standard models rely on the simplistic assumption that binary evolution below the period gap is driven by gravitational radiation alone \citep{2023A&A...679L...8S}. However, observational studies consistently find that the shortest period systems are near $\sim82$ minutes \citep{2012MNRAS.421.2414W, 2009MNRAS.397.2170G}. A possible explanation for the discrepancy between the predicted and observed minimum period is the presence of additional AML, potentially arising from the residual magnetic braking in fully convective donor stars \citep{2023A&A...679L...8S,2011ApJS..194...28K}.

Furthermore, standard models also predict that a large fraction of all CVs, roughly 38\%--75\%, should have evolved past their orbital period minimum and become period bouncers \citep{2023A&A...679L...8S, 2023MNRAS.525.3597I}. In contrast, observational studies indicate that these post-period minimum systems constitute only a small fraction of the CV population, perhaps $\sim\,2\%$~\citep{2023MNRAS.525.3597I}. This suggests that more observations are required to constrain evolutionary models of binary systems in the late evolutionary stages. However, CV systems near the minimum orbital period have low mass transfer rates and therefore low accretion luminosities, making them faint in quiescence and difficult to detect.

While the number of confirmed period bouncers remains small, another class of accreting white dwarf binaries occupies orbital periods well below the CV minimum period. These are the hydrogen deficient AM~CVn binaries, which represent the ultra-compact end of accreting white dwarf evolution. \footnote{It is worth noting that not all systems found below the standard CV period minimum are AM~CVns. For instance, SDSS~J1507 ($P_{\rm orb} = 66.61$~min) is a short period CV featuring an unevolved, substellar donor that likely formed directly from a detached white/brown dwarf binary \citep{2007MNRAS.381..827L}. In contrast, systems like EI~Psc \citep{2002ApJ...567L..49T} and V485~Cen \citep{1996A&A...311..889A} host evolved, hydrogen-depleted donors and are considered systems en route to becoming AM~CVns.} These ultra-compact systems, in which a white dwarf primary accretes from a helium-rich donor star, have $P_{\rm orb} \simeq 5-70$ min~\citep{2025A&A...700A.107G, 2025arXiv251118008R, 1967AcA....17..255S, 1967AcA....17..287P}. Their formation is thought to proceed through three primary channels: (1) stable mass transfer in double white dwarf systems, (2) helium star donors emerging from a common-envelope phase, and (3) evolved main-sequence donors that lose most of their hydrogen before contact \citep{2025A&A...700A.107G, 2023A&A...678A..34B,1996MNRAS.280.1035T,1979AcA....29..665T}.

To date, only about 100 AM~CVns have been identified \citep{2025A&A...700A.107G, 2026PASA...43...52K}, representing a very small fraction compared to the known CV population. \cite{2025A&A...700A.107G} showed that the observed cumulative distance distribution of AM~CVn systems is consistent with an empirically calibrated local space density within approximately 300 pc. However, the observed sample becomes increasingly incomplete beyond this distance. While approximately 50 AM~CVn systems are expected within 500 pc, only about 30 are currently known in this region~\citep{2025A&A...700A.107G}. Similar to evolved CVs, the mass transfer rates in AM~CVns are low, making them faint and difficult to identify.

\subsection{CVs and AM~CVns as transient sources}
Dwarf nova CVs and AM~CVns with orbital periods longer than 20 min show episodes of increased brightness known as outbursts (hereafter referred to as normal outbursts, NO) and superoutbursts (SO). These events are thought to be triggered by thermal-viscous and tidal instabilities, and are usually explained by the disk instability model \citep[DIM; see review by][]{2020AdSpR..66.1004H, 2005PJAB...81..291O}. SOs are longer-duration and higher-luminosity events than NOs, during which tidal instabilities at the 3:1 resonance drive disk precession and produce periodic brightness variations known as \textit{superhumps} \citep{2021A&A...650A.114H, 2005PJAB...81..291O, 1975MNRAS.170..219W, vogt1974photometric}. More recent studies of AM~CVn systems suggest that enhanced mass transfer (EMT) from the donor also contributes to the observed outburst behaviour \citep{Rivera_Sandoval_2022, Rivera_Sandoval_2021, 2020ApJ...900L..37R, 2012A&A...544A..13K}. 

Large-scale surveys such as the Sloan Digital Sky Survey \citep[SDSS;][]{2000AJ....120.1579Y}, the Palomar Transient Factory \citep[PTF;][]{Law_2009}, the Zwicky Transient Facility \citep[ZTF;][]{Bellm_2018}, the All-Sky Automated Survey for Supernovae \citep[ASAS-SN;][]{2017PASP..129j4502K}, the Gaia mission \citep{2016A&A...595A...1G} and The Dark Energy Spectroscopic Instrument \citep[DESI; ][]{levi2019darkenergyspectroscopicinstrument} among others have provided extensive data for studying accreting white dwarf binaries through spectroscopic properties, photometric colours, and orbital periods, thus substantially improving our knowledge. However, observational limitations such as cadence, magnitude depth, sky coverage, and seasonal visibility strongly affect the detectability of AM~CVns and CVs, particularly of short-period systems. 

More recently, the continuous high-cadence observations of the Transiting Exoplanet Survey Satellite \citep[TESS;][]{2015JATIS...1a4003R}, have significantly improved the characterisation of accreting white dwarf variability discovering features that would otherwise be missed by ground-based telescopes \citep{2026kepler,2026arXiv260303539D, 2025ApJS..279...48B, 2021MNRAS.508.3275P}. For example, TESS light curves have been used to characterise the AM~CVn outburst morphology, including the precursor, rise, plateau, and decay phases, and to capture re-brightenings (echo-outbursts), distinguish between NOs and SOs, measure outburst durations and recurrence timescales \citep[e.g.][]{2026arXiv260104323S, 2025ApJS..279...48B, 2021MNRAS.508.3275P, 2021MNRAS.502.4953D}. Furthermore, when TESS observations are combined with multi-wavelength data from other facilities, they provide a powerful framework for studying short periods accreting white dwarfs \citep{2026PASA...43...52K, Inight_2023}.

In this work, we present the \textit{Compact Objects and the Physics of Accretion Survey (COPAS)}, a survey dedicated to identifying and studying accreting white dwarfs through their outbursts by exploiting the continuous monitoring of TESS. In this first manuscript, we report a catalogue of systems below the CV minimum period ($P < 83$ min). We emphasize that the survey is focused on the optical properties of the systems. The X-ray properties will be discussed elsewhere. The remainder of this paper is organised as follows. In Section~\ref{sec:observation}, we describe the high-cadence TESS observations, and the follow-up spectroscopic observations for some targets. In Section~\ref{sec:Timeanalysis}, we describe our time-series analysis, red-noise modelling and our uncertainty estimation methods. In Section~\ref{sec:results}, we present our findings, including periodicities, outburst behaviour, recurrence timescales, spectroscopic classifications, and mass-ratio estimates. In Section~\ref{sec:discussion} we discuss the colour--magnitudes and colour-colour diagrams, as well as observational biases affecting the survey. Lastly, Section~\ref{sec:end} presents our conclusions.  

\section{Observations} \label{sec:observation}
\subsection{TESS Light Curve Analysis}
Since its launch in 2018, TESS has delivered nearly continuous, high-cadence optical observations well suited for monitoring transient stellar variability such as SOs and NOs. The telescope consists of four wide-field cameras, each covering a field of view (FOV) of 24$^\circ$ $\times$ 24$^\circ$. Observations are divided into sectors, each lasting approximately 27.4 days, equivalent to two spacecraft orbits. The overlap between sectors in certain regions of the sky enables extended observation periods for individual targets. 

\begin{figure}[t]
    \centering
    \includegraphics[width=\columnwidth]{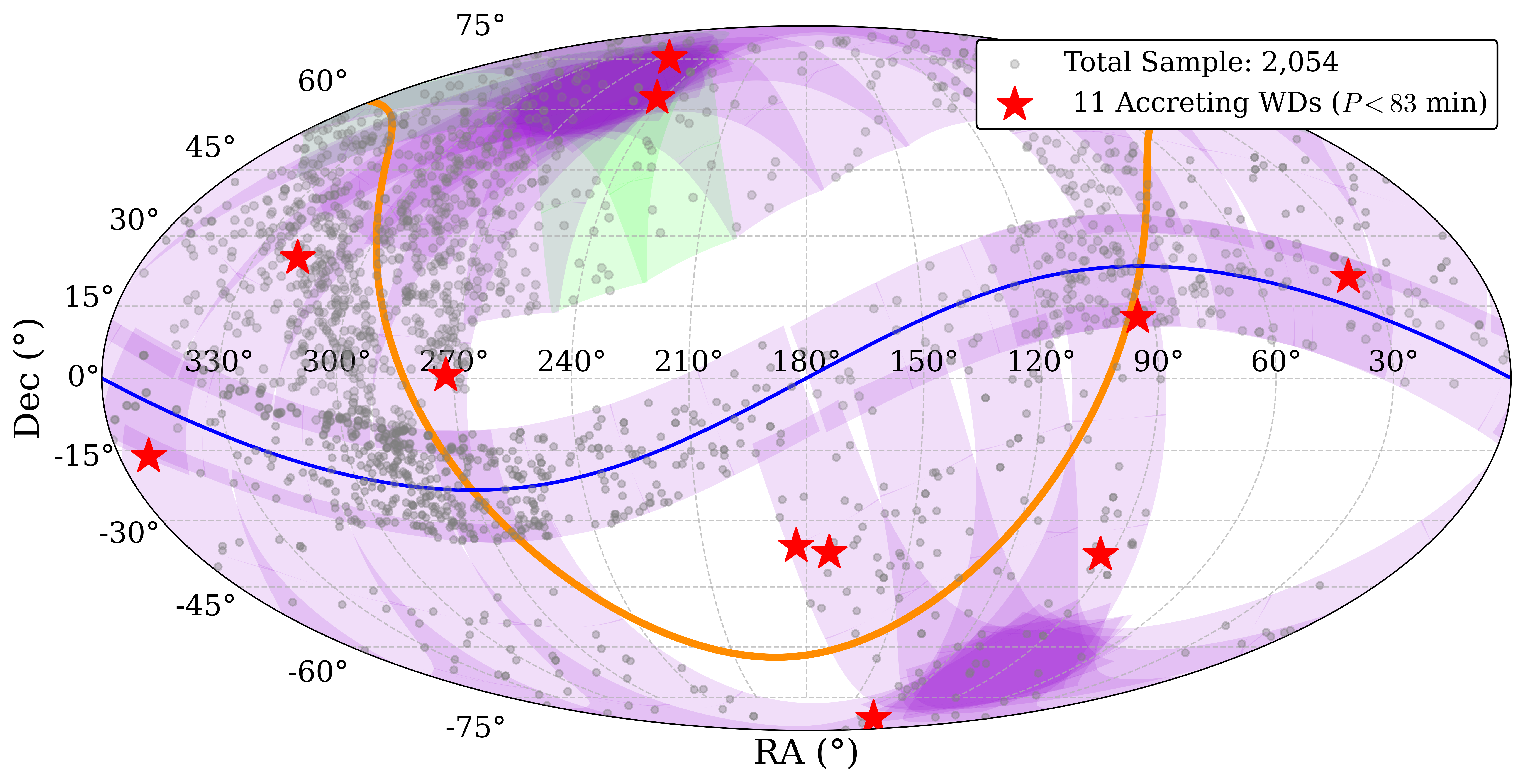}
    \caption{Sky distribution of the $2\,054$ (grey points) total sources of the COPAS survey. We highlight the 11 short period accreting white dwarfs reported in this work (red stars) overlaid on the TESS sector coverage (purple). The Galactic plane (orange) and ecliptic (blue) are shown for reference. Regions with darker purple shading indicate overlapping TESS sectors, thus providing extended temporal coverage. The shaded lime-green regions mark Sectors 77 and 78, which were partially affected by spacecraft Safe Hold.}
    \label{fig:Geomap}
\end{figure}

The observations for this study were obtained during TESS Cycles 6 and 7 (programs G06152 and G07145; P.I. Rivera Sandoval) using 2-minute cadence data. We conducted a targeted search of $2\,054$ transients previously classified as CV candidates using surveys such as Gaia, ZTF, and ASAS-SN. TESS Cycle 6 observations correspond to Sectors 70--83 (14 sectors), covering approximately $\sim 38.7\%$ of the sky from 2023 September 20 on the ecliptic through 2024 October 1, in the Northern Hemisphere. During this cycle, observations from Sectors 77 and 78 were partially affected by spacecraft Safe Hold events triggered by instrument malfunctions \citep{TESS_DRN_Sector_77, TESS_DRN_Sector_78}. Cycle 7 continued from 2024 October 1 in the Northern Hemisphere, through 2025 September 15 in the Southern Hemisphere, spanning Sectors 84--96 (12 sectors) and covering $\sim 47.3\%$ of the sky. The two cycles overlap by $\sim 8.9\%$ of the sky, so the unique combined coverage is $\sim 77\%$ of the sky. The full sky distribution of the sample and the corresponding TESS sectors are illustrated in Figure~\ref{fig:Geomap}. 

\begin{figure}[t]
    \centering
    \includegraphics[width=\columnwidth]{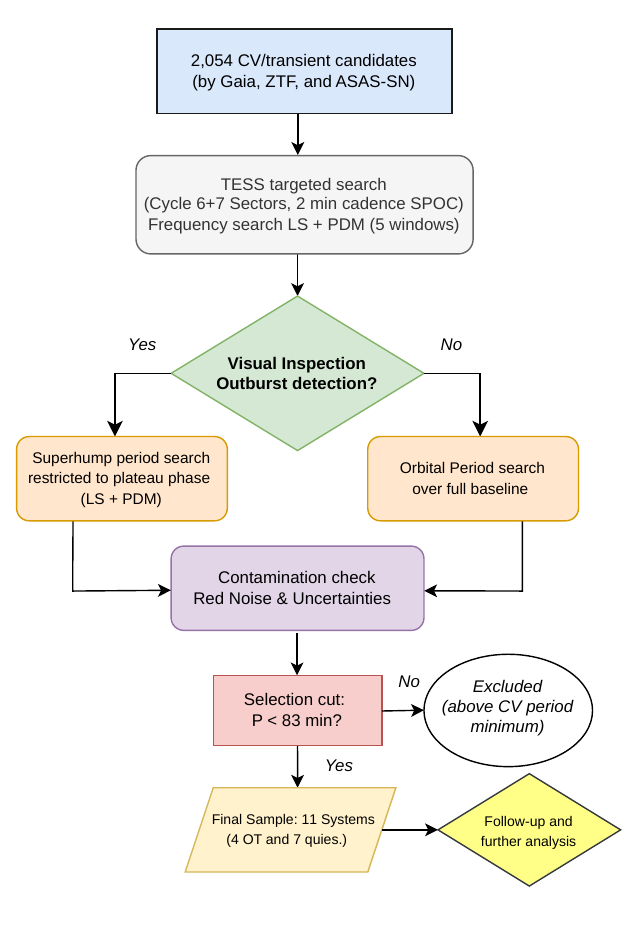}
    \caption{Flowchart summarising the sample selection pipeline, from the initial $2\,054$ candidates through outburst or quiescence state dependent period searches (Section~\ref{sec:Timeanalysis}). Systems with detected periods shorter than 83 min were selected for further analysis. The diagram was created using draw.io.}
    \label{fig:flowchart}
\end{figure}

We used the Python package Lightkurve \citep{2018ascl.soft12013L} to download 2-minute cadence Target Pixel Files (TPFs) from the Mikulski Archive for Space Telescopes (MAST) and extract light curves. These TPFs contain calibrated postage-stamp images centred on each target, and processed through the TESS Science Processing Operations Center (SPOC) pipeline \citep{2016SPIE.9913E....C, 2016SPIE.9913E..3EJ}. For each object with TPF coverage, we used the SPOC-provided simple aperture photometry (SAP) light curves produced by the SPOC pipeline. These light curves are generated from target and background aperture masks, which correspond to minimally processed pipeline photometry. We retained only high-quality measurements by selecting data points with \texttt{QUALITY = 0}. The time is expressed as Barycentric TESS Julian Date $(\text{BTJD} - 2457000)$, and flux is measured in electrons per second ($e^{-}\,s^{-1}$). For visualisation purposes, we applied a moving-median smoothing filter with a window size of 0.05 d to the light curves. This improved the visibility of transient features such as SOs and NOs and of the overall outburst structure. The sample selection and the overall analysis workflow is summarised in Figure~\ref{fig:flowchart}.

\subsection{Spectroscopic Observations and Analysis}
We obtained follow-up spectroscopic observations for three objects in our study. The observations were obtained with the 8.1-m telescopes at the Gemini Observatory located on Maunakea in Hawai'i (objects ZTF18aaxuusk and ASASSN-21eo) and Cerro Pachón in Chile (object Gaia21akb). Spectra were taken under the programs GN-2025B-Q-105 and GS-2025B-Q-105 (P.I. Kára) with the GMOS spectrograph equipped with a GG455 broadband filter and an R150 low-resolution grating ($R \sim 600$). The targets were observed in $4\times2$ binning and only the central portion of the CCD was read out. We obtained two spectra for each target with two different central wavelengths which allowed us to minimise the effects of the gaps between the CCDs on the final spectrum. The spectra cover the wavelength range between $4\,700-10\,000\,\mathrm{\text{\AA}}$ for Gaia21akb, and $5\,800-10\,000\,\mathrm{\text{\AA}}$ for ZTF18aaxuusk and ASASSN-21eo, due to the low brightness of these two targets. The spectra were reduced using the DRAGONS data reduction software \citep{DRAGONS, 2023RNAAS...7..214L}. The observing log and continuum signal-to-noise ratios (SNRs) are listed in Table~\ref{T:GEMINI}. 

\begin{table}[]
\caption{Log of spectroscopic observations, number of exposure time and number of exposure time of individual spectra  ($N_{\text{exp}} \times t_{\text{exp}}\text{ [s]}$), and signal-to-noise ratio (SNR).}
\label{T:GEMINI}
\centering
\begin{tabular}{l l l l c}
\hline
Object       & Date        & Instrument & $N_{\text{exp}} \times t_{\text{exp}}\text{ [s]}$ & SNR \\
\hline
ZTF18aaxuusk & 2025 Jul 14 & GMOS-N     & $2\times480$  &  $17$               \\
ASASSN-21eo  & 2025 Nov 11 & GMOS-N     & $2\times690$  &  $5$               \\
Gaia21akb    & 2025 Oct 24 & GMOS-S     & $2\times240$  &  $32$              \\
\hline
\end{tabular}
\end{table}

\section{Periodicity Analysis} \label{sec:Timeanalysis}
For our time series analysis, we utilised two methods for period detection: Lomb-Scargle (LS; \cite{1982ApJ...263..835S, 1976Ap&SS..39..447L}) and Phase Dispersion Minimization \citep[PDM;][]{2019ascl.soft06010C, 1978ApJ...224..953S}. The LS periodograms were computed using the \textit{Astropy} package~\footnote{https://www.astropy.org}, while PDM analysis was performed with \textit{PyAstronomy}~\footnote{https://pyastronomy.readthedocs.io}. LS and PDM searches were both conducted in frequency space, with peak frequencies converted to periods in minutes for visualisation purposes.

Since the objects studied in the \textit{COPAS} survey had unknown periods \textit{a priori}, we analysed each light curve over five overlapping frequency windows (reported here as equivalent period ranges): $5-180$ min, $160-600$ min, $580-2\,880$ min, $2\,800-7\,200$ min, and $7\,000-14\,400$ min. The shortest range ($5-180$ min) targets AM~CVns and CV systems below the period gap ( e.g., \ref{Appendix:A}). The $160-600$ min range ($\lesssim$10 h) covers orbital timescales for most CVs. The longer ranges ($580-14\,400$ min) were included to identify variability on multi-day timescales, which also helps to classify the transients. The division into these overlapping ranges was used only to facilitate the inspection of variability on different timescales. Candidate signals were cross-checked between overlapping ranges, allowing us to distinguish fundamental periods from higher harmonics (e.g., $P/2$, $P/3$, or $P/4$), spin periods, and aliases through comparison of periodograms and phase-folded light curves \citep{VanderPlas_2018}. Given the long \textit{TESS} datasets and the short periods we search, any candidate periodicity is sampled over a large number of cycles, which helps distinguish coherent periodic signals from longer timescales stochastic variability \citep{1978ComAp...7..103P}. 

\begin{figure}[t]
    \centering
    \includegraphics[width=\columnwidth]{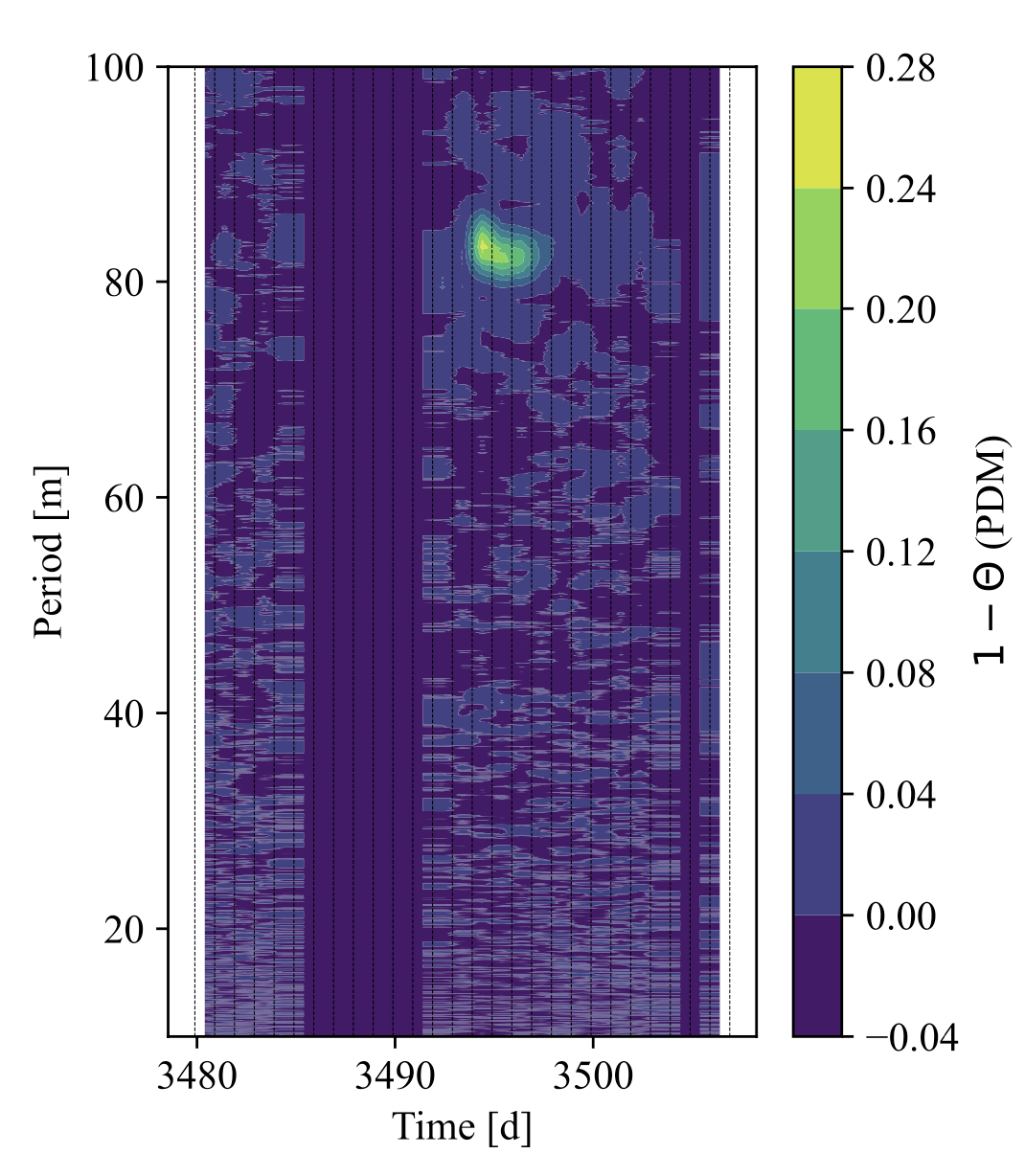}
    \caption{The PDM dynamical map of ASASSN-18rd (TIC: $18136935$, Sector 80) was measured over trial periods of $10-100$ min. The colour scale represents $1 - \Theta$, where the yellow-region corresponds to lower dispersion and indicate stronger periodic signals near $80–83$ min during the plateau phase.}
    \label{fig:2D}
\end{figure}

We also computed PDM as an independent consistency check. PDM identifies periodic variability by dividing the folded light curve into phase bins and computing the flux variance within each bin \citep{1978ApJ...224..953S}. While LS is better for sinusoidal variability, PDM performs well for the non-sinusoidal signals common in superhumps and irregular outbursts. We inspected individual PDM periodograms for each object to confirm that the LS candidate period corresponded to a clear minimum in the PDM statistic. For selected systems, we build a dynamical spectrum by obtaining the sliding-window periodograms and mapping $1 - \Theta$, where $\Theta$ is the PDM statistic, as a function of time and period. The temporal evolution of the signal is shown in Figure~\ref{fig:2D}.
  
To account for temporal signal evolution in systems showing transient activity (such as SOs and NOs), we visually inspected each light curve to identify distinct outburst stages: rise, precursor, plateau, decay, and rebrightening (see Section~\ref{sec:AMCVN}). Although all outburst stages were studied, we primarily restricted the period search to the plateau phase, where superhumps are strongest, and therefore a superhump period is easier to detect. Superhump periods are just a few percent away from the orbital period, making them useful for classifying the type of accreting white dwarf detected in our survey. By restricting the search for the superhump period to the plateau, we minimise phase smearing and improve the precision of the measured superhump period, as these signals can evolve during the later stages of the SO \citep{smak2016superhumpsevolutionsuperoutbursts}. 

In systems where we did not observe transient behaviour, we searched for orbital periodic signals in quiescence. To determine whether a detected signal was consistent with the orbital period, we performed visual inspections of the light curves, periodograms, and phase-folded light curves. For sources with multi-sector coverage, the same period had to appear across all available sectors. 

Due to the large pixel size of TESS ($21~\arcsec\,\mathrm{pixel}^{-1}$), flux contamination from nearby stars can lead to a blended light curve, making it highly possible to attribute photometric variability to the wrong star, particularly for systems in quiescence. We utilised the \texttt{Lightkurve.interact\_sky} tool to identify nearby sources using Gaia positions and magnitudes. We cross-checked them using the International Variable Star Index (VSX) database~\footnote{https://vsx.aavso.org/} and Aladin Sky Atlas \citep{2000A&AS..143...33B} to measure the angular separation of nearby sources. When possible, we extracted the LS periodograms from individual pixels to determine the periodic signal originated from the target or a nearby source.

\subsection{Red Noise Modelling} \label{sec:rednoise} 
We used the Python package \textit{RedNoiseFALs} \citep{ejaz2026rednoisebasedfalsealarm}~\footnote{https://zenodo.org/records/15881590} to calculate the expected False-Alarm Levels (FALs) of each periodogram while also taking into account possible red-noise. \textit{RedNoiseFALs} fits three noise models to the LS power spectrum ($\hat{S}^{\rm LS}(f)$) by minimising the Whittle negative log-likelihood (NLL): a white noise model (WN), an autoregressive model of order 1 [AR(1)], and a power law model (PL). We performed the necessary renormalization of the time steps by the median sampling cadence prior to executing \textit{RedNoiseFALs} fits. We then executed $10\,000$ realisations of each noise model (WN, AR(1), and PL) to produce three distinct distributions of the Whittle NLL using the same search intervals ($5-180$~min) as the previous analysis. We then selected the model with the lowest mean Whittle NLL as the best-fit noise model, and we used this model to calculate the frequency dependent FALs thresholds of $5\%$, $1\%$, and $0.1\%$ so we could evaluate the statistical significance of the periodogram peaks. See Figure~\ref{fig:rednoise} for the results of this analysis.  

\subsection{Uncertainties} \label{sec:uncertainties}
\begin{figure}[H]
    \centering
    \includegraphics[width=\columnwidth]{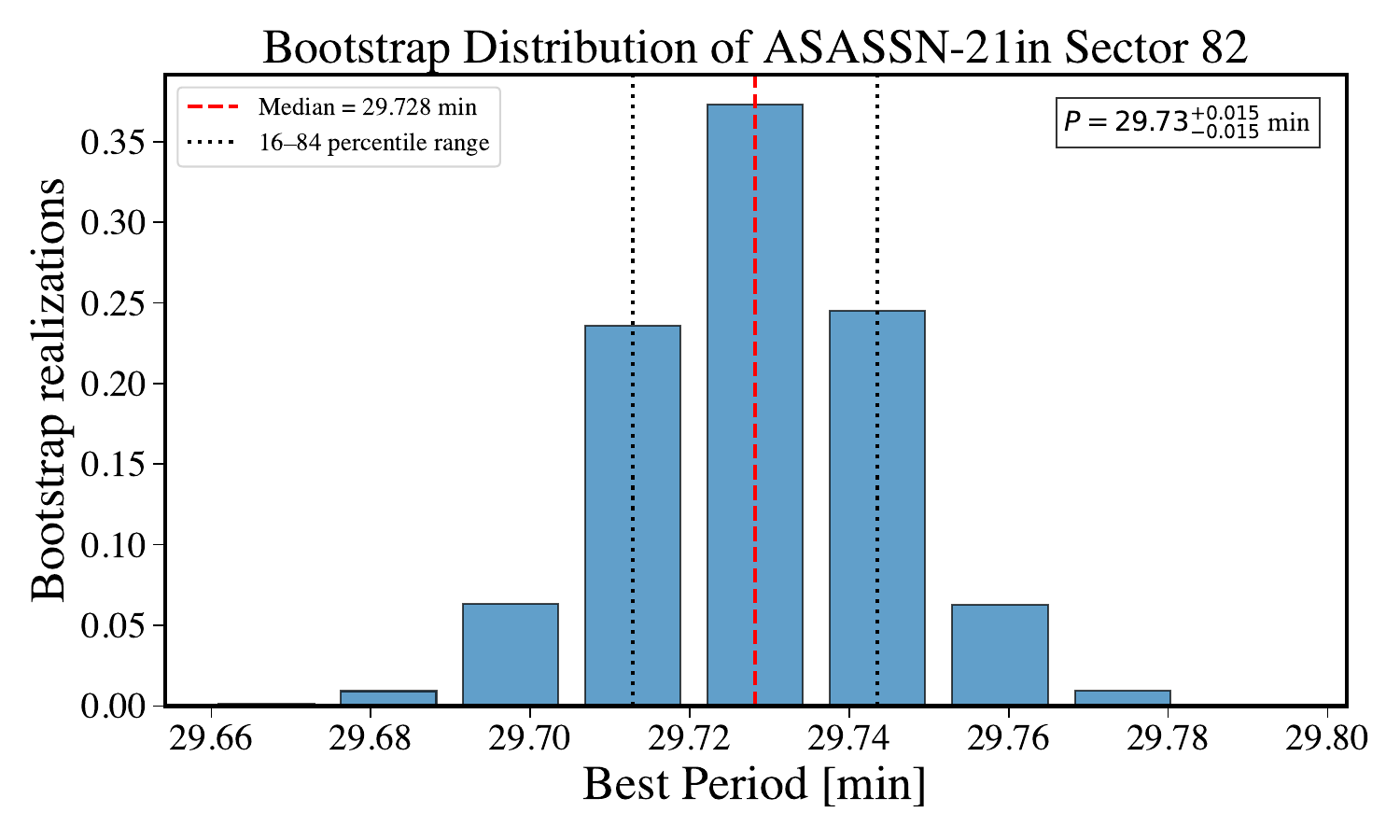}
    \caption{Bootstrap residual distribution of ASASSN-21in (TIC: 452684182); The red dashed line indicates the median of the distribution, while the shaded region represents the 16th--84th percentile interval of the dominant peak, corresponding to the $\pm 1\sigma$ uncertainty.}
    \label{fig: Uncertainty}
\end{figure}

We used a non-parametric residual bootstrap to estimate the uncertainty of the detected periods from the LS periodograms. This process is independent from the \textit{RedNoiseFALs} analysis described in Section~\ref{sec:rednoise}. Unlike a standard bootstrap, which resamples the data as pairs and can alter sampling times, a residual bootstrap keeps the original observation times fixed and only resamples the residuals. This preserves the spectral window of the unevenly sampled time series and produces a more reliable uncertainty estimate for the period \citep{2022A&A...662A..82G, Freedman1981, efron1979bootstrap}. We calculated the residuals by subtracting the corresponding best fit sinusoidal model, evaluated at the peak frequency determined by the analysis described in Section \ref{sec:Timeanalysis}, from the observed flux. We then generated $N_{\text{boot}} = 10\,000$ residual bootstrap realisations by resampling the residuals with replacement while keeping the original observation times fixed. For each realisation, we recalculated the LS periodogram on a fixed-frequency grid and selected the period with the peak frequency. The final determined period was calculated based on the median peak of the bootstrap distribution, and the final uncertainty was defined as the 16th and 84th percentiles of the distribution to establish the $\pm1\sigma$ confidence interval of the period. An example of a resulting distribution is shown in Figure~\ref{fig: Uncertainty} for ASASSSN-21in.

\section{Results} \label{sec:results} 
We present results for 11 objects with periods $P < 83$ min that exhibit characteristics consistent with accreting white dwarf binaries. For systems in outburst, we report the superhump period ($P_{\rm SH}$) measured during the plateau phase. For quiescent systems, we report the best candidate orbital period ($P_{\rm orb}$); while two of these objects have been spectroscopically confirmed, the remaining were identified using periodogram analysis, phase-folded light curves, and consistency with previous classifications in the literature. The findings are summarized in Table~\ref{tab:table1}. The Lomb-Scargle periodograms are shown in Figure~\ref{fig:periodograms}, the phase-folded light curves are shown in Figure~\ref{fig:phase_plots2} and the power-law red-noise FALs in Figure~\ref{fig:rednoise}. 

\begin{table*}[!t]
\centering 
\captionsetup{width=\linewidth}
\caption{Summary of accreting white dwarf binaries observed in TESS Cycles 6 and 7. Columns list the object name, TESS Input Catalogue (TIC), J2000 coordinates (RA and Dec in degrees), object type, observed states (OT = outburst; quies. = quiescence), TESS sectors, and the measured period in minutes with its corresponding uncertainties.} 
\begin{tabular}{l c c c c c c c} 
\hline
Name        & TIC ID     & Coordinates & Type                  & State  & TESS Sectors       & Period & Comments\\
            &            & (J2000)     &                       &        &                    & (min)  &                                      \\
\hline 
ASASSN-21in & 452684182  & 318.3118,   & AM~CVn$^{*}$            & OT     & 82                 & ${29.73^{+0.015}_{-0.015}}$   & First $P_{\rm SH}$ and $T_{\rm rec}$  (this work)  \\
            &            & +25.2335     &                  &        &                    &        &  3~Nearby stars (6--11.32")  \\
            &             &             &                        &       &                    &         &  $^{\dagger}$ZTF18abjgzxt or MGAB-V3719   \\
            &             &             &                        &       &                    &         &                             \\

ZTF18aaxuusk & 1401083226 & 264.0748,  & AM~CVn            & OT     & 73, 74, 76, 77, 79,          & ${30.94^{+0.020}_{-0.020}}$  & $P_{\rm SH}$ consistent with \cite{SalazarManzano2023} \\
           &            & +75.3563     &                   &        & 82, 83  &        & Spec. confirmed and $T_{\rm rec}$ (this work)          \\
           &             &             &                        &       &                    &         & $^{\dagger}$MASTER OT J173618.08+752123.3  \\
            &             &             &                        &       &                    &         &                                   \\
ASASSN-19ct & 941094762  & 173.3138,   & AM~CVn & quies. & 90                 & ${30.96^{+0.003}_{-0.003}}$  & $P_{\rm SH}$ reported by \cite{2026PASA...43...52K} \\
           &            & -37.1724    &                  &        &                    &        &  First $P_{\rm orb}$ and Mass Ratio (this work)                   \\
            &             &             &                        &       &                    &         &    $^{\dagger}$Gaia20afu \\
                        &             &             &                        &       &                    &         &                          \\
Gaia23asm   & 951008167  & 182.99108,  & AM~CVn$^{*}$ & quies. & 90             & ${47.01^{+0.005}_{-0.005}}$  & First $P_{\rm orb}$ and $T_{\rm rec}$ (this work)                                                                                                      \\
           &            & -35.69812   &                   &       &                    &                                  & Possible Candidate \\
            &             &             &                        &       &                    &         & Nearby star (4.33")   \\
            &             &             &                        &       &                    &         &      $^{\dagger}$ASASSN-18ej                \\
            &             &             &                        &       &                    &         &                          \\
ASASSN-14cn & 1201247611 & 242.8917,   & AM~CVn & quies. & 74--79, 81--86      & ${49.71^{+0.002}_{-0.002}}$  &  $P_{\rm orb}$ consistent with \cite{2021MNRAS.508.3275P}\\
            &            & +63.1423     &                  &        &                    &        &   $^{\dagger}$Gaia14aae or ZTF18aaplouo                        \\
            &             &             &                        &       &                    &              &                          \\
ASASSN-21eo & 437887678  & 94.04238,   & AM~CVn & quies. & 71, 72, 87         & ${52.42^{+0.008}_{-0.008}}$  & First $P_{\rm orb}$, $T_{\rm rec}$ and Spec. confirmed (this work) \\
            &            & +12.72211    &                   &        &                    &        & 3~Nearby stars (7-11")  \\
            &             &             &                        &       &                    &         & $^{\dagger}$ZTF21aarlmkd or TCP J06161010+1243200  \\
            &             &             &                        &       &                    &         &                          \\
Gaia23axx   & 620274558  & 35.52300,   & AM~CVn$^{*}$ & quies. & 70, 71             & ${53.46^{+0.010}_{-0.010}}$  &  First $P_{\rm orb}$ and $T_{\rm rec}$ (this work)                                     \\
            &            & +21.10014    &                       &        &                    &       &  Possible Candidate  \\
            &             &             &                        &       &                    &         &    Nearby star (6.37") \\
            &             &             &                        &       &                    &         &   $^{\dagger}$MGAB-V3470 or ZTF18acalseu  \\
                        &             &             &                        &       &                    &         &                          \\
Gaia21akb   & 328010691  & 352.35867,  & CV                    &  quies. & 70, 96             & ${75.71^{+0.001}_{-0.001}}$  & $P_{\rm orb} = 77.55$ min reported by \cite{littlefield2026tesslightcurvesnew} \\
            &            &     -16.27629                      & &        &                    &       & Spec. confirmed (this work) \\
            &             &             &                        &       &                    &         &      $^{\dagger}$ZTF20acqpkxj or 1RXS J232928.0-161654             \\
                        &             &             &                        &       &                    &         &                          \\
Gaia19ekt    &  705387284    & 93.15156 & CV$^{*}$     &  quies. & 87 & ${82.32^{+0.019}_{-0.019}}$ & First $P_{\rm orb}$ (this work) and Candidate \\
                       &                         & -37.61568 &          &             &        &     & 3 Nearby stars (9.22--10.42")   \\
                                   &             &             &                        &       &                    &         &   $^{\dagger}$WD J061236.33-373656.69   \\
                        &             &             &                        &       &                    &         &                          \\
ASASSN-18rd & 18136935   & 272.1162,   & CV$^{*}$                     & OT     & 80                 & ${82.75^{+0.034}_{-0.034}}$   &  First $P_{\rm SH}$ and $T_{\rm rec}$ (this work)        \\
            &            & +0.5916      &                       &        &                    &        & Crowded stellar field  \\
            &             &             &                        &       &                    &         &    \\            
ASASSN-15ev & 764400107  & 114.581,    & CV$^{*}$                    & OT     & 93, 94             & ${83.48^{+0.034}_{-0.034}}$  &  $P_{\rm SH}$ consistent with~\cite{2016PASJ...68...65K} \\
            &            & -82.8439              &                       &        &                    &                               & $P_{\rm orb}$ predicted by~\cite{2019MNRAS.486.2422P}   \\   
            &            &             &                       &        &                    &        &   measured $T_{\rm rec}$ (this work)  \\

\hline
\end{tabular}
    \caption*{Note: superhump period ($P_{\rm SH}$), orbital period ($P_{\rm orb}$), and measured recurrence timescales ($T_{\rm rec}$). $^{*}$ Not spectroscopically confirmed.  $^{\dagger}$ other designation(s).}
    \label{tab:table1}
%\end{rotatetable}
\end{table*} 

%Lightcurves examples 
\begin{figure*}
    \centering
    \includegraphics[width=0.90\linewidth]{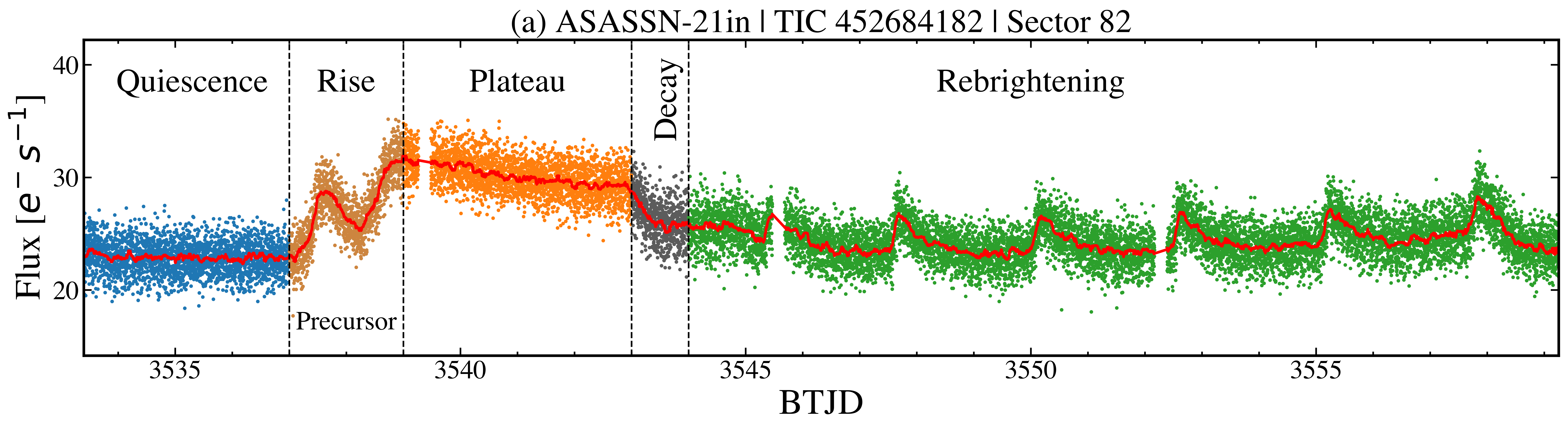} 
    \includegraphics[width=0.90\linewidth]{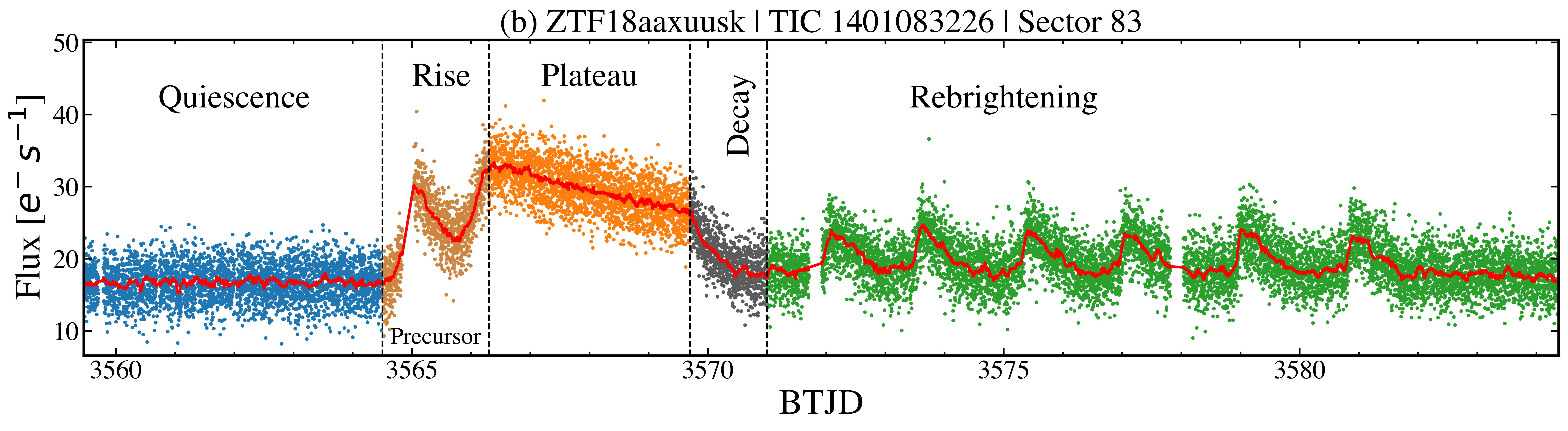}
    \includegraphics[width=0.90\linewidth]{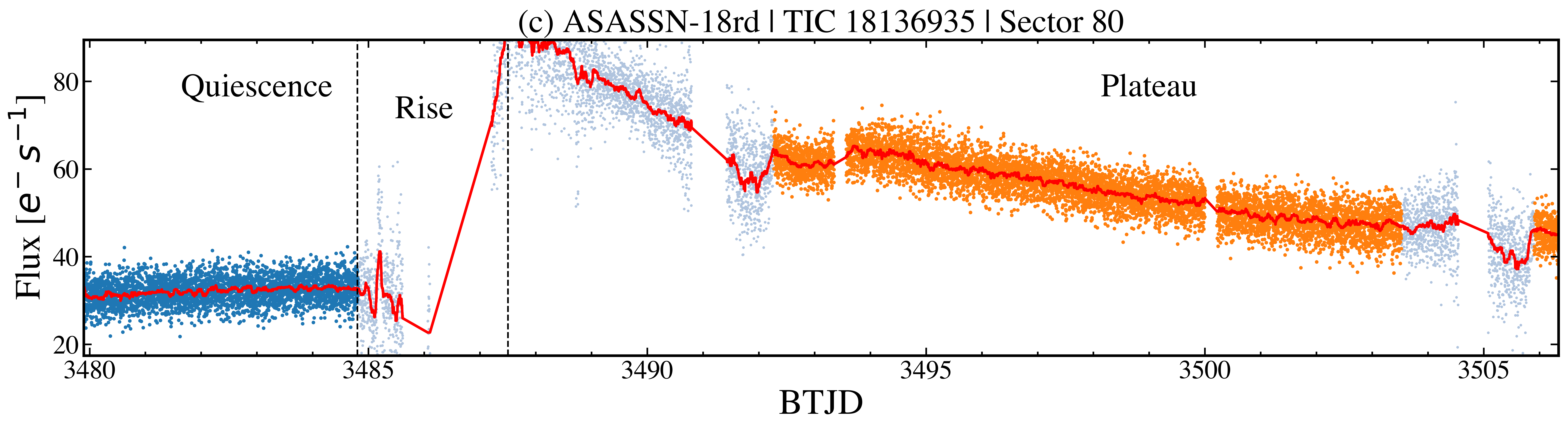} 
    \includegraphics[width=0.90\linewidth]{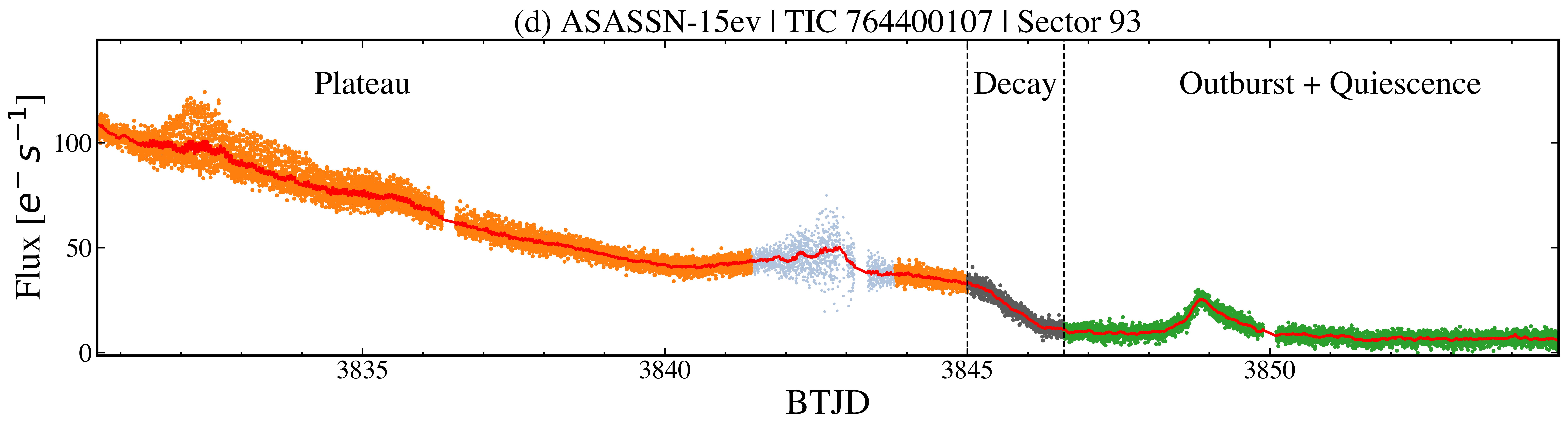}
    \includegraphics[width=0.90\linewidth]{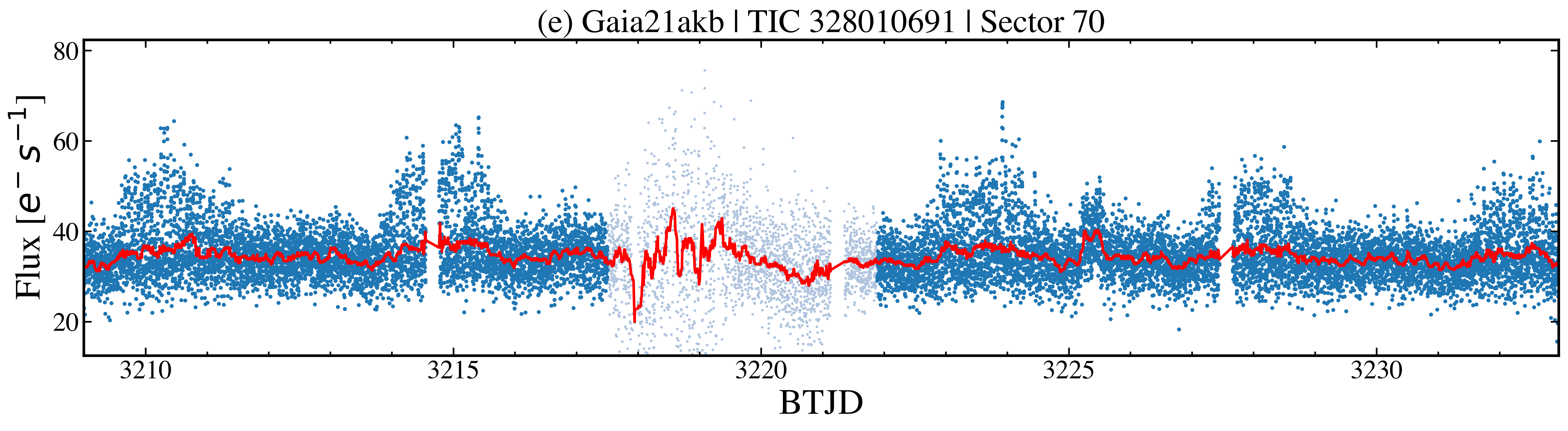}
    
    \caption{TESS light curves of systems showing outburst behaviour and quiescence. The plots highlight distinct evolutionary phases, including the precursor, rise, plateau, decay, and re-brightening, which are colour-coded to distinguish the stages of the outburst. The measurements with low quality flags are shown in light blue. The red line represents the 0.05-day moving median. The 1-2 day gaps correspond to TESS Low-Altitude Housekeeping Operations (LAHO) for data downlink to Earth.}
    \label{fig:LCS_OT}
\end{figure*}  

\subsection{Superhump Periods in AM~CVns} \label{sec:AMCVN}
\subsubsection*{ASASSN-21in}
The object ASASSN-21in  was previously classified as a SU Uma-type dwarf nova CV by \citet{2019TNSTR1295....1N}. In TESS Sector 82, we detected a SO (see Figure~\ref{fig:LCS_OT}~(a)) and measured a superhump period of $P_{\rm SH} = 29.73$ min ($f_{\rm best}=48.43$~d$^{-1}$). The signal was detected during the plateau phase (Figure~\ref{fig:periodograms} and \ref{fig:phase_plots2}~(a)). The corresponding peak exceeds 0.1\% FAL, indicating highly significant detection (Figure~\ref{fig:rednoise}~(a)). This period is well below the CV period minimum, supporting the AM~CVn classification. Three nearby stars (6-11" separation) were identified as potential contaminants using Aladin. However, the outburst behaviour and long-term ground-based observations confirm that the variability originated from the target (see Section~\ref{sec:recurrence}). 

\subsubsection*{ZTF18aaxuusk}
A similar photometric behaviour was observed in ZTF18aaxuusk, which was initially classified by ZTF as a CV of type DN. \cite{Kojiguchi2020} (vsnet-chat 8472) suggested that this object could be a V803 Cen star-like system due to its cycle length of 250 days. In TESS Sector 83, we detected a superhump period of $P_{\rm SH}= 30.94$ min ($f_{\rm best}=46.53$~d$^{-1}$) (see Figure~\ref{fig:periodograms} and \ref{fig:phase_plots2}~(b)). The peak lies above the 0.1\% FAL, as shown in Figure~\ref{fig:rednoise}~(b). This period is consistent with the superhump period previously reported by \cite{SalazarManzano2023}, who analysed TESS Sectors 10, 23, 47, and 54. We confirm its AM~CVn nature through spectroscopic analysis (see Section \ref{spectralresults}), which reveals a lack of Hydrogen and abundance Helium which is the signature of AM~CVns.

\subsection{Orbital Periods in AM~CVns} 
\subsubsection*{ASASSN-21eo}
We also detected periodic variability in systems observed in quiescence without evidence of outburst activity. The object ASASSN-21eo, was previously suggested to be an AM~CVn candidate based on its rapid fading behaviour after the outburst \citep{2021PASJ...73.1375K}. Using the LS and PDM methods across three TESS Sectors, we detected a periodic signal in Sector 71, corresponding to an orbital period of $P_{\rm orb} = 52.42$ min ($f_{\rm best}=27.47$~d$^{-1}$) exceeding 0.1\% FAL (see Figure~\ref{fig:periodograms}, \ref{fig:rednoise}, and \ref{fig:phase_plots2}~(f)). 

By carefully inspecting our object using the AAVSO variable star plotter, we identified three nearby sources (7-11" separation). The nearest star (Gaia DR3 3332047452287378176) with $G_{\rm mag} = 16.80$ lies within the TESS aperture, raising the possibility of photometric contamination in both the light curve and the periodic signal. However, the spectroscopic follow-up we carried out confirms the AM~CVn nature of this system (see Section~\ref{spectralresults}).

\subsubsection*{Gaia23axx}
Another candidate revealing a periodic signal in quiescence is object Gaia23axx. This object was initially identified as a CV candidate following an outburst on 2023-03-03, reported as a Gaia transient alert with a discovery magnitude change from 18.1 to 21.0. We observed this object in TESS Sectors 70 and 71, where a consistent periodic signal of $P_{\rm orb}= 53.46$ ~min ($f_{\rm best}=26.93$~d$^{-1}$) was observed, exceeding the 0.1\% FAL (see Figure~\ref{fig:periodograms}, \ref{fig:rednoise}, and \ref{fig:phase_plots2}~(g)). However, a nearby star (Gaia DR3 99677532286826240), located 6.37" away with $G_{\rm mag}=20.83$, may contribute to contamination of the TESS aperture. Given the short detected period and colours (see Section~\ref{sec:discussion}), we classify this system as an AM CVn candidate. 

\subsubsection*{Gaia23asm}
Similarly, Gaia23asm was previously considered a CV candidate, exhibiting an outburst amplitude of approximately 3 magnitudes, as reported by Gaia. In Sector 90, we detected a periodic signal of $P_{\rm orb}= 47.01$ min ($f_{\rm best}=30.63$~d$^{-1}$) and a second peak at 78.11 min both reaching the 0.1\% FAL (see Figure~\ref{fig:periodograms} and \ref{fig:rednoise}~(d)). The two periods are not related by a simple integer ratio. Due to the small modulation at 47.01 min (see Figure~\ref{fig:phase_plots2}~(d)) and given that the period falls within the 5--70 minute range typical of ultra-compact binaries, we classify this system as an AM~CVn. We note that the second peak could also be related to the orbital period of a short period CV, although phase-folding does not reveal a coherent modulation. A nearby star (Gaia DR3 3461950868453873280), located 4.33" away with $G_{\rm mag}=20.5$, lies within the TESS aperture and may contribute to flux contamination. Spectra are needed to clarify the nature of the source. 

\subsubsection*{ASASSN-19ct and ASASSN-14cn}
In contrast to candidates that require further verification, the objects ASASSN-19ct and ASASSN-14cn have been reported to be AM~CVns based on their short-periods and spectroscopy. In our TESS analysis, we detected that the orbital period of ASASSN-19ct is $P_{\rm orb} = 30.96$ min ($f_{\rm best}=46.52$~d$^{-1}$; 0.1\% FAL) in Sector 90 (see Figure~\ref{fig:periodograms} and \ref{fig:phase_plots2}~(c)). This is highly consistent with the superhump period ($P_{\rm SH} = 30.94\pm0.21$ min) reported by \cite{2026PASA...43...52K} measured during the plateau phase of a SO in Sector 37. For this system, we estimated a mass-ratio based on our $P_{\rm orb}$ and previous $P_{\rm SH}$ detections (see Section~\ref{Massratio}). We also detected ASASSN-14cn in multiple sectors ($74-79$ and $81-86$). The orbital period of $P_{\rm orb} = 49.71$ min ($f_{\rm best}=28.97$~d$^{-1}$; 0.1\% FAL), with the first harmonics of $P_{\rm orb} = 24.85$ min (0.1\% FAL), consistent with \citet{2021MNRAS.508.3275P}. 

\subsection{Superhump Periods in CVs}
\subsubsection*{ASASSN-18rd}
ASASSN-18rd is a dwarf nova, previously reported by \cite{Kawash_2021}, that reached a peak brightness of approximately 16.1 magnitude during outburst. In TESS Sector 80, an incomplete plateau phase is captured that lasted roughly nine days (BTJD 3494--3506), after which the system declined (see Figure~\ref{fig:LCS_OT}~(c)). Due to the crowded stellar field surrounding the target (as indicated by AAVSO), we performed a pixel-level analysis by extracting light curves and computing periodograms for individual TESS pixels. This approach confirms that both the outburst and the periodic signal are strongest at the target position. We detected a superhump period of $P_{\rm SH} = 82.75$~min ($f_{\rm best}=17.41$~d$^{-1}$) (see Figure~\ref{fig:periodograms}~(j)). The peak power at the corresponding frequency remain above the 0.1\% FAL despite the sloped red noise continuum, confirming that our detected signal is statistically significant (see Figure~\ref{fig:rednoise}~(j)). 

\begin{figure}[H]
    \centering
    \includegraphics[width=\columnwidth]{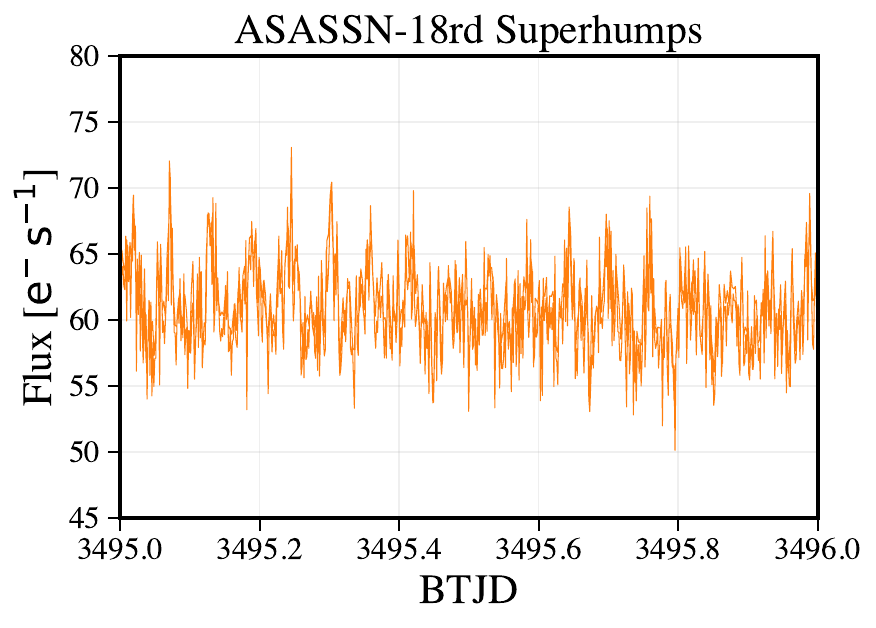}
    \caption{Zoomed-in TESS light-curve of ASASSN-18rd showing superhump brightness variations during the plateau phase in Sector~80.}
    \label{fig: superhump}
\end{figure}

The dynamical PDM map also showed a concentrated region of low dispersion near 80--83 min during the plateau phase, indicating a stable and time localised periodic signal (Figure~\ref{fig:2D}). The corresponding phase folded light curve in Figure~\ref{fig:phase_plots2}~(j)) and the zoomed light curve as shown Figure~\ref{fig: superhump} revealed coherent superhump brightness variations. Using the empirical relation $P_{\rm orb} = 0.9162(52)P_{sh} + 5.39(52)$, derived by \cite{G_nsicke_2009}, we estimate an orbital period of $P_{\rm orb} = 81.18 \pm 2.0 $ min, assuming a conservative error margin of 2-min to account for small drifts in the superhump period often observed during dwarf nova outbursts.

\subsubsection*{ASASSN-15ev}
\citet{2016PASJ...68...65K} reported a superhump period of $P_{\rm SH} = 83.48 min$ for ASASSN-15ev. Our TESS analysis of Sector 93 recovered the same period ($P_{\rm SH} = 83.48$) during the plateau phase (see Figure~\ref{fig:periodograms} and \ref{fig:phase_plots2}~(k)). Although the periodogram continuum contains strong red noise, the peak at $f_{\rm best}=17.26$~d$^{-1}$ remains significant, rising well above the 0.1\% FAL, as shown in Figure~\ref{fig:rednoise} (k). Our estimate orbital period $P_{\rm orb} = 81.87 \pm 2.0 $ (0.056847~d) matches the prediction calculated by \cite{2019MNRAS.486.2422P}. Although this system is known, its recurrence time remains uncertain. In this work, we estimate a recurrence timescale of 1.85 years, discussed in Sections~\ref{sec:recurrence}.  

\subsection{Orbital Period in CVs}
\subsubsection*{Gaia21akb}
Gaia21akb was observed in TESS Sector 70 (see Figure~\ref{fig:LCS_OT}~(e)). Previously,~\cite{littlefield2026tesslightcurvesnew} classified it as an  asynchronous polar CV with a dominant period signal at $75.70$ min, which they interpreted as a  $2\omega - \Omega$ sideband frequency, inferring an orbital period of $77.55$ min. Our independent LS and PDM analysis recovers this same periodic signal of $P_{\rm orb} = 75.71$~min ($f_{\rm best}=19.02$~d$^{-1}$) which is supported by the corresponding phase-folded light curve (see Figure~\ref{fig:periodograms}, and \ref{fig:phase_plots2}~(h)). Figure \ref{fig:rednoise} (h) shows a clear red-noise continuum, with the detected peak exceeding the 0.1\% FAL, confirming the statistical significance of this period. In addition, our work presents a spectroscopic analysis of this system (see Section \ref{spectralresults}).

\subsubsection*{Gaia19ekt}
Lastly, Gaia19ekt triggered a transient alert on 2019-09-02 and was classified as a candidate CV after Gaia reported a brightness increase of approximately $2$ mag. In Sector 87, we detected a first harmonic near $P/2\simeq 41.16$~min and orbital period of  $P_{\rm orb} = 82.32$~min ($f_{\rm best}=17.49$~d$^{-1}$) both exceeding near the 0.1\% FAL (see Figure~\ref{fig:periodograms}, \ref{fig:rednoise} and \ref{fig:phase_plots2}~(i)). However, three nearby stars are located 9.59--11.28" from the object, the closest of which has $G_{\rm mag}=17.26$. Since potential contamination from these stars may have affected the period analysis, further observations are required to confirm the nature and period of this source.

\subsection{Outburst Behaviour and Recurrence Timescales} \label{sec:recurrence}
To characterise the outburst recurrence time of our sample, we used TESS as the primary photometric baseline and complemented it with long-term ground-based observations from ZTF \citep[ALeRCE;][]{2021AJ....161..242F} and Asteroid Terrestrial-impact Last Alert System \citep[ATLAS;][]{2018PASP..130f4505T}. The TESS coverage for each system is summarised in Table~\ref{tab:recurrence}. 
  
\begin{table*}[ht]
\centering
\caption{The table includes the object classification, TESS sector coverage where SO occur, SO duration, plateau duration, number of SO detected, recurrence timescale ($T_{\rm rec}$), and the survey used for the recurrence analysis.}
\label{tab:recurrence}
\begin{tabular}{lccccccc}
\hline
Object            & Type  & TESS Sector   & SO Duration & Plateau Duration & Number of SO & $T_{\rm rec}$ & Survey \\
       &      &             & (days)      & (days)            &            &      (years)     &  \\
\hline
ASASSN-21in      & AM~CVn & 82            & $6$  & $4$ & $6$ & $\sim1.01$  & ZTF \\
ZTF18aaxuusk     & AM~CVn & 83            & $6$  & $4$ & $7$ & $\sim1.04$  & ZTF \\
Gaia23asm        & AM~CVn candi. & -      & unclear    &   unclear & 1 &  $ > 3.177$   & ATLAS  \\ 
ASASSN-21eo      & AM~CVn & -             &  unclear  & unclear &   1  &  $ > 4.99$  & ATLAS/ZTF \\ 
Gaia23axx        & AM~CVn candi. & -        & unclear  &  unclear   & $\sim3$   &  $\sim2.92$ & ATLAS/ZTF \\ 
ASASSN-18rd      & CV            & 80       & $23.7$ & $19.14$& $3$ & $\sim1.96$ & ATLAS \\
ASASSN-15ev      & CV             & 93       & $21.3$ & $15.73$ &  $2$ & $\sim1.85$ & ATLAS \\  
\hline
\end{tabular}
\caption*{Note: The recurrence timescale of ASASSN-19ct has already been measured in previous studied, with estimates ranging from approximately 180-365 days (roughly 0.5-1 years; \citealt{2026PASA...43...52K,2025PASJ...77.1126K}). Objects like ASASSN-14cn, Gaia21akb and Gaia19ekt are excluded from this table because they remain in quiescence and have uncertain observational gaps.}
\end{table*} 

\subsubsection{Outburst Behaviour in AM~CVns}
For AM~CVn \textit{ASASSN-21in} ($P_{\rm SH}=29.73$ min), we observed a single SO in Sector 82 with a duration of approximately seven days. The TESS light curve provided a detailed outburst profile featuring a precursor, rise, four-day plateau, decay, and six re-brightenings (echo-outbursts), as shown in Figure~\ref{fig:LCS_OT}~(a). The long-term ZTF light curve identifies the SO event between MJD 60535--60560, but the individual re-brightenings blend into the main outburst (see Figure~\ref{fig:Longtermspart1}). To measure the recurrence timescale, we complemented the TESS observations with the long-term ZTF light curve, which captures six distinct SO at MJD 59121, 59356, 59733, 60100, 60538, and 60972. We calculated an average SO recurrence time of approximately 1.01 yr (370.2 d). 

Similarly, the AM~CVn \textit{ZTF18aaxuusk} ($P_{\rm SH}=30.94$ min) has the most extensive TESS coverage in our sample, with observations spanning 7 sectors (see Table~\ref{tab:table1}). TESS detected only one SO event, in Sector 83, while the remaining sectors captured the system in quiescence. The TESS light curve reveals a precursor, rise, four-day plateau, decay, and six re-brightenings (echo-outbursts), with a total SO duration of approximately seven days (see Figure~\ref{fig:LCS_OT}~(b)). The single TESS outburst detection is insufficient to constrain the recurrence timescale. We therefore complemented it with long-term ZTF light curve, which captures seven distinct SO events in MJD 58271, 58685, 58940, 59379, 59789, 60141, and 60559 (see Fig.~\ref{fig:Longtermspart1}). From these events, we calculated an average recurrence time of 1.04 yr (381.3 d). 

It is interesting to note that the outburst profiles of these two systems are remarkably similar and closely resemble the morphology of AM CVns such as PTF J0179+4858 ($P_{\rm SH}=26.8$ min) observed by \cite{2021MNRAS.508.3275P}. These recurring patterns of precursors and echo-outbursts imply that the mechanism shaping the light-curve is highly dependent on EMT as shown in previous studies \citep{2021MNRAS.508.3275P}. The recurrence timescale of these AM~CVn systems is consistent with previous observational and theoretical results, demonstrating a strong empirical correlation between the orbital period and the outburst recurrence timescales \citep{2015MNRAS.446..391L,2025arXiv251118008R}. In particular, systems with shorter orbital periods $<$ 40 min typically have much higher mass transfer rates, leading to relatively short observable recurrence times of $\approx$ 1 yr \citep{2025PASJ...77.1126K}. For example, the known system CSS 150211:091017-200813 has a recurrence timescale of 335 days and an orbital period of 29.66 min, comparable to the values observed in our systems.

\subsubsection{Outburst Behaviour in CVs}
The CV system \textit{ASASSN-18rd} ($P_{\rm SH}=82.75$ min) in Sector 80 reveals an incomplete duration of a SO and a plateau phase that could be affected by instrumental systematics and quality flag filtering during its evolution (see Figure~\ref{fig:LCS_OT}~(c)). By combining the continuous high-cadence TESS observations and the long-term ATLAS light curve, we were able to constrain the SO behaviour (see Figure~\ref{fig:Longtermsart2}). We estimated a total SO duration of approximately 23.7 d, from the TESS detected quiescence at MJD 60484.3 to the ATLAS measurements at MJD 60508. The plateau phase extended from MJD 60487.21--60506.35, with a duration of 19.14 d. The long-term light curve captures three SO events at MJD 58337, 59075, 60487.4, from which we measured an average recurrence timescale of 1.96 years (716.8 d). 

Another similar system, \textit{ASASSN-15ev} ($P_{\rm SH}=83.48$ min), for which combining TESS observations from Sector 93 with the long-term ground-based ATLAS light-curve (see Figure~\ref{fig:Longtermsart2}), revealed an SO duration of 21.3 d from the ATLAS detection onset at  MJD~60824 to TESS resolved decline ending at MJD~60845.3 and a plateau phase of 15.73 d (MJD 60829.27--60845). This is notably longer than the $\le$ 13 d plateau phase limit reported by \cite{2016PASJ...68...65K}. We estimated an average recurrence timescale of 1.85 years (674 days) based on the two observed SO events.  

These observed outburst behaviour, short-periods, and 3-week SO durations aligned well with the characteristics of WZ~Sge dwarf nova (a subclass of short-period CVs). These highly evolved, short-period systems are characterised by low mass-transfer and near-total absence of normal outburst activity \citep{tampo2025wzsgetypedwarfnovae}. A classical WZ~Sge stars typically show much longer recurrence timescales, with a median recurrence interval of approximately 11.5 yr \citep{2015PASJ...67..108K}. Systems like \textit{ASASSN-18rd} and \textit{ASASSN-15ev} show significantly shorter recurrence cycles, placing them at the short cycle end of the WZ~Sge population, comparable to systems such as AL Com, with a recurrence timescale of 450 d and a superhump period of 0.057293 d \citep[82.5 min,][]{2016PASJ...68L...2K}.

\subsubsection{Unconstrained Recurrence Timescales}
A few of the observed TESS quiescence systems have unclear behaviour due to gaps and poor coverage in the long-term photometric data, making it difficult to determine the exact properties, duration, or frequency of their outbursts. For systems with only one outburst, we estimated a lower limit on the recurrence timescale based on the time elapsed since the last detected outburst.

For instance, in the case of the the AM~CVn \textit{ASASSN-21eo}, due to the gaps in the ground-based long-term light curve (see Figure~\ref{fig:Longtermsart3}) it remains unclear whether we see a normal outburst or a superoutburst that is partially covered around MJD $\approx$ 59500. This also makes it impossible to estimate its duration. We established a lower limit on the recurrence timescale of $T_{\rm rec} > 1824$ d (4.99 yr). Another system with only one outburst detection around MJD $\approx$ 59990 is the AM~CVn candidate \textit{Gaia23asm}. We detected a lower limit recurrence timescale of $T_{\rm rec} > 1159$ d (3.17 yr). A system with unclear outburst behaviour is the AM~CVn candidate \textit{Gaia23axx}. We identified three main partial outburst events around MJD~58737, 60292, and 60867. From these peaks, we estimated an average recurrence timescale of approximately 2.92~yr (1065~d).

\subsection{Spectral Results} \label{spectralresults}
\begin{figure*}
    \centering
    \includegraphics[width=0.99\linewidth]{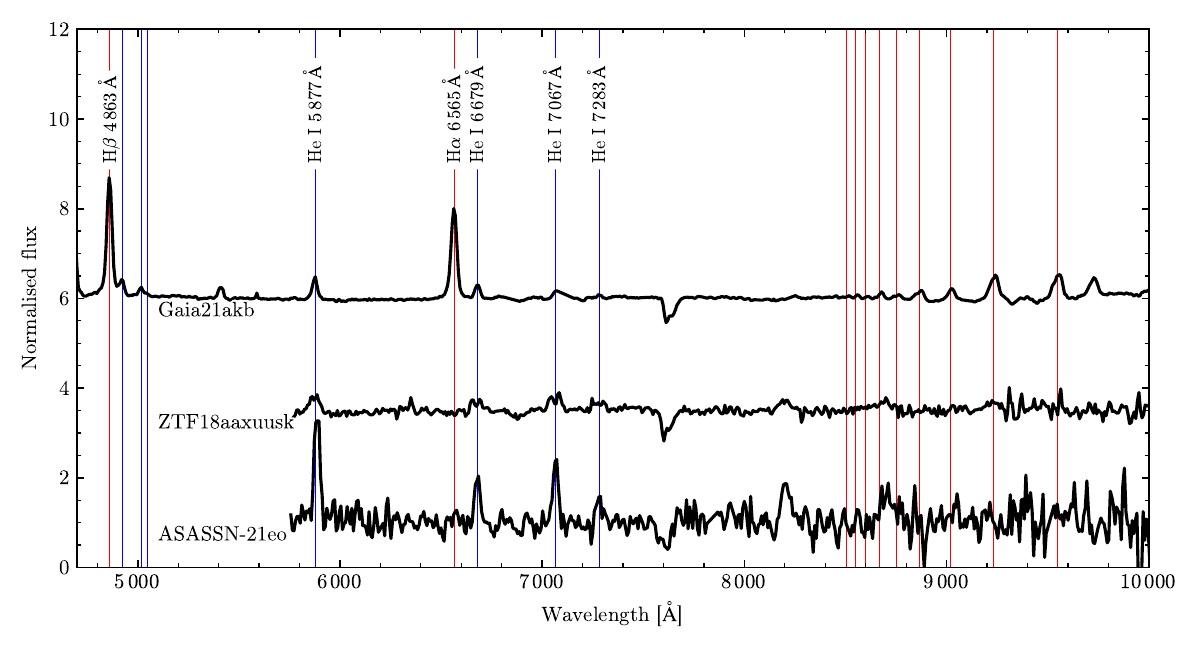}
    \caption{Spectra of Gaia21akb, ZTF18aaxuusk, and ASASSN-21eo obtained at the Gemini observatory. Prominent hydrogen and helium lines are marked by red and blue vertical lines, respectively. The spectra have been normalized to the continuum and vertically offset for clarity.}
    \label{fig:spectra}
\end{figure*}

Figure~\ref{fig:spectra} shows the normalised Gemini  spectra of three targets identified in this survey. The spectrum of \textit{ZTF18aaxuusk} is characterised by prominent Helium emission features, including He~I 5877~\AA, He~I 6678~\AA, He~I 7067~\AA, and He~I 7283~\AA. No Balmer emission is apparent. All Helium lines show double-peaked profiles, indicating a high inclination of this system. This Helium-dominated spectrum is consistent with an AM CVn system. 

In contrast, \textit{Gaia21akb} displays strong hydrogen emission, most notably H$\beta$ 4861~\AA\ and H$\alpha$ 6565~\AA, together with several He~I lines; all emission lines are single-peaked. The presence of prominent Balmer emission lines indicates that this object is consistent with a low inclination Hydrogen-rich CV. We calculated a line flux ratio He~I~5877/H$\alpha = 0.25$, which agrees with the typical values observed in CVs \citep{2012MNRAS.425.2548B, 2022ApJ...925L..22L}.

Although the spectrum of \textit{ASASSN-21eo} has a low $\mathrm{SNR = 5}$, it shows several single-peaked helium emission lines and no sign of hydrogen lines. 
The helium-dominated spectrum, together with the short photometric period, this strongly supports the AM~CVn classification.  

\subsection{Mass-Ratios} \label{Massratio}
We calculated the mass ratio for the AM~CVn system ASASSN-19ct, to help constrain the mass of the donor star. To estimate the mass ratio $q = M_2/M_1$, we used the superhump excess $\epsilon = (P_{\rm sh} - P_{\rm orb})/P_{\rm orb}$, and the updated empirical relations of \cite{2019MNRAS.486.5535M}, based on the earlier calibrations of \cite{2006MNRAS.373..484K, 2005PASP..117.1204P, 2001PASP..113..736P}. During a SO event, superhumps evolve through three distinct stages: Stage A is the start of the SO, where the superhumps period is long and stable, Stage B is the middle part of the SO, where the superhump period becomes shorter and unstable, and Stage C is the final stage of the SO, where the superhump period reaches its shortest length and becomes stable again \citep{2019MNRAS.486.5535M, 2013PASJ...65..115K}.  

For the AM~CVn system ASASSN-19ct our detected orbital period ($P_{\rm orb} = {30.96^{+0.003}_{-0.003}}$  min) is nearly identical to the superhump period reported by \cite{2026PASA...43...52K} ($P_{\rm SH}=30.94\pm 0.2$ min). The orbital and superhump periods are not expected to be identical; this similarity and the larger uncertainty likely arise because the measurement was derived only from the observed fraction of the SO plateau. Consequently, the resulting superhump excess $\epsilon$ would be consistent with zero and cannot be used to reliably constrain $q$. Instead, we rely on the superhump period reported by \cite{vsn-a-23109} ($P_{\rm SH}=31.2048\pm0.112$ min), detected toward the end of the SO, and we calculated $\epsilon = 0.008$. Following the approach of \cite{green_2026_18842123} and \cite{2026PASA...43...52K}, if we assume that the superhump period corresponds to stage C, then $q_{\rm C} = 0.05 \pm 0.02$. Using the precisely estimated primary mass of $M_1 = 0.81 \pm 0.05 M_{\odot}$ derived by \cite{van_Roestel_2021}, we estimate the donor mass to be $M_{2,{\rm C}} \approx  0.04 \pm 0.018 M_{\odot}$. This value is very similar to the one of the previously known AM~CVn system YZ LMi \citep[$P_{\rm orb} \sim 28.3$ min, $M_{2} \sim 0.035 \pm 0.003 M_{\odot}$;][]{van_Roestel_2021, 2011MNRAS.410.1113C}, and matches what is expected for a degenerate companion in an AM CVn. 

For comparison, we also estimated the mass of a Roche-lobe-filling, zero-temperature degenerate donor at the measured orbital period of ASASSN-19ct. Using Kepler's third law and Eggleton Roche-lobe relation \citep{1983ApJ...268..368E}, we estimated $M_{2} \sim 0.024 M_{\odot}$ and the corresponding mass ratio of $q_{\rm RLOF} \approx 0.029$. However, as noted by \cite{green_2026_18842123}, these superhump derived mass-ratios should be considered with caution, as the empirical relations are not well calibrated for helium-dominated accretion disks.

\section{Discussion} \label{sec:discussion}

\begin{table*}
    \centering
    \caption{Gaia DR3 properties of the accreting white dwarf binaries analysed in this work. Columns list the object name, type, Gaia DR3 source identifier, apparent magnitudes (G mag), $BP-RP$ colours and parallaxes taken from~\citet{2023A&A...674A..39G}, while the distance (pc) is obtain from~\citet{2021yCat.1352....0B}.}
    \label{tab:hr_table}
    \begin{tabular}{l c c c c c c c l}
    %\begin{tabular}{l c c c c c c c }
    \hline
    Name & Gaia DR3 ID & Type & Spec. Confirmed & Period & $G$ & ${BP}-{RP}$   &  Parallax  & Distance [pc] \\
         &             &     &      & (min) & (mag) & (mag)  &  & \\  
    \hline
    ASASSN-21in  & 1841455206549674624 & AM CVn & -- &   $29.73^{(a)}$ & 20.03 & 0.72  & $0.62 \pm 0.34$ & $2527^{+2840}_{-1138}$
 \\

    ZTF18aaxuusk & 1655973607896450944 & AM CVn & Y  & $30.94^{(a)}$ & 20.74 & 0.38   &  $1.86 \pm 0.89$
 & $1051^{+697}_{-383}$   \\ 

    ASASSN-19ct  & 5386114565161537920 & AM CVn & L & $30.96^{(b)}$ & 17.44 & 0.01  & $4.20 \pm 0.08$ & $238^{+5}_{-5}$  \\ 

    Gaia23asm    & 3461950864159780480 & AM CVn cand. & -- & $47.01^{(b)}$ & 19.35 & 0.73    & $0.64 \pm 0.31$ & $1858^{+1417}_{-632}$  \\

    ASASSN-14cn  & 1629388752470472704 & AM CVn & L  & $49.71^{(b)}$& 18.30 & 0.19   & $3.88 \pm 0.12$ & $258^{+8}_{-8}$ \\ 

    ASASSN-21eo  &       --            & AM CVn & Y & $52.42^{(b)}$ & -- &    --  &      --         &        --      \\

    Gaia23axx    & 99677532286825984   & AM CVn cand. & -- &  $53.46^{(b)}$ &20.44 & 0.33   & $0.98 \pm 0.89$ & $2885^{+958}_{-1499}$ \\ 

    Gaia21akb    & 2395305769240905600 & CV  & Y    & $75.71^{(b)}$  & 18.32 & 0.64  & $2.77 \pm 0.12$ & $358^{+13}_{-12}$  \\

    Gaia19ekt   & 2884625449040056704 &  CV cand. & -- & $82.32^{(b)}$ & 19.08  & 0.27     & $3.87 \pm 0.14$ & $259^{+9}_{-11}$   \\

    ASASSN-18rd  &       --            & CV     & -- &  $82.75^{(a)}$   & -- &  --     &      --         &     --         \\
       
    ASASSN-15ev  & 5194659693498442112 & CV    & --  & $83.48^{(a)}$    & 19.50 & 0.68 &             $2.17 \pm 0.21$ & $464^{+54}_{-39}$  \\

    \hline
    \end{tabular} 
     \caption*{Note: $^{(a)}$ superhump period; $^{(b)}$ orbital period. Column "Spec. Conf." indicates spectroscopic confirmation status: Y = confirmed Gemini follow-up in this work (Section~\ref{spectralresults}); L = confirmed spectroscopically in the literature \citep[ASASSN-19ct;][]{2026PASA...43...52K} and \citep[ASASSN-14cn;][]{2015MNRAS.452.1060C}}.
\end{table*} 

\subsection{Colour-Magnitude Diagram} \label{HRD}
\begin{figure*}
    \centering
    \includegraphics[width=0.90\linewidth]{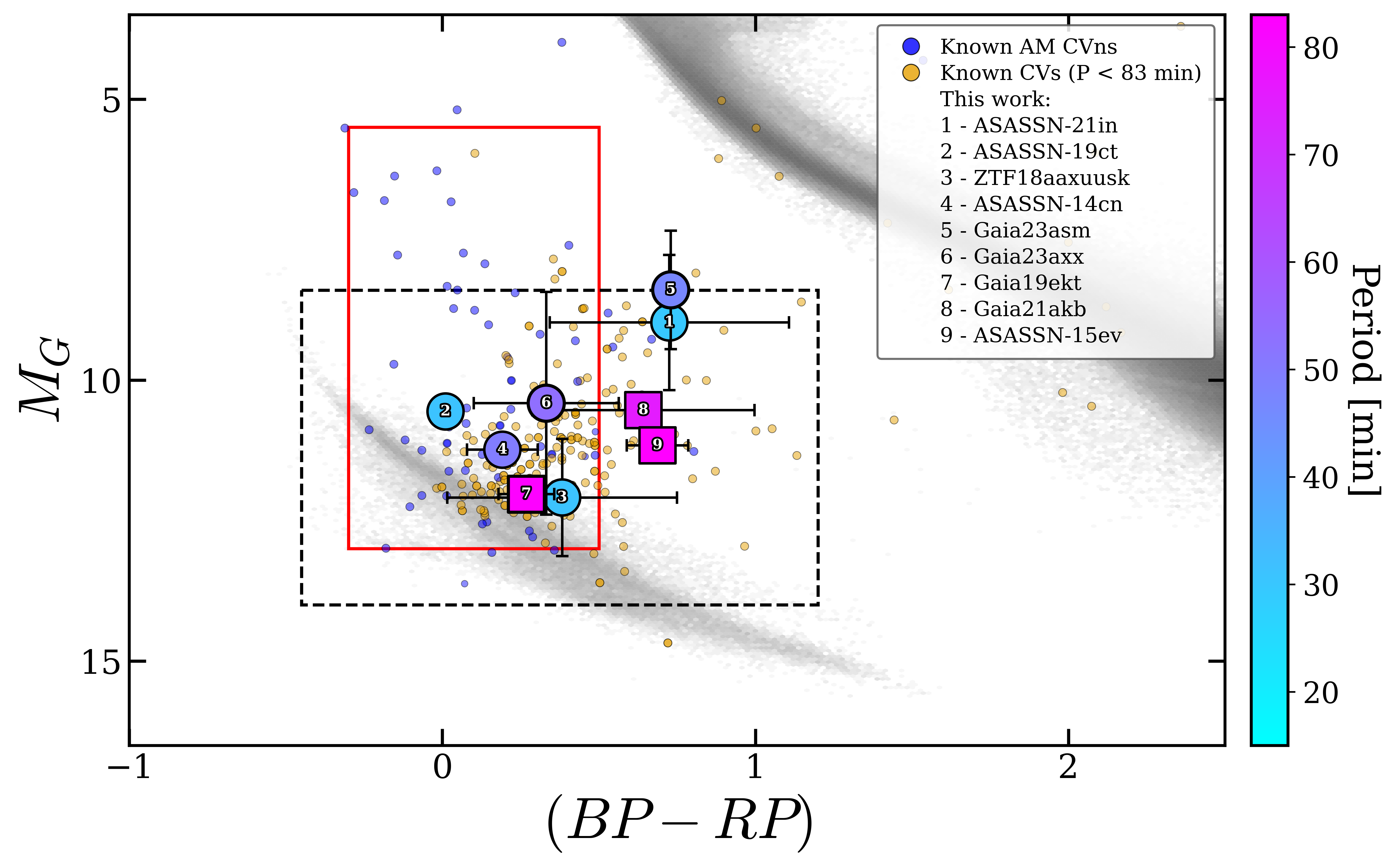}
    \caption{CMD of the systems identified in this work compared to known AM~CVns (blue circles), period bouncer and short-period CVs ($P < 83$ min; gold circles). Along with the Gaia DR3 source distribution within 200 pc (grey background). The red rectangle encloses the region occupied by 89\% of known AM CVn systems \citep{2026PASA...43...52K}. The black dashed rectangle marks the period bouncers and CVs near the minimum period region defined by \citet{2026ApJ...998..153X}. Error bars include photometric and distance uncertainties propagated from Gaia parallaxes. Nine of the 11 systems are shown; AM~CVns are represented by circles and CVs by squares. ASASSN-21eo and ASASSN-18rd are excluded as they were not detected in Gaia DR3.}
    \label{fig:HR}
\end{figure*}

The Gaia BP-RP versus G colour magnitude diagram (CMD, Figure~\ref{fig:HR})  
provides complementary information to classify and characterize our targets. 
We cross-match our objects with the Gaia DR3 catalogue using a 2-arcsec search radius. However, ASASSN-21eo and ASASSN-18rd were not found in the Gaia catalogue, likely due to their faintness during quiescence and therefore are not displayed on the figure. The CMD was created using the Gaia properties summarised in Table~\ref{tab:hr_table}. For each source, the absolute magnitude was derived from the G-band magnitude and parallax measurements using the standard distance modulus. The Gaia DR3 extinction parameters are unavailable for our 11 sources in our sample. Therefore, no reddening correction was applied to the Gaia CMD, and no parallax quality cuts were applied to retain the full sample. However, several objects with a low parallax signal-to-noise ratio (S/N $<3$) have large distance uncertainties, which propagate into significant uncertainties in their absolute magnitudes. 

For comparison, we included the known AM~CVn systems from \citet{green_2026_18842123} and \citet{2026PASA...43...52K}, which typically appear bluer and closer to the white dwarf sequence than the general CV population. We also included known period bouncers and CVs near the orbital period minimum ($P_{orb}< 83$ min) from \citet{Mu_oz_Giraldo_2024} and \citet{2003A&A...404..301R}. These systems generally occupy the region between the main sequence and the white dwarf sequence.   

Among the confirmed AM CVn systems, ASASSN-19ct and ASASSN-14cn fall within the WD cooling sequence within the region occupied by 89\% of known AM CVns \citep{2026PASA...43...52K} as illustrated in Figure~\ref{fig:HR}. The remaining confirmed and candidate systems required more careful interpretation in light of their parallax uncertainties. ASASSN-21in, ZTF18aaxuusk, Gaia23asm, and Gaia23axx all have parallax $S/N < 3$, propagating into $M_{G}$ uncertainties of $>$ 2 mag, making their CMD position unreliable. However, despite their large uncertainties, ZTF18aaxuusk and Gaia23asm remain consistent with the region occupied by known AM CVn systems. Gaia23asm and Gaia23axx (quiescence) remain as AM~CVn candidates due to their poorly constrained distances (Table~\ref{tab:hr_table}) which prevent a reliable distinction between AM~CVns and CVs. We note that the system Gaia23asm is much redder than other AM CVns with similar periods, which could support the interpretation of a short period CV. Because of that, specifically for this system we looked for archival X-ray information in order to obtain additional hints for a more accurate classification but no information was available. 

As shown in Figure~\ref{fig:HR}, the CV candidate Gaia19ekt lies along the WD cooling sequence within the period bouncer region (black rectangle) defined by \citet{2026ApJ...998..153X}, which encompasses all known period bouncers and high-probability candidates. Its reliable parallax ($S/N =27$) means that its position in the CMD is genuinely meaningful rather than an artifact of distance uncertainty. The confirmed CVs Gaia21akb and ASASSN-15ev also fall within this region, consistent with their short periods near the CV minimum period. This placements closely aligns with the evolutionary framework proposed by \cite{2020arXiv201112253A}, which demonstrates that as CVs evolve toward the period minimum, the donor star dims significantly, causing the system to migrate closer to the white dwarf sequence as the primary WD dominates the system's colour.

\subsection{Colour-Colour Diagram}
\begin{figure}[H]
    \centering
    \includegraphics[width=\columnwidth]{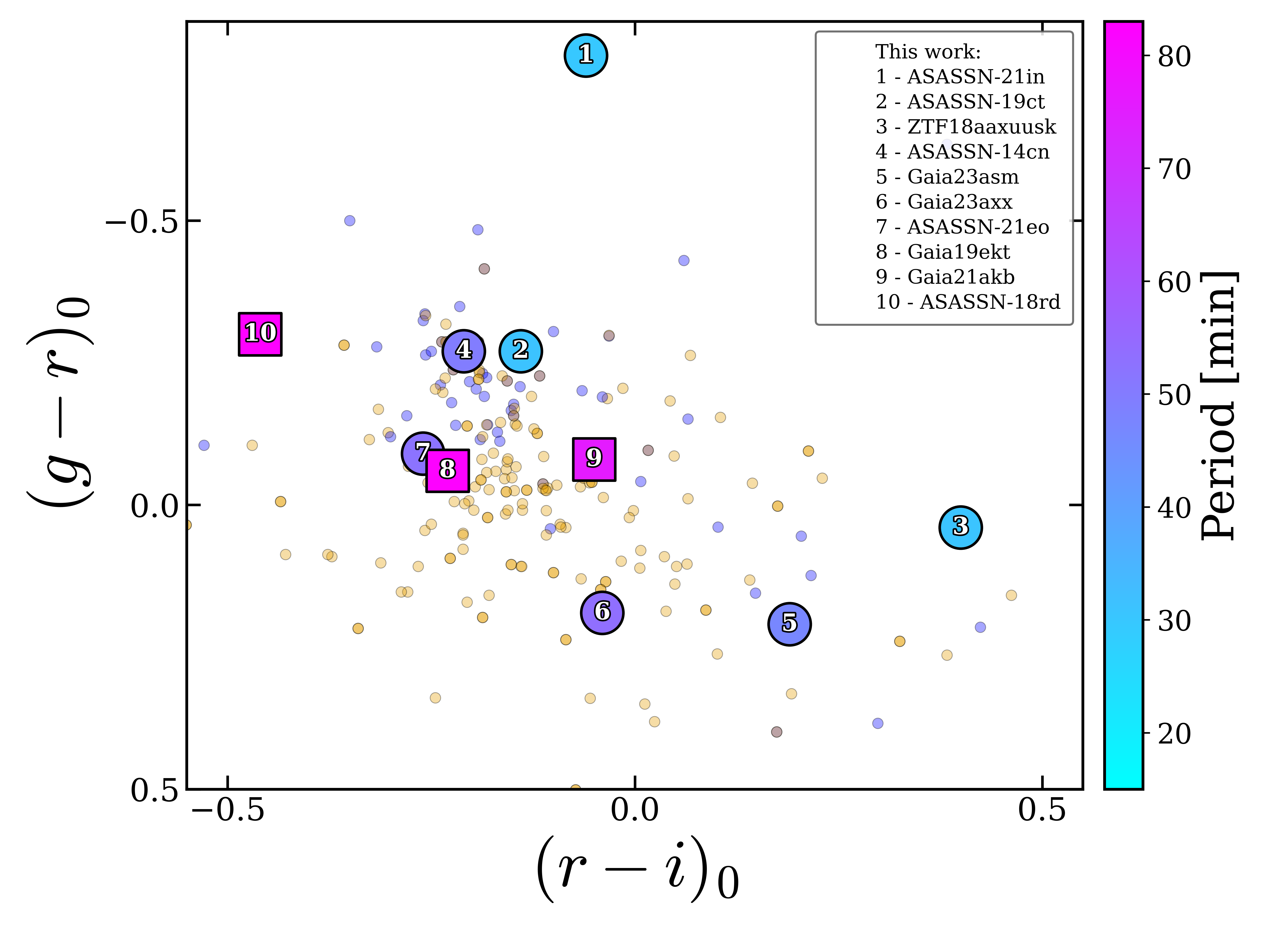}
    \caption{The numbers and symbols in this figure correspond to the same sample shown in Figure~\ref{fig:HR}. ASASSN-15ev is excluded due to incomplete filter coverage, while ASASSN-21eo~(7) and ASASSN-18rd~(10) are included here. The background markers indicate known AM~CVns (blue) and known CVs/period bouncers (gold), from the catalogues discuss in Section~\ref{HRD}.} 
    \label{fig:colorcolor}
\end{figure}

The colour-colour diagrams (CCD) shown in Figure~\ref{fig:colorcolor} compares our objects with the same sample of known period bouncers, CVs near the minimum period, and AM~CVn systems presented in Section~\ref{HRD} using their $g$, $r$, and $i$ photometry retrieved from SDSS. In our sample, all colours were corrected for interstellar reddening using the coefficients of \citet{2011ApJ...737..103S} to derive the intrinsic colours $(g-r)_0$ and $(r-i)_0$, where the subscript $0$ indicates reddening corrected, measured during quiescence. ASASSN-15ev is excluded from the CCD due to incomplete multi-band filter coverage; high-speed observations were taken unfiltered due to its faintness in quiescence ($r \approx 18.0\text{-}20.3$~mag; \citealt{2019MNRAS.486.2422P}). 

In Figure~\ref{fig:colorcolor}, the majority of our systems fall in the central region of the diagram, where the known AM~CVns (blue circles) and short period CVs (gold circles; $P_{orb} < 83$ min) overlap. For example, AM CVn ASASSN-21eo~(7)  lies among the known AM~CVns central region. The AM CVn object ASASSN-21in~(1) is the bluest in $(g-r)_0$, and lies well above the main distribution, while the CV ASASSN-18rd~(10) is the bluest in $(r-i)_0$ sitting out on the left edge, away from the known short CVs. The confirmed AM~CVn ZTF18aaxuusk~(3) instead lies in the region populated by a mix of CVs and AM~CVns; despite this photometric ambiguity, spectroscopy confirms its classifications. In contrast, our CV candidate Gaia19ekt~(8), and confirmed CV Gaia21akb~(9) sit in the central region of the CV near the period minimum, consistently with our classification. 

\subsection{Distribution and biases} \label{se}
\begin{figure}[H]
    \centering
    \includegraphics[width=\columnwidth]{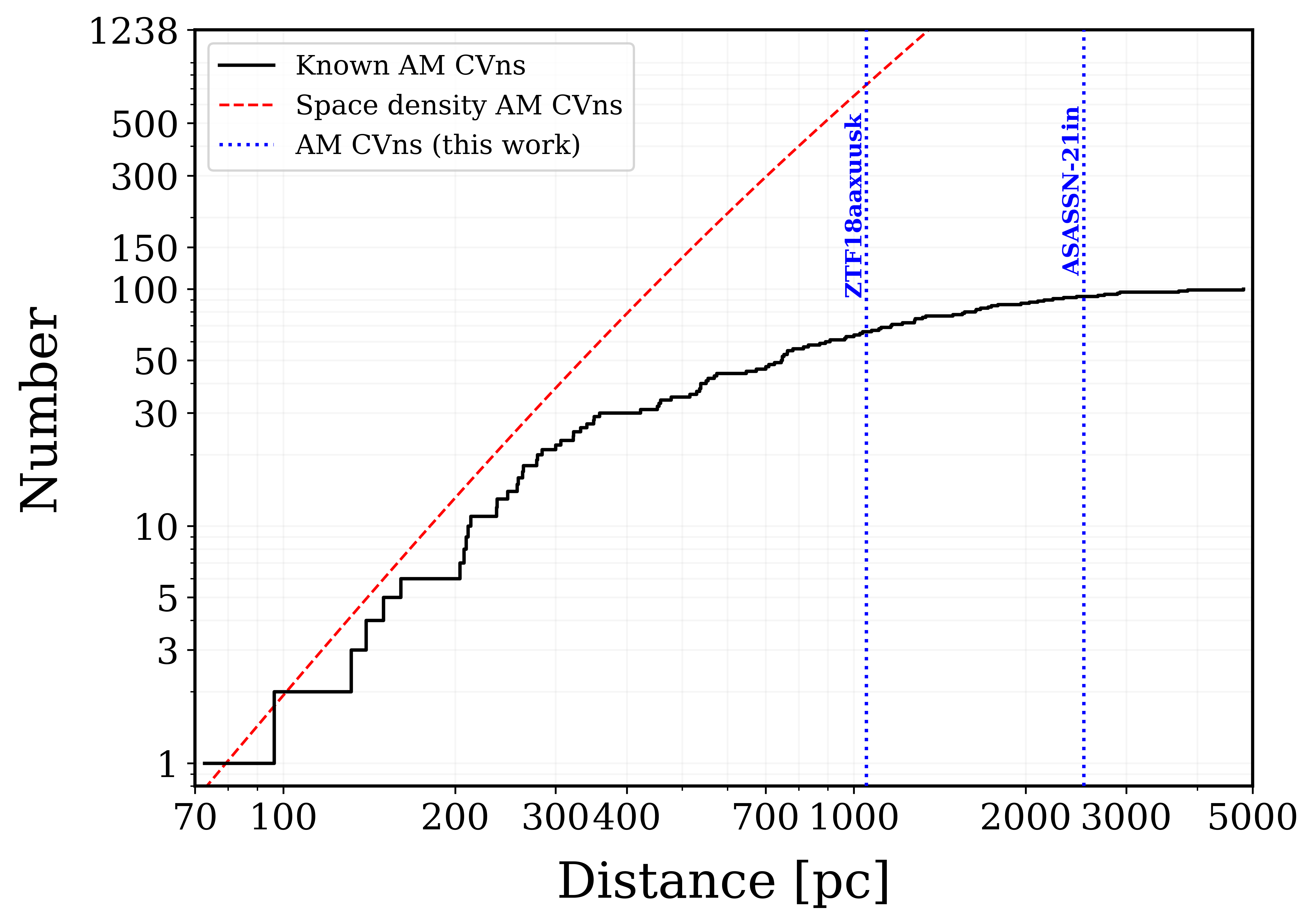}
    \caption{Cumulative distance distribution of known AM~CVns \citep{green_2026_18842123, 2026PASA...43...52K} within 5000 pc (black solid line) compared to the empirically derived local AM~CVn space density of \cite{2025PASP..137a4201R}, assuming an exponential disk model (red dotted line). The two blue vertical lines represent the confirmed AM~CVn systems identified in this work.}
    \label{fig:Space Density}
\end{figure}

Figure~\ref{fig:Space Density} shows an updated cumulative distance distribution of known AM~CVns \citep{green_2026_18842123, 2026PASA...43...52K} compared to the spatial distribution of an empirically-calibrated exponential Galactic disk model, adopting a space density of $\rho_0 = 5.5 \times 10^{-7} \text{ pc}^{-3}$ and a scale height of $h=205$ pc \citep{2025PASP..137a4201R}. The observed AM CVn distribution closely follows the spatial distribution within $\sim$300 pc, but it falls significantly below it at larger distances, indicating that the known AM~CVn population is substantially incomplete beyond this limit. This could be a combination of observational biases and/or an intrinsically lower space density at larger distances.

It is important to note, however, that the discrepancy between observations and models is not unique to AM~CVns. Current models predict that between $40\%$ to $80\%$ of the CVs should be period bouncers \citep{2020MNRAS.491.5717B}, but, despite recent multi-wavelength surveys \citep[e.g.][]{2026ApJ...998..153X, Mu_oz_Giraldo_2024}, the large majority of period bouncers is still missing. In fact, of the 2,054 CV candidates studied in our survey, 128 accreting white dwarf binaries were observed in outburst (6.23\% of the total sample). Of these, we report only four systems in outburst and seven in quiescence below the $P < 83$~min with sufficient S/N. These short period outbursting systems correspond to 0.195\% of the total sample.  

However, it is important to consider the observational biases in our survey which can help understand these results. The first bias arises from the technical limitations of TESS itself, including its shallow magnitude detection limit. The TESS limiting magnitude is typically $T_{mag} \lesssim 16$, although sources as faint as but reaching  $T_{mag} \sim 18–19$ can be detected under very favourable conditions, such as low background levels, uncrowded fields and sufficient SNR ratio. Short period accreting white dwarfs have small mass transfer rates making them intrinsically faint in quiescence, typically above 19 mags \citep{green_2026_18842123, Mu_oz_Giraldo_2024}, well below TESS limit. Therefore, if some of these systems showed a small amplitude outburst, they would still be challenging to detect with TESS and can be confused with background noise. This naturally leads to a bias in the detection of systems that show superoutbursts whose amplitudes can reach 8 mags \citep{2026PASA...43...52K}. The large pixel size of the TESS camera (21-arcsec) causes flux contamination from nearby stars, as evidenced by the neighbouring sources identified for several of our candidates. 

A second bias comes from the mismatch between the outburst recurrence timescale of these systems and the TESS observation window (see Section~\ref{sec:recurrence}). For AM CVn systems with orbital periods $>$ 40 min, the DIM predicts recurrence timescales $>$ 1 yr \citep{2015ApJ...803...19C}, implying a very low probability of capturing an outburst within a single TESS sector of $\sim$27.4 d. This interpretation is consistent with the results of \citet{2021MNRAS.508.3275P}, whose observations showed that it is difficult to detect a SO during a limited observation window. This suggests that a significant fraction of AM CVn systems in our sample may have been observed by TESS in quiescence, with their outbursts occurring undetected between TESS sectors. The probability of detecting the systems during outburst could be quantified by considering their outburst duration and recurrence timescale together with available TESS coverage \citep[e.g.][]{2015MNRAS.446..391L}.

On the other hand, the discrepancy may also indicate that binary population synthesis models systematically overestimate the formation efficiency or lifetimes of these long-lived evolutionary phases. Uncertainties in common-envelope evolution, angular momentum loss prescriptions, donor star evolution, and the stability of mass transfer can all significantly affect the predicted present-day space density of AM~CVns and short period CVs. Distinguishing between observational incompleteness and shortcomings in the evolutionary models therefore requires substantially more complete, volume-limited samples. The Vera C. Rubin Observatory Legacy Survey of Space and Time (LSST) will repeatedly image the southern sky with unprecedented depth, enabling the discovery of faint AM CVn binaries and short period CVs over much larger Galactic volumes than previously possible \citep{buckley2025discoveringcataclysmicvariablesrubin}. In parallel, upcoming gravitational-wave observations by LISA will provide an independent census of ultracompact binaries, including many AM CVn systems that are electromagnetically faint or obscured. The combined electromagnetic and gravitational-wave samples will offer powerful constraints on AM CVn formation channels, binary evolution models, and the true Galactic space density \citep[e.g.,][]{2026arXiv260505308K}.

\section{Conclusion} \label{sec:end} 
In the first phase of COPAS, a survey to study accreting white dwarfs through their outbursts and periods, we analyse 2,054 candidate accreting white dwarf binaries observed during TESS Cycle 6 and 7. By combining TESS photometry with long-term monitoring from ZTF and ATLAS, together with Gemini spectroscopy, we identified and characterised 11 short-period accreting white dwarf binaries below $P < 83 \text{ min}$. The primary findings of the COPAS survey at this stage are summarised as follows: 

\begin{itemize}

    \item We reclassified 5 cataclysmic variables as AM CVns systems, demonstrating the relevance of continuous monitoring. 

    \item We detected superhump periods for two AM CVns in outburst (ASASSN-21in, $P_{\rm SH} = 29.73$~min; ZTF18aaxuusk, $P_{\rm SH} = 30.94$~min). In quiescence, we extracted orbital periods for three confirmed AM~CVns (ASASSN-19ct, $P_{\rm orb} = 30.96$~min; ASASSN-14cn, $P_{\rm orb} = 49.71$~min; ASASSN-21eo, $P_{\rm orb} = 52.42$~min) and two AM~CVn candidates (Gaia23asm, $P_{\rm orb} = 47.01$~min; Gaia23axx, $P_{\rm orb} = 53.46$~min). 
    
    \item We measured superhump periods for two CVs during outburst plateau (ASASSN-18rd, $P_{\rm SH} = 82.75$~min; ASASSN-15ev, $P_{\rm SH} = 83.47$~min). We also determined the orbital periods for Gaia21akb ($P_{\rm orb} = 75.71$~min), confirming it as a hydrogen-rich CV sitting at the theoretical period minimum in a high state and a CV candidate in quiescence Gaia19ekt ($P_{\rm orb} = 82.32$~min).
    
    \item We obtained Gemini follow-up spectroscopy for three objects. ZTF18aaxuusk displays prominent double-peaked helium emission lines with no strong Balmer emission, confirming it as a high-inclination AM~CVn. Despite its lower signal-to-noise ratio, ASASSN-21eo also shows helium emission without hydrogen, supporting its AM~CVn classification. In contrast, Gaia21akb exhibits strong H$\alpha$ and H$\beta$ emission, definitively identifying it as a hydrogen-rich CV.

    \item Combining high-cadence TESS photometry with long-term ZTF and ATLAS light curves, we characterized the superoutburst evolution and recurrence times of four systems. The AM~CVns ASASSN-21in and ZTF18aaxuusk exhibit nearly identical outburst profiles with six rebrightenings and recurrence times of $\sim$1 year. The short-period CVs ASASSN-18rd and ASASSN-15ev show longer superoutbursts (20--28 days) with recurrence times of 1.85--1.96 years, placing them at the short end of the WZ Sge recurrence-time distribution.
    
    \item Using superhump excess measurements, we estimated the mass ratio and donor mass of ASASSN-19ct. We estimated a superhump Stage C mass ratio of $q_C = 0.05 \pm 0.02$ and a heavily degenerate donor mass of $M_{2,C} \approx 0.04 \pm 0.018 M_\odot$. 

\end{itemize}

\section{Acknowledgments}
WM, LRS, JK and RJO acknowledge support from NASA grants NNH22ZDA001N-6152 and 80NSSC24K0638. MPM is partially supported by the Swiss National Science Foundation IZSTZ0\_216537 and by UNAM PAPIIT-IG101224. YC acknowledges support from the grant RYC2021-032718-I, funded by MCIN/AEI/10.13039/501100011033 and the European Union NextGenerationEU/PRTR. This paper includes data collected with the TESS mission, obtained from the MAST data archive at the Space Telescope Science Institute (STScI). Funding for the TESS mission is provided by the NASA Explorer Program. STScI is operated by the Association of Universities for Research in Astronomy, Inc., under NASA contract NAS 5–26555. Based on observations obtained at the international Gemini Observatory, a program of NSF NOIRLab, which is managed by the Association of Universities for Research in Astronomy (AURA) under a cooperative agreement with the U.S. National Science Foundation on behalf of the Gemini Observatory partnership: the U.S. National Science Foundation (United States), National Research Council (Canada), Agencia Nacional de Investigaci\'{o}n y Desarrollo (Chile), Ministerio de Ciencia, Tecnolog\'{i}a e Innovaci\'{o}n (Argentina), Minist\'{e}rio da Ci\^{e}ncia, Tecnologia, Inova\c{c}\~{o}es e Comunica\c{c}\~{o}es (Brazil), and Korea Astronomy and Space Science Institute (Republic of Korea). Based on observations obtained with the Samuel Oschin 48-inch Telescope at the Palomar Observatory as part of the Zwicky Transient Facility project. ZTF is supported by the National Science Foundation under Grant No. AST-1440341 and a collaboration including Caltech, IPAC, the Weizmann Institute for Science, the Oskar Klein Center at Stockholm University, the University of Maryland, the University of Washington, Deutsches Elektronen-Synchrotron and Humboldt University, Los Alamos National Laboratories, the TANGO Consortium of Taiwan, the University of Wisconsin at Milwaukee, and Lawrence Berkeley
National Laboratories. Operations are conducted by COO, IPAC, and UW. The national facility capability for SkyMapper has been funded through ARC LIEF grant LE130100104 from the Australian Research Council, awarded to the University of Sydney, the Australian National University, Swinburne University of Technology, the University of Queensland, the University of Western Australia, the University of Melbourne, Curtin University of Technology, Monash University and the Australian Astronomical Observatory. SkyMapper is owned and operated by The Australian National University's Research School of Astronomy and Astrophysics. The survey data were processed and provided by the SkyMapper Team at ANU. The SkyMapper node of the All-Sky Virtual Observatory (ASVO) is hosted at the National Computational Infrastructure (NCI). Development and support of the SkyMapper node of the ASVO has been funded in part by Astronomy Australia Limited (AAL) and the Australian Government through the Commonwealth's Education Investment Fund (EIF) and National Collaborative Research Infrastructure Strategy (NCRIS), particularly the National eResearch Collaboration Tools and Resources (NeCTAR) and the Australian National Data Service Projects (ANDS). The Pan-STARRS1 Surveys (PS1) and the PS1 public science archive have been made possible through contributions by the Institute for Astronomy, the University of Hawaii, the Pan-STARRS Project Office, the Max-Planck Society and its participating institutes, the Max Planck Institute for Astronomy, Heidelberg and the Max Planck Institute for Extraterrestrial Physics, Garching, The Johns Hopkins University, Durham University, the University of Edinburgh, the Queen's University Belfast, the Harvard-Smithsonian Center for Astrophysics, the Las Cumbres Observatory Global Telescope Network Incorporated, the National Central University of Taiwan, the Space Telescope Science Institute, the National Aeronautics and Space Administration under Grant No. NNX08AR22G issued through the Planetary Science Division of the NASA Science Mission Directorate, the National Science Foundation Grant No. AST–1238877, the University of Maryland, Eotvos Lorand University (ELTE), the Los Alamos National Laboratory, and the Gordon and Betty Moore Foundation. 

\bibliographystyle{aa_link}
\bibliography{References.bib}

\appendix
\renewcommand{\thefigure}{\thesection.\arabic{figure}}
\renewcommand{\thetable}{\thesection.\arabic{table}}

%\pagebreak
\section{Periodograms and Phase-Folded Light Curves} \label{Appendix:A}
In this appendix, we present the periodograms and the phase-folded light curves for the 11 systems discussed in Section~\ref{sec:results} (see also Table~\ref{tab:table1}).

\begin{figure*}[h]
    \centering
    \setlength{\tabcolsep}{1pt}
    \renewcommand{\arraystretch}{1}
    
    \begin{tabular}{ccc}
        % Row 1
        \begin{subfigure}{0.33\textwidth}
            \centering
            \includegraphics[width=\linewidth]{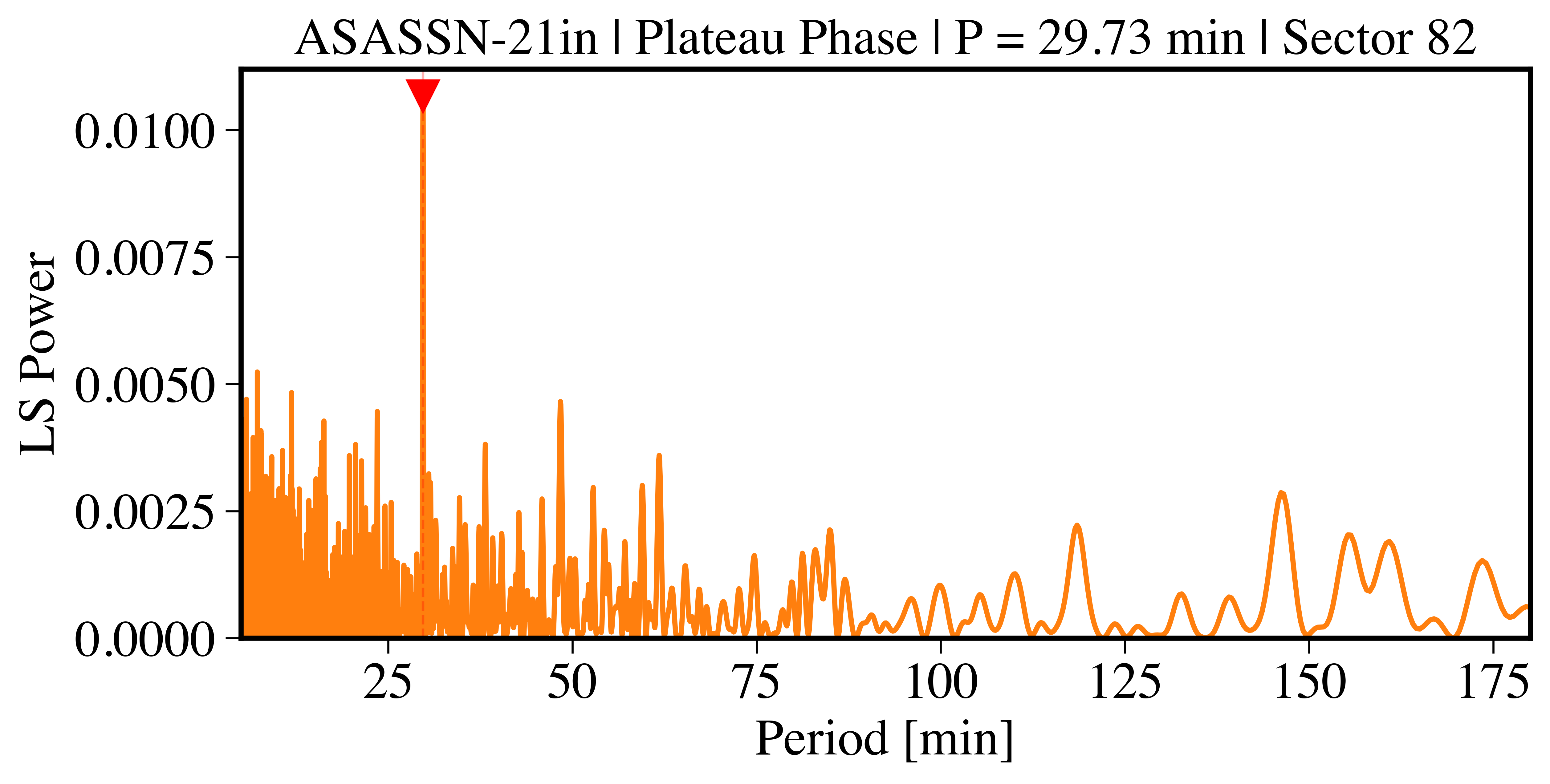}
            \caption{}
        \end{subfigure} &
        \begin{subfigure}{0.33\textwidth}
            \centering
            \includegraphics[width=\linewidth]{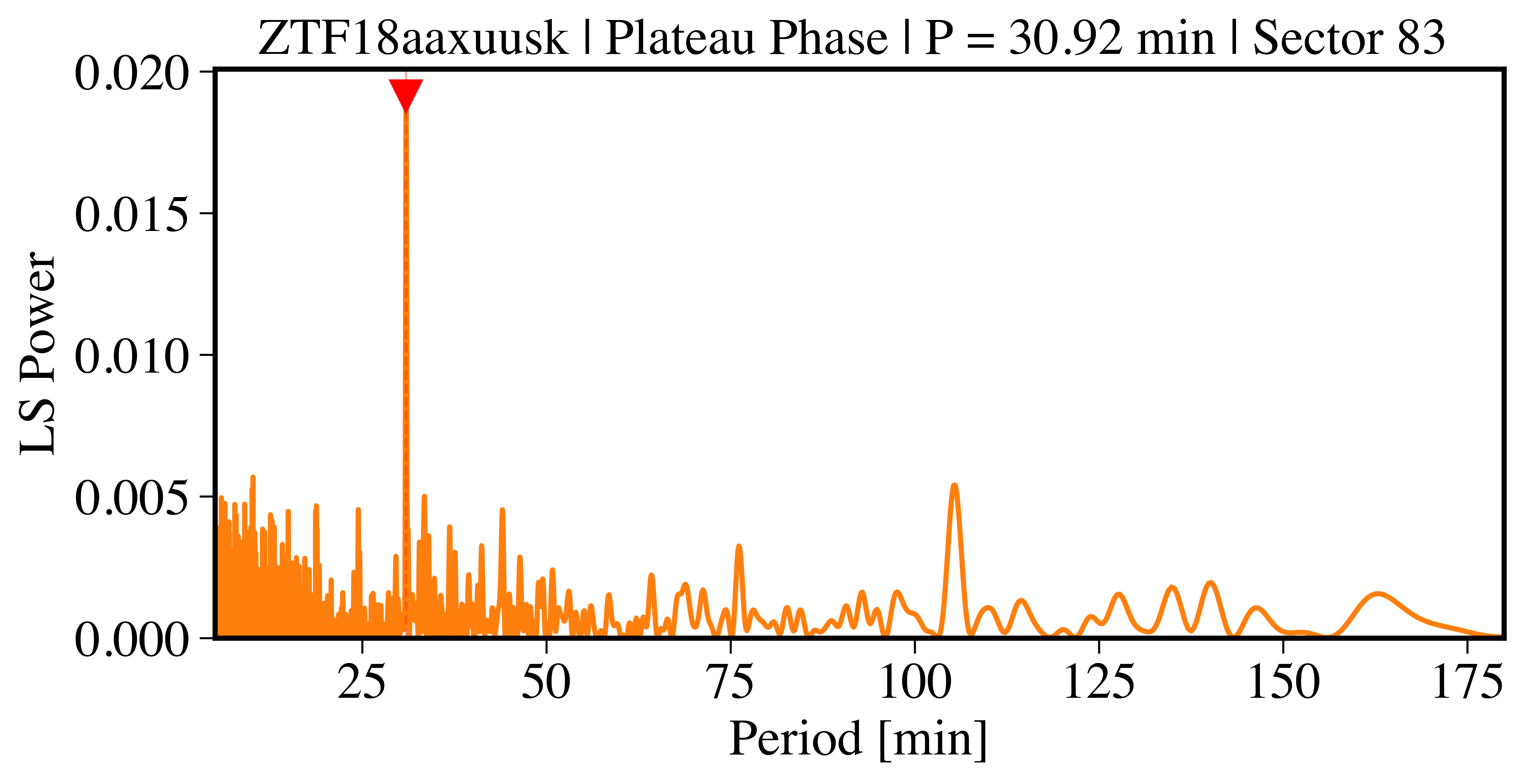}
            \caption{}
        \end{subfigure} &
        \begin{subfigure}{0.33\textwidth}
            \centering
            \includegraphics[width=\linewidth]{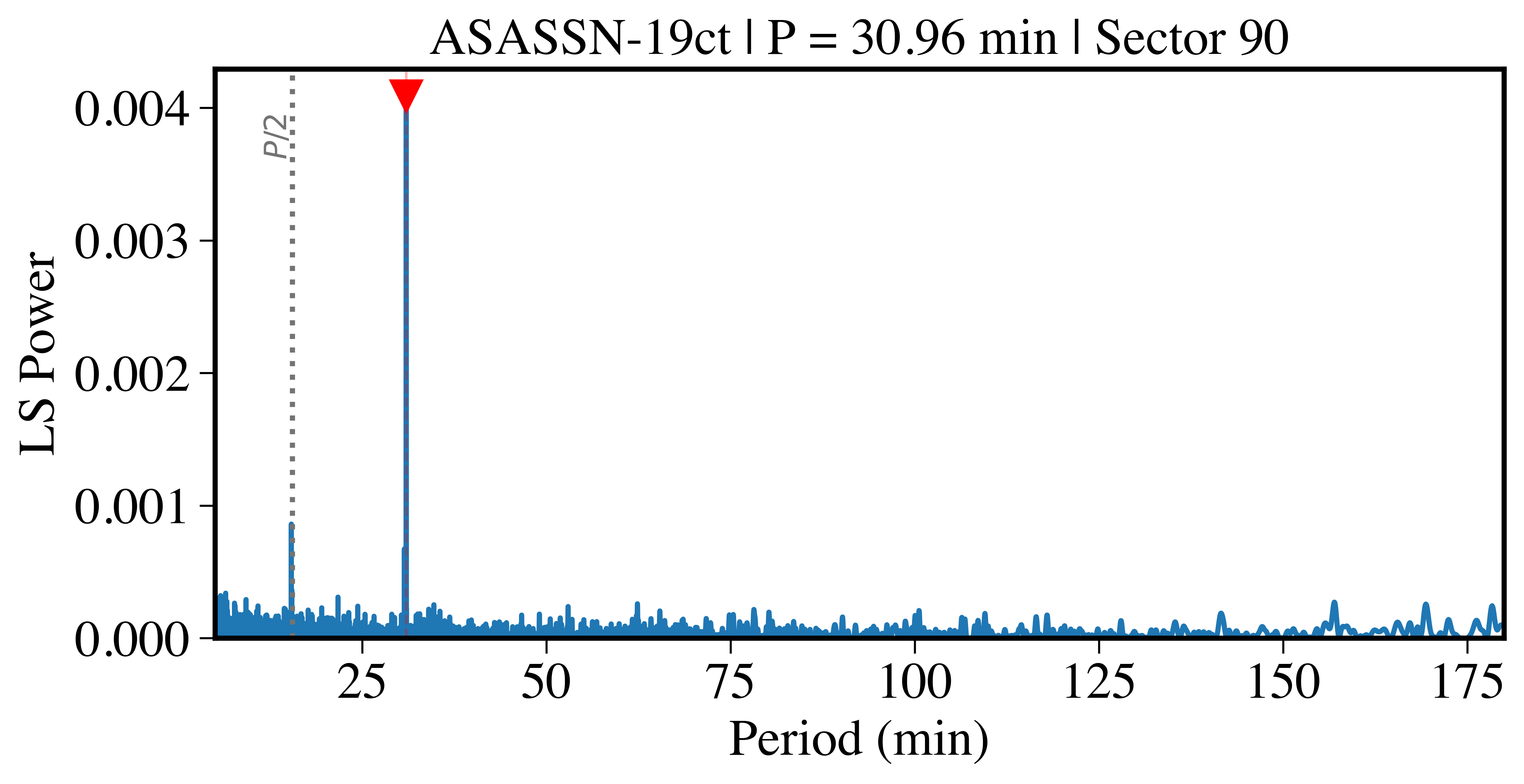}
            \caption{}
        \end{subfigure} \\
        
        % Row 2
        \begin{subfigure}{0.33\textwidth}
            \centering
            \includegraphics[width=\linewidth]{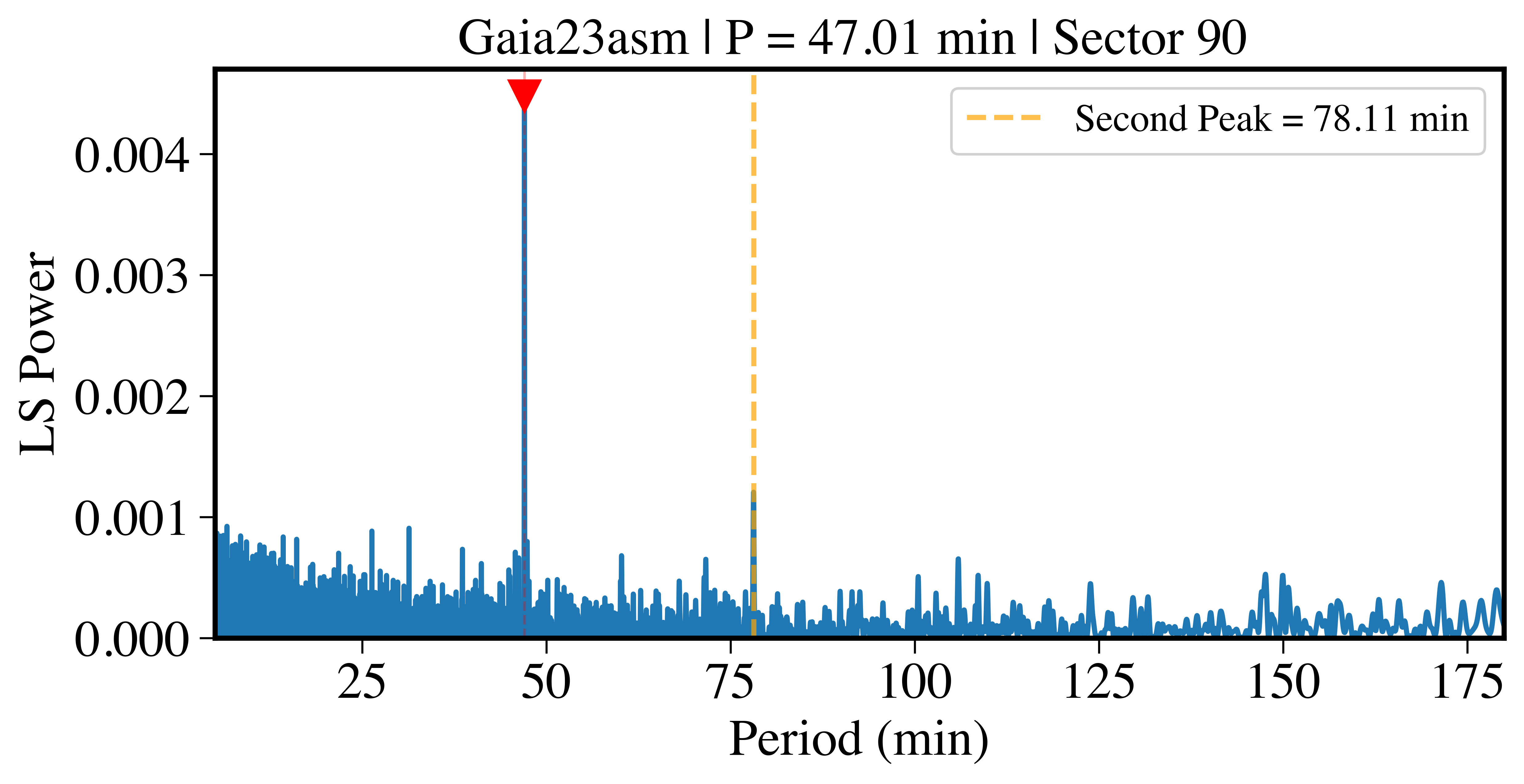}
            \caption{}
        \end{subfigure} &
        \begin{subfigure}{0.33\textwidth}
            \centering
            \includegraphics[width=\linewidth]{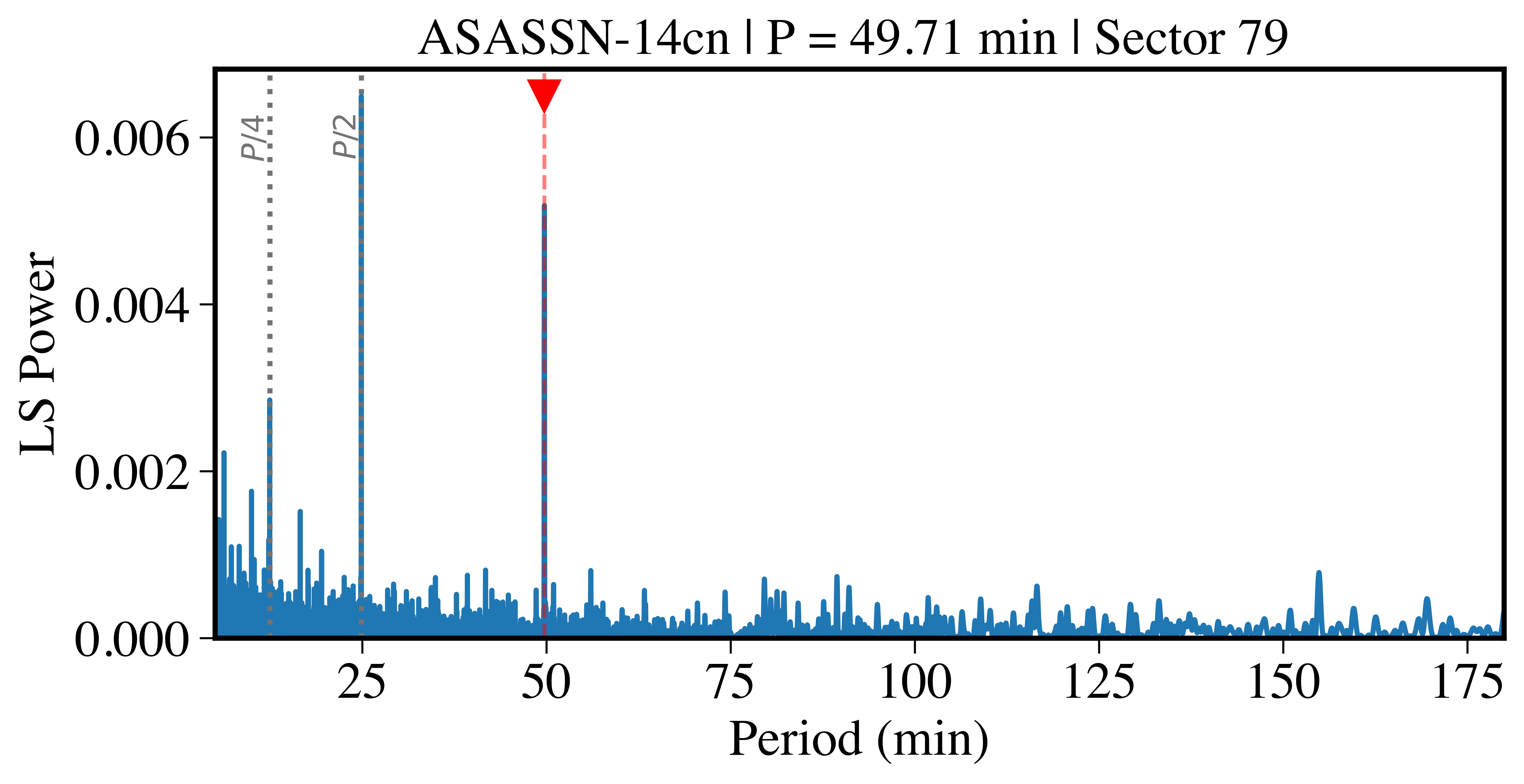}
            \caption{}
        \end{subfigure} &
        \begin{subfigure}{0.33\textwidth}
            \centering
            \includegraphics[width=\linewidth]{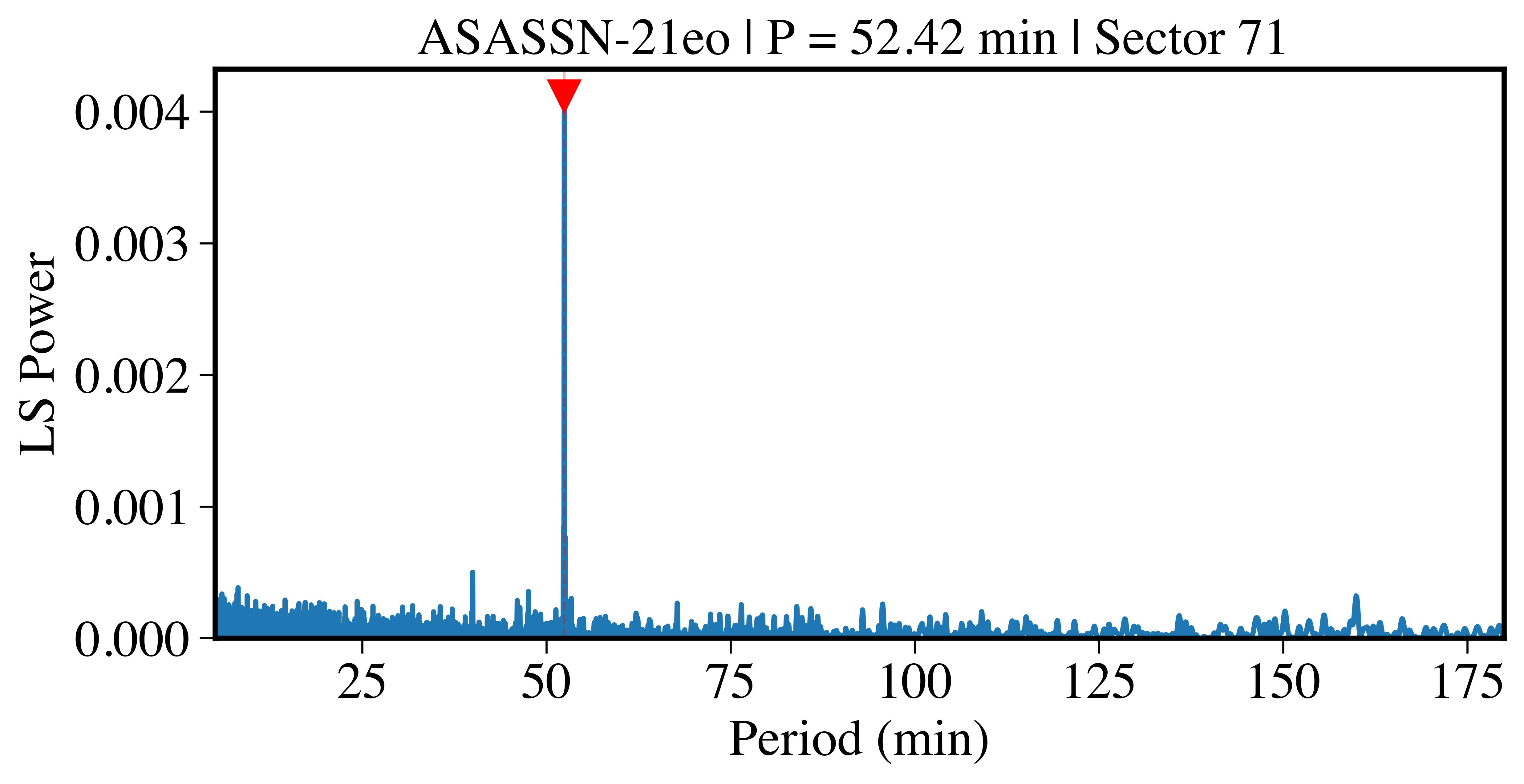}
            \caption{}
        \end{subfigure} \\
        
        % Row 3
        \begin{subfigure}{0.33\textwidth}
            \centering
            \includegraphics[width=\linewidth]{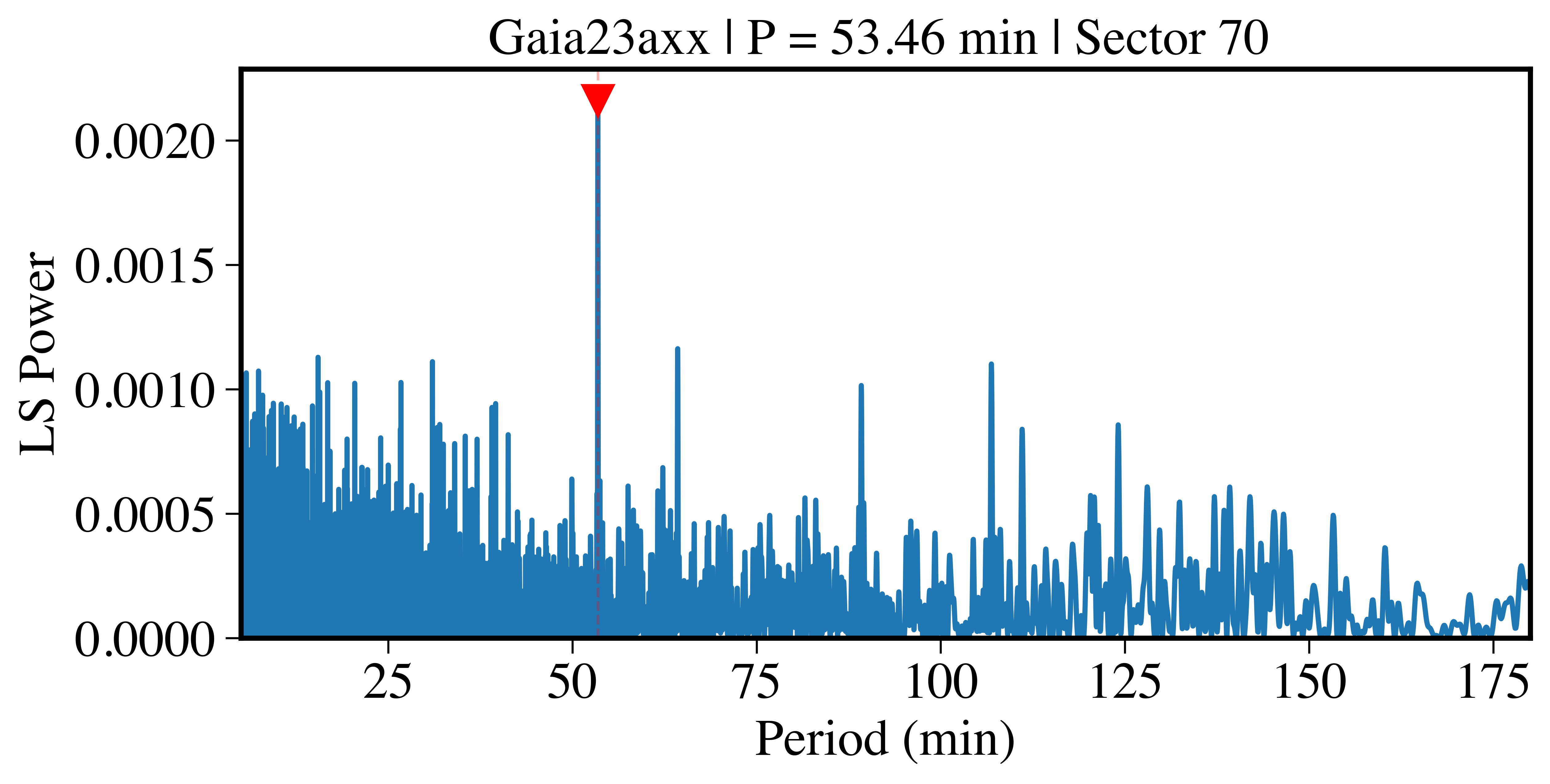}
            \caption{}
        \end{subfigure} &
        \begin{subfigure}{0.33\textwidth}
            \centering
            \includegraphics[width=\linewidth]{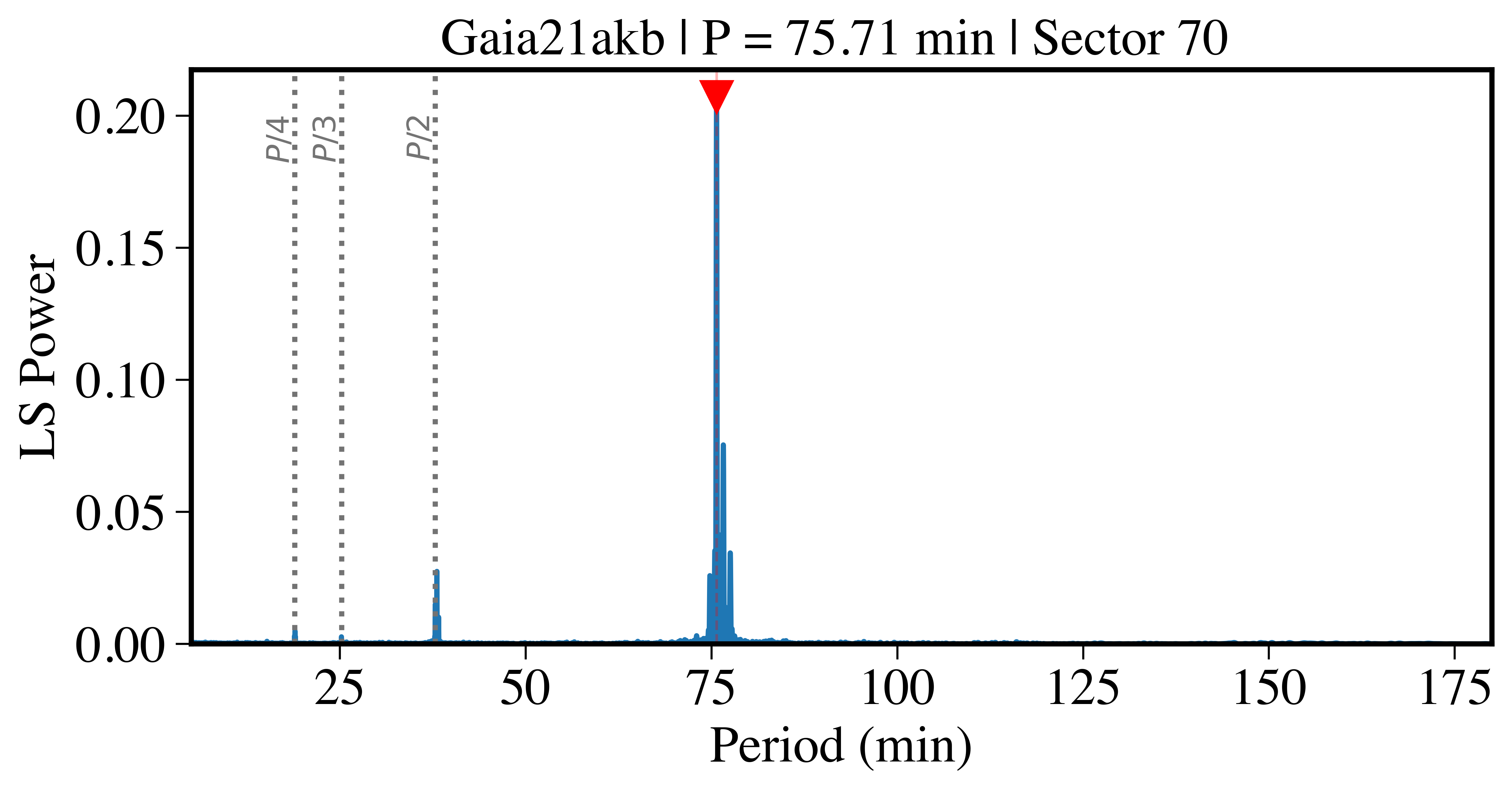}
            \caption{}
        \end{subfigure} &
        \begin{subfigure}{0.33\textwidth}
            \centering
            \includegraphics[width=\linewidth]{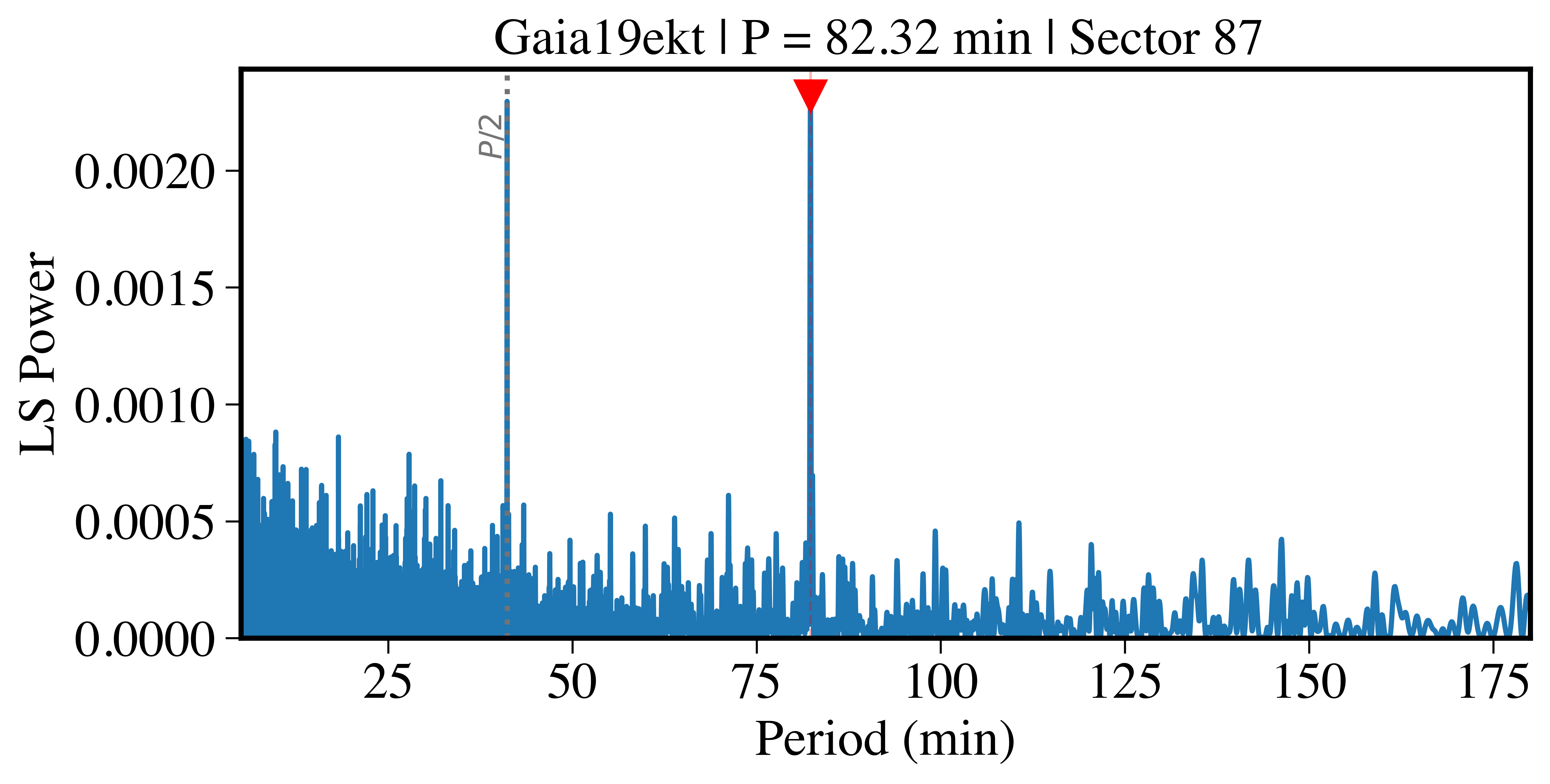}
            \caption{}
        \end{subfigure} \\
        
        % Row 4
        \begin{subfigure}{0.33\textwidth}
            \centering
            \includegraphics[width=\linewidth]{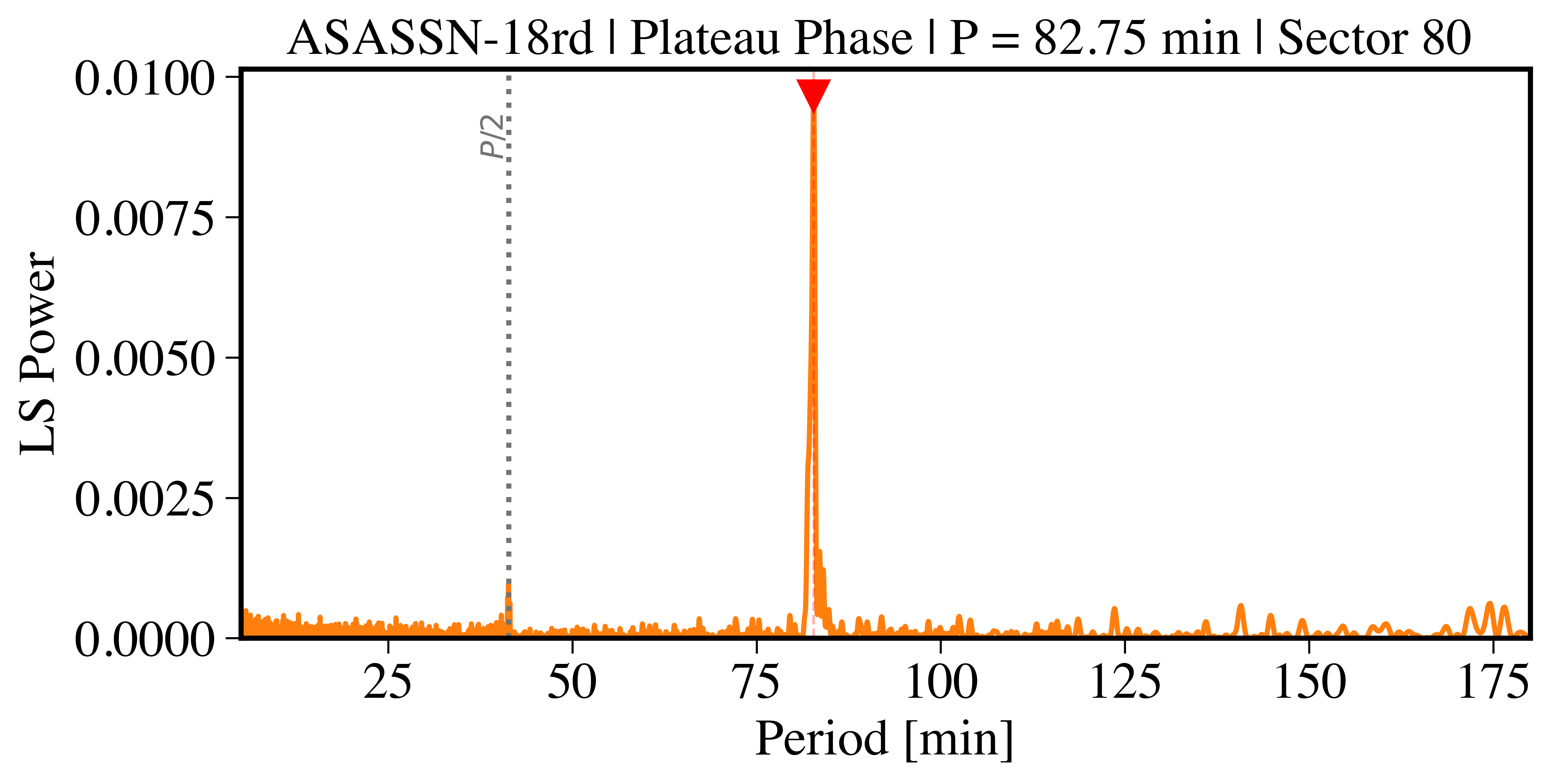}
            \caption{}
        \end{subfigure} &
        \begin{subfigure}{0.33\textwidth}
            \centering
            \includegraphics[width=\linewidth]{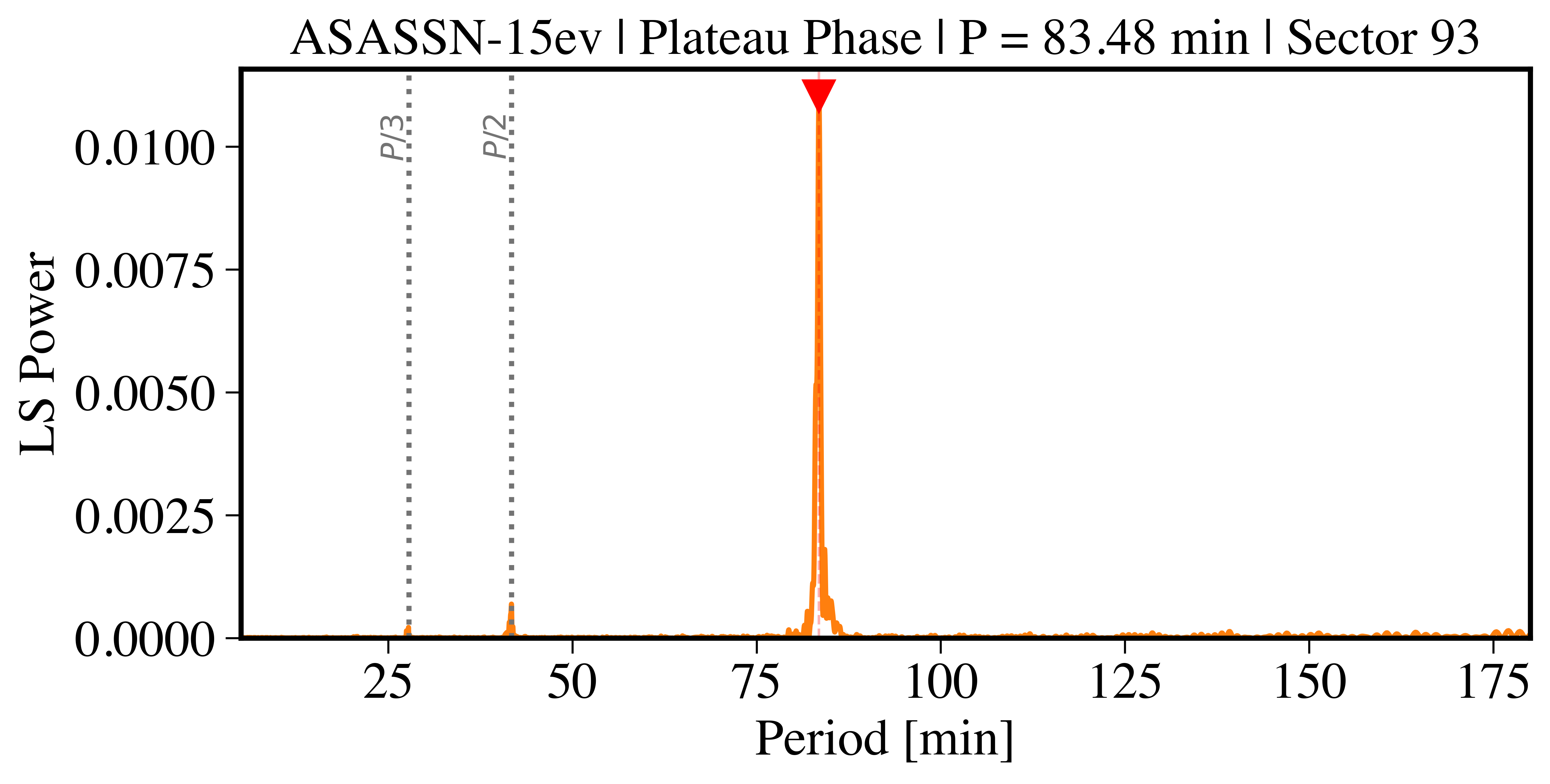}
            \caption{}
        \end{subfigure} &
        \multicolumn{1}{c}{} \\

    \end{tabular}
    
    \caption{Lomb-Scargle periodograms of the 11 accreting white dwarfs binaries discussed in Section~\ref{sec:results}. Red triangles indicate the best period corresponding to superhump ($P_{\rm SH}$) and orbital ($P_{\rm orb}$) periods reported in this works. In selected panels, the grey dashed vertical lines indicate the harmonics of the periods. Panels shown in orange correspond to $P_{\rm SH}$ measured during the plateau phase, while panel shown in blue correspond to periods  $P_{\rm orb}$ measured in quiescence.}
    \label{fig:periodograms}
\end{figure*}

%\clearpage % Optional: prevents Figure B from floating above Section B title
%\section{Phase Folded Light Curves} 
\begin{figure*}[!ht]
    \centering
    \setlength{\tabcolsep}{2pt}
    \renewcommand{\arraystretch}{1}
    
    \begin{tabular}{ccc}
        % Row 1
        \begin{subfigure}{0.32\textwidth}
            \centering
            \includegraphics[width=\linewidth]{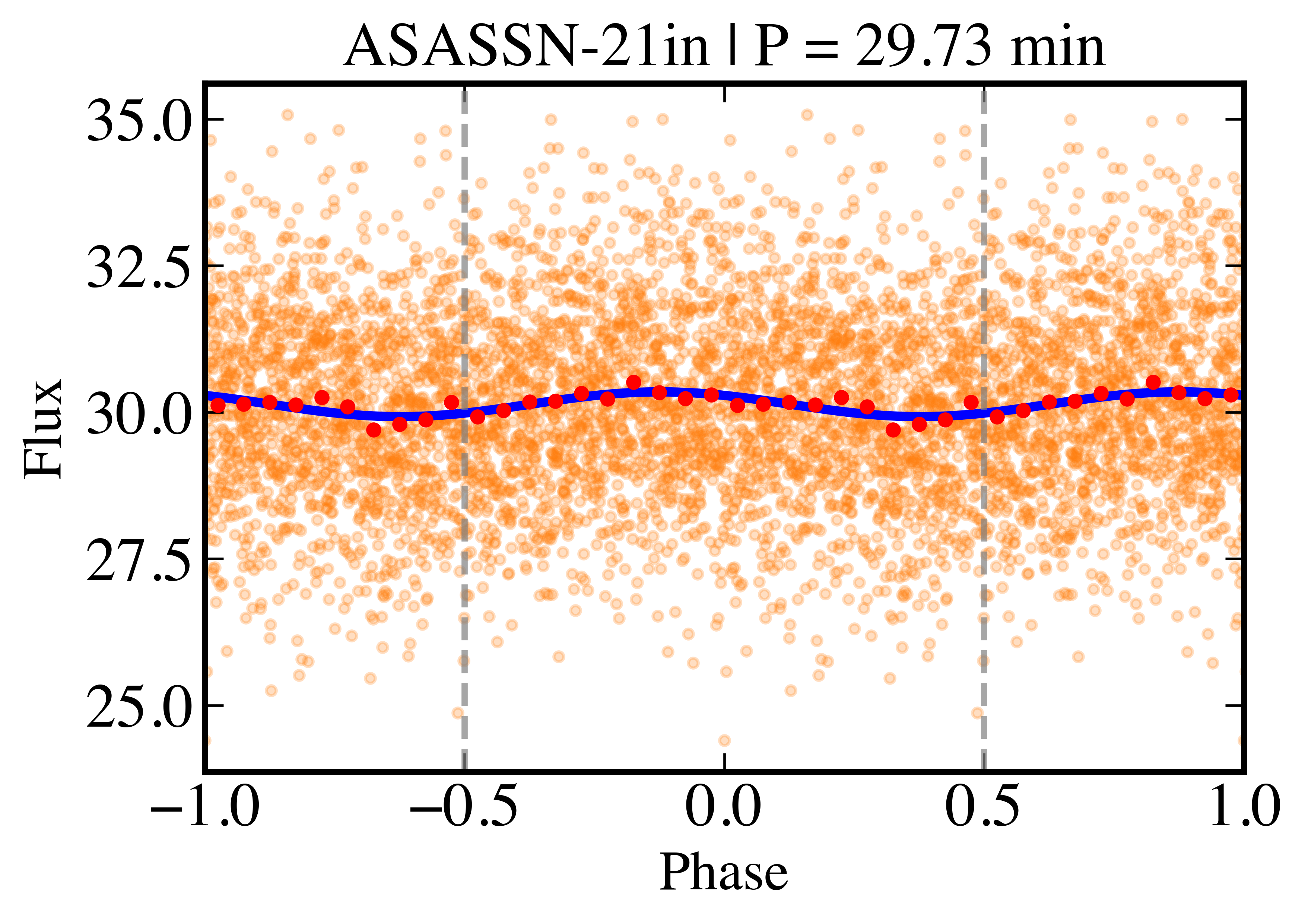}
            \caption{}
        \end{subfigure} &
        \begin{subfigure}{0.32\textwidth}
            \centering
            \includegraphics[width=\linewidth]{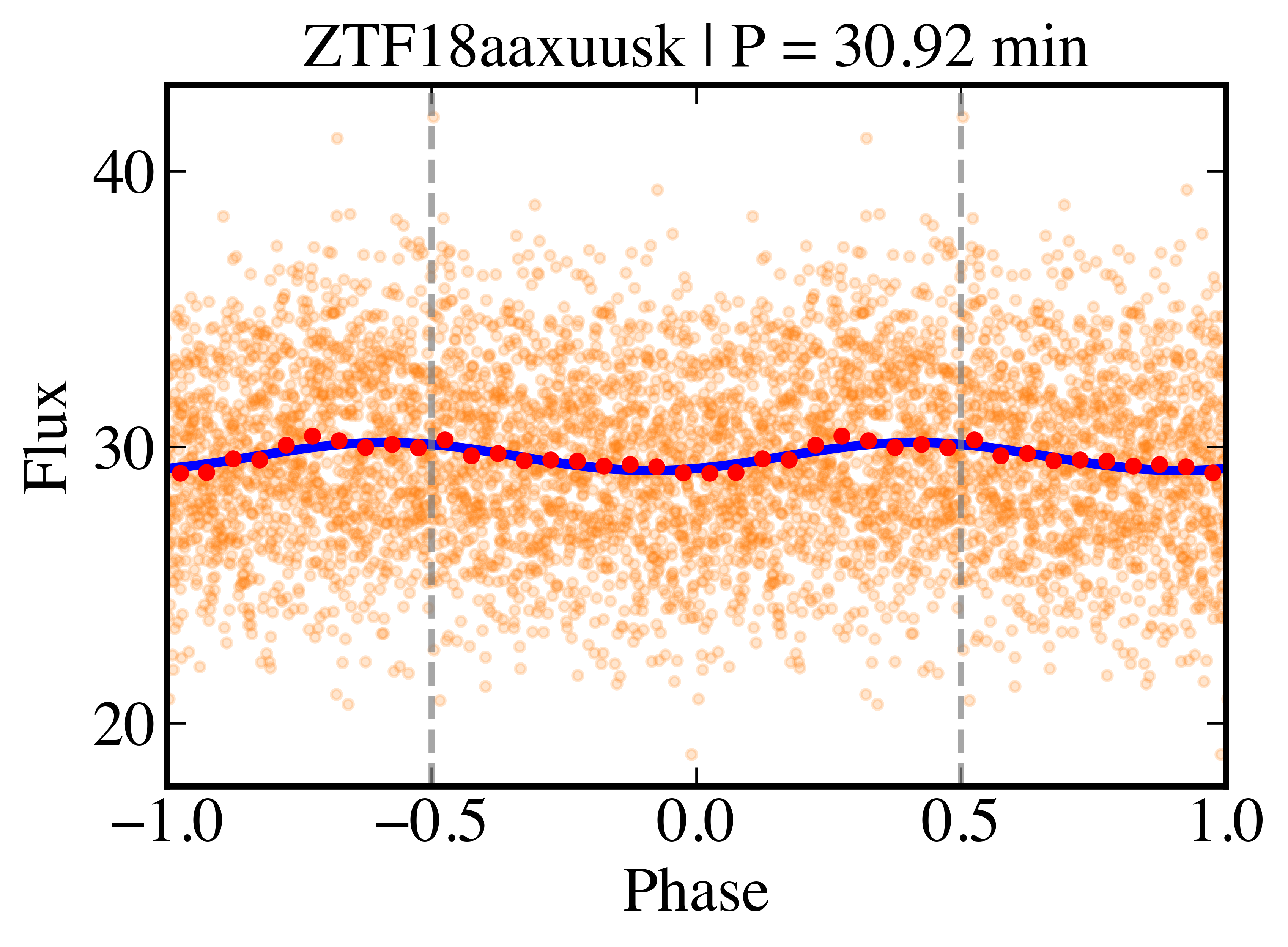}
            \caption{}
        \end{subfigure} &
        \begin{subfigure}{0.32\textwidth}
            \centering
            \includegraphics[width=\linewidth]{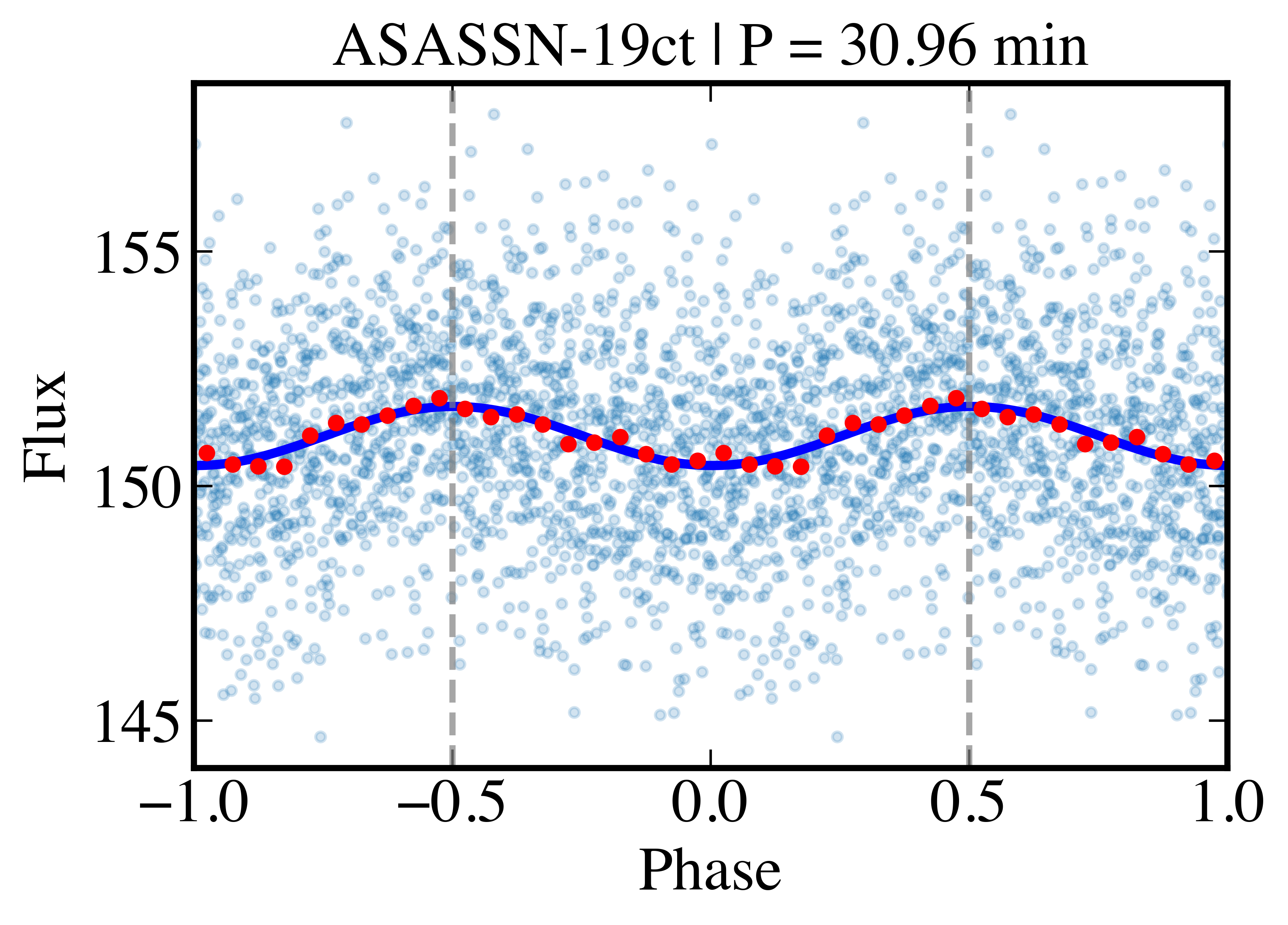}
            \caption{}
        \end{subfigure} \\
        
        % Row 2
        \begin{subfigure}{0.32\textwidth}
            \centering
            \includegraphics[width=\linewidth]{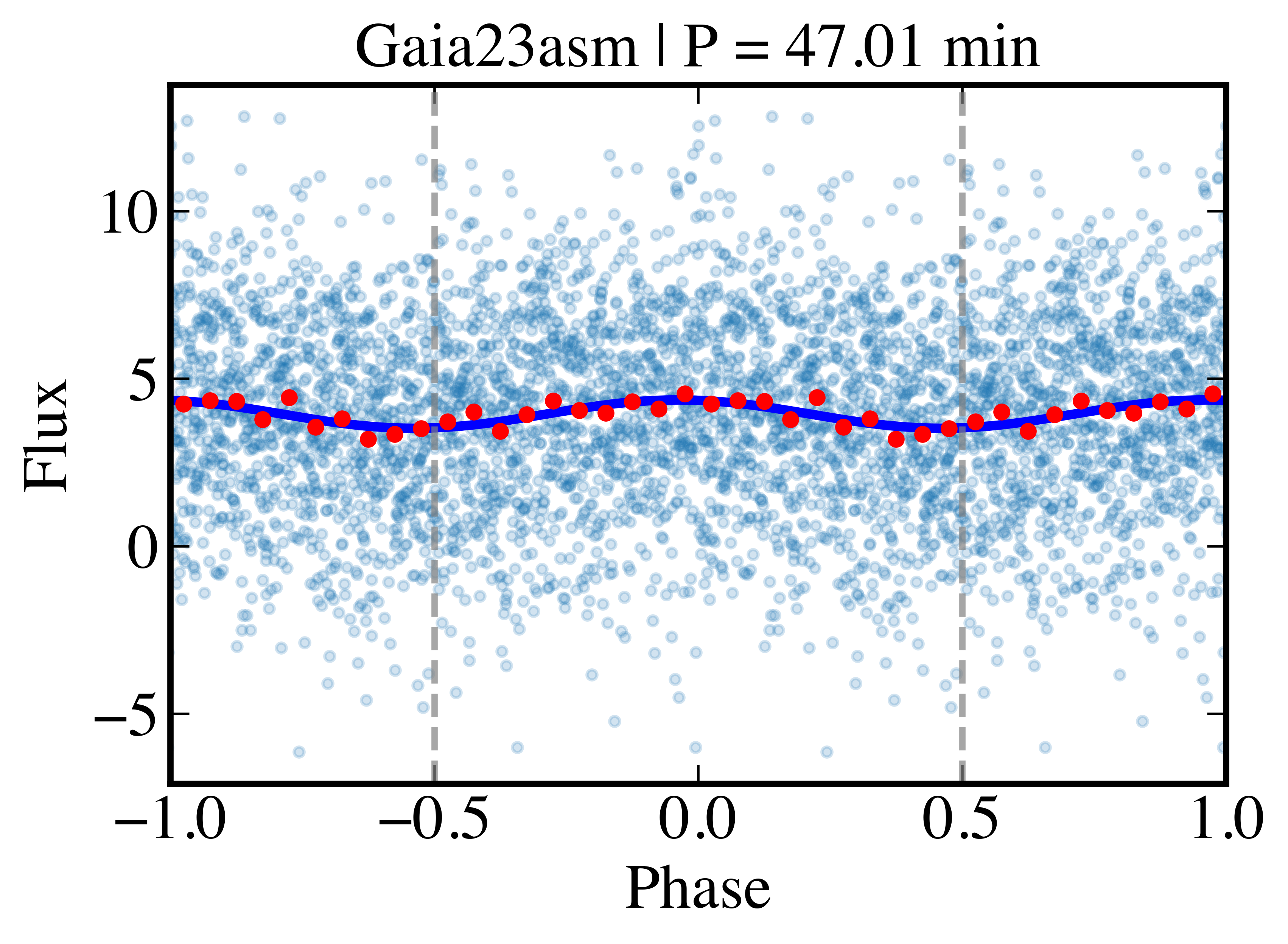}
            \caption{}
        \end{subfigure} &
        \begin{subfigure}{0.32\textwidth}
            \centering
            \includegraphics[width=\linewidth]{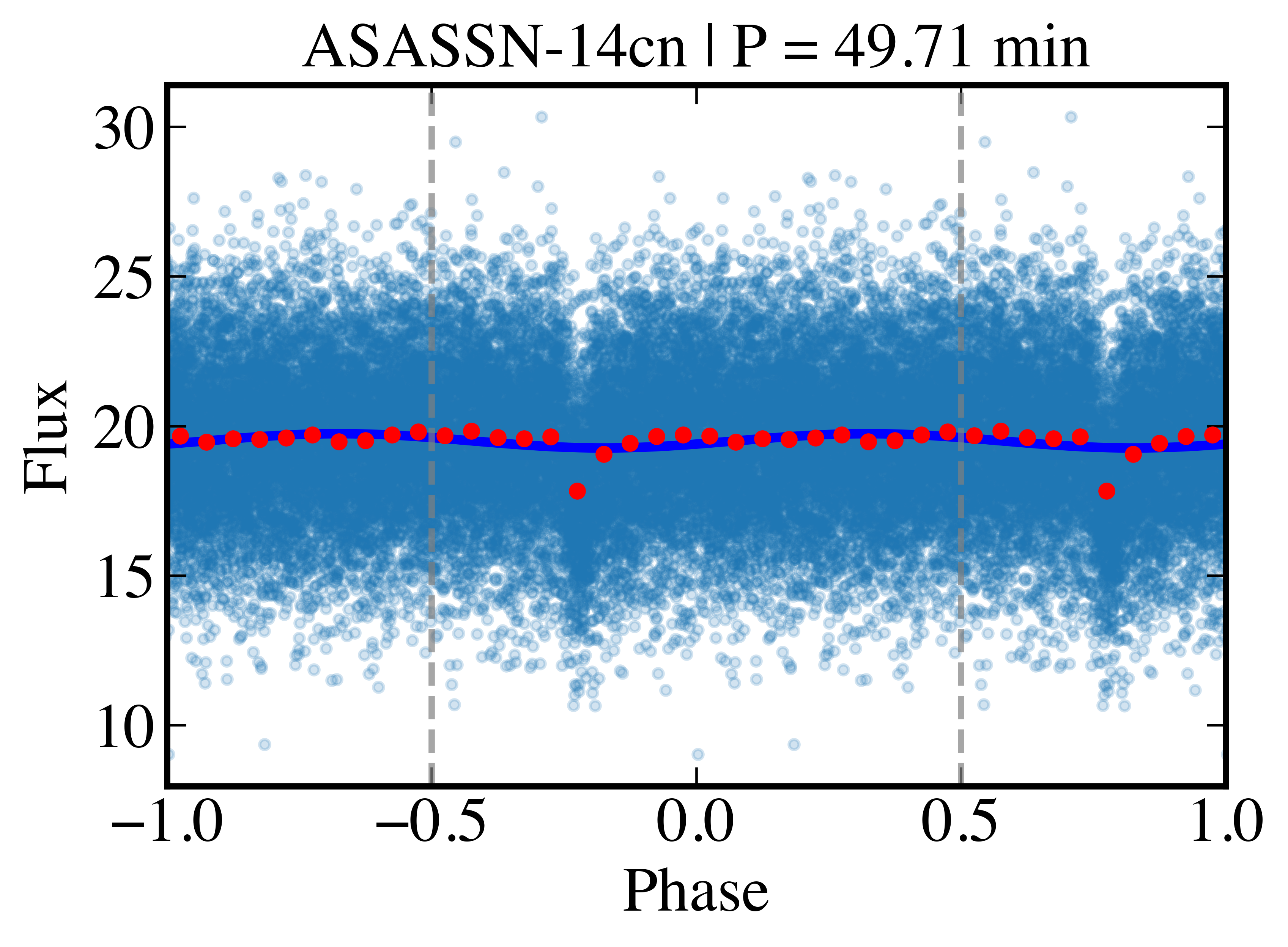}
            \caption{}
        \end{subfigure} &
        \begin{subfigure}{0.32\textwidth}
            \centering
            \includegraphics[width=\linewidth]{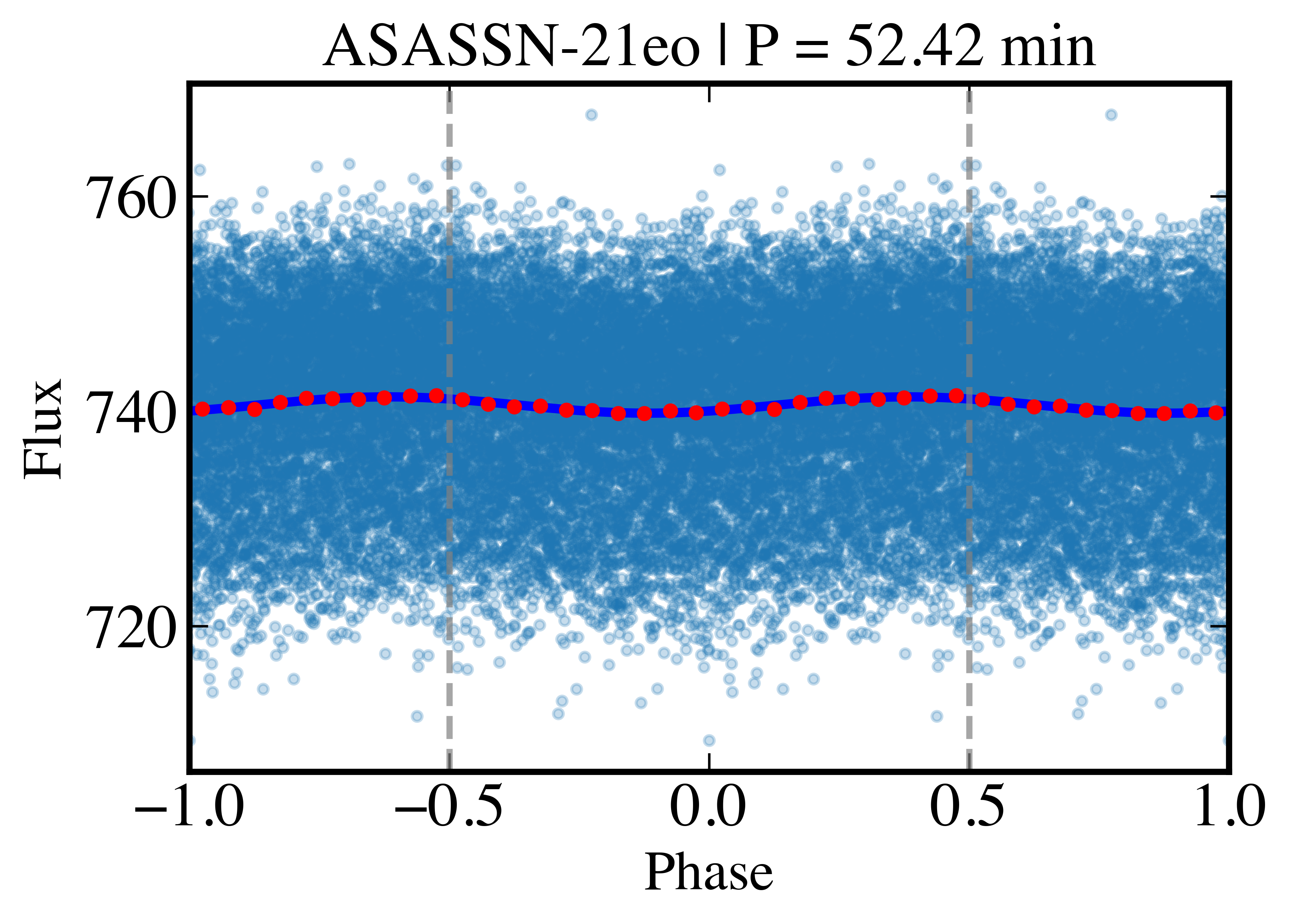}
            \caption{}
        \end{subfigure} \\
        
        % Row 3
        \begin{subfigure}{0.32\textwidth}
            \centering
            \includegraphics[width=\linewidth]{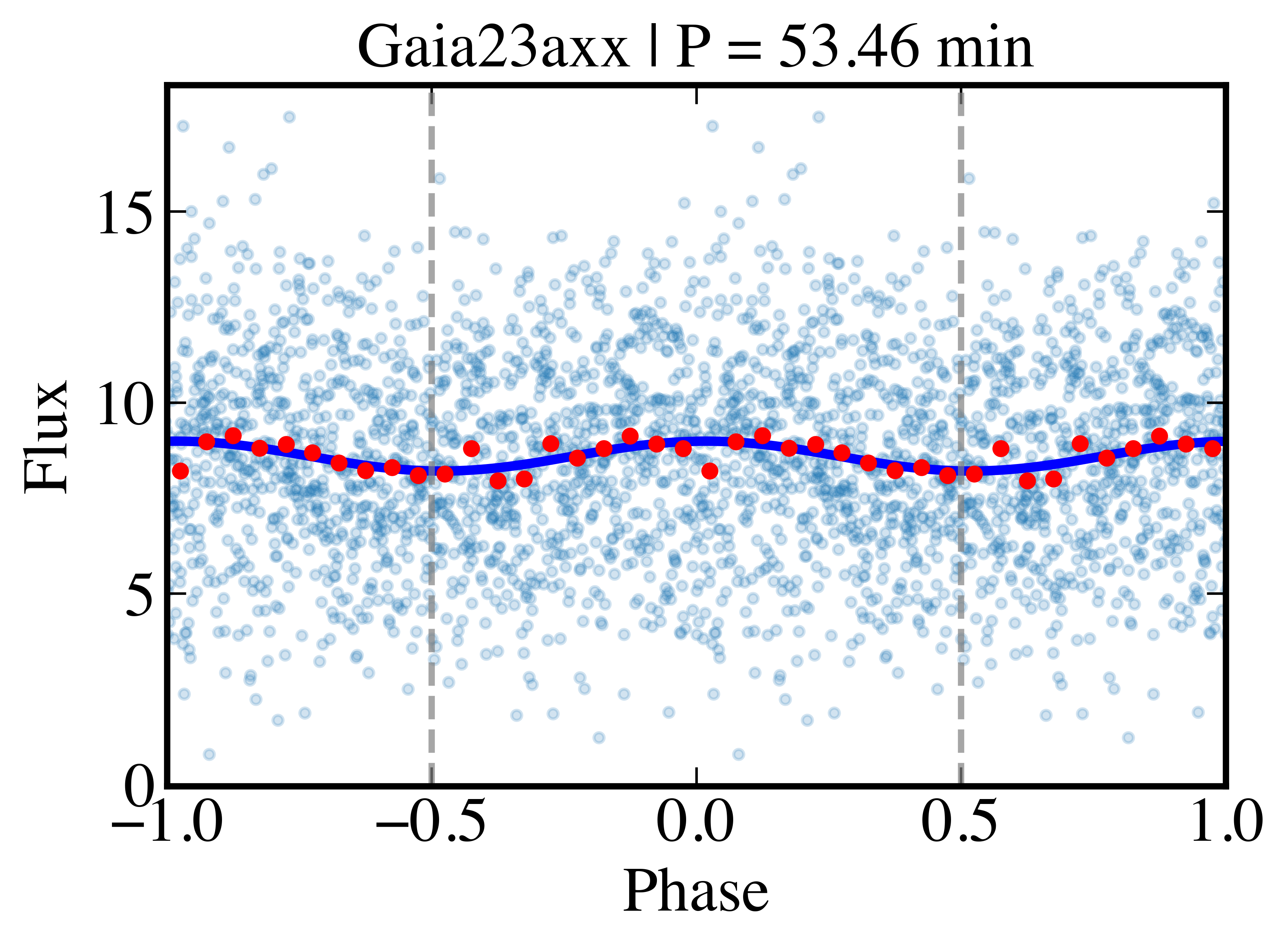}
            \caption{}
        \end{subfigure} &
        \begin{subfigure}{0.32\textwidth}
            \centering
            \includegraphics[width=\linewidth]{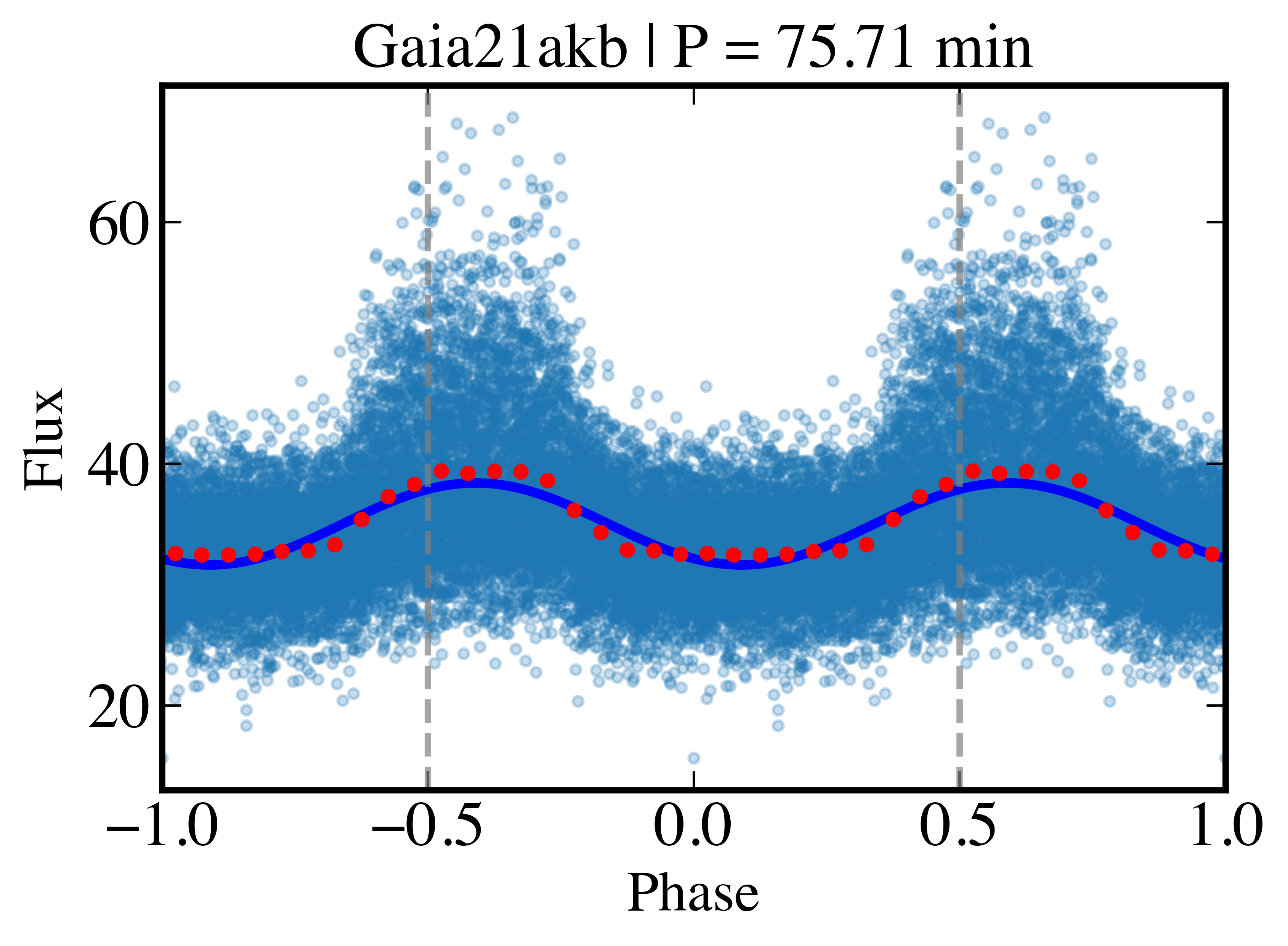}
            \caption{}
        \end{subfigure} &
        \begin{subfigure}{0.32\textwidth}
            \centering
            \includegraphics[width=\linewidth]{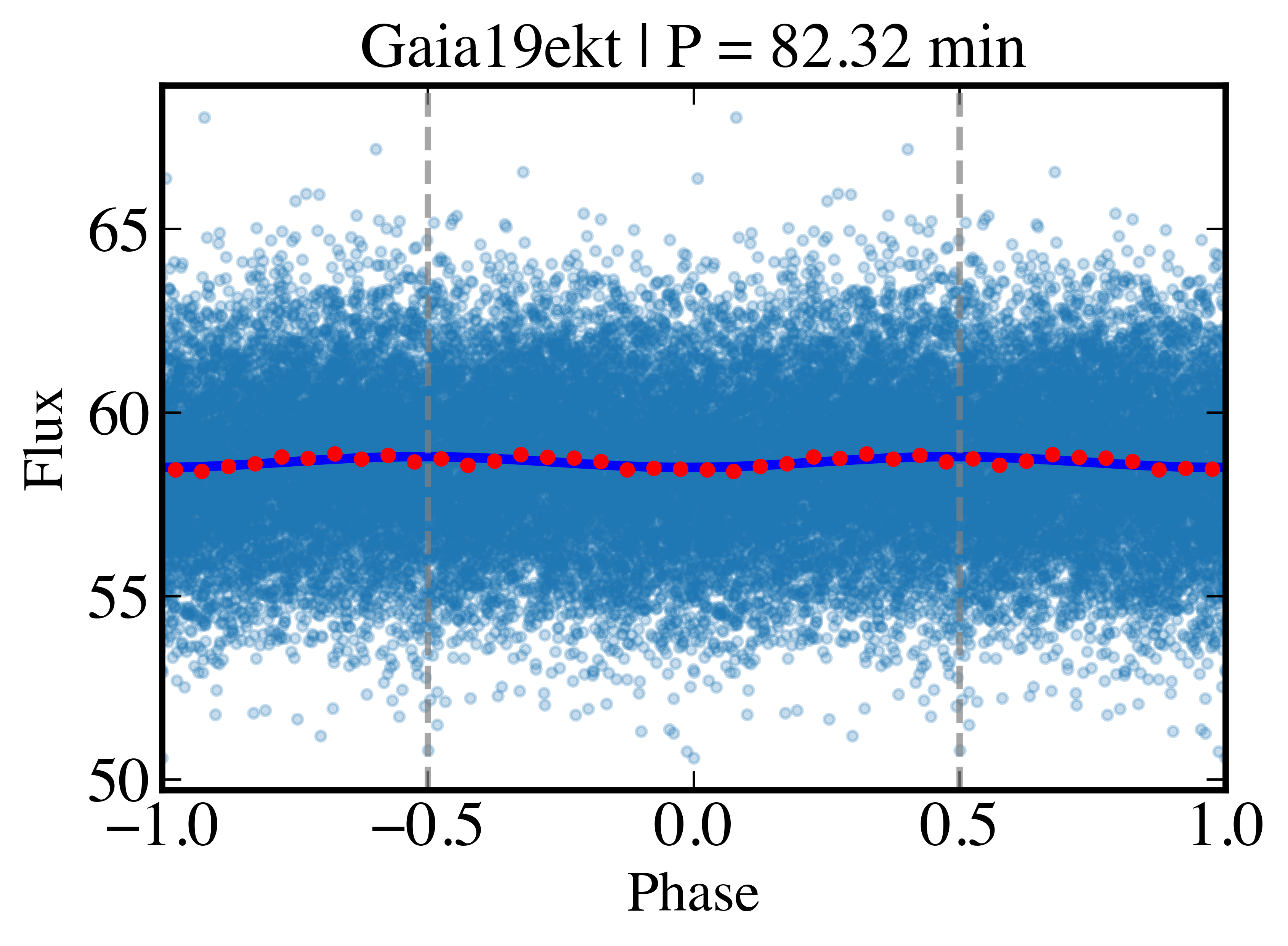}
            \caption{}
        \end{subfigure} \\
        
        % Row 4
        \begin{subfigure}{0.32\textwidth}
            \centering
            \includegraphics[width=\linewidth]{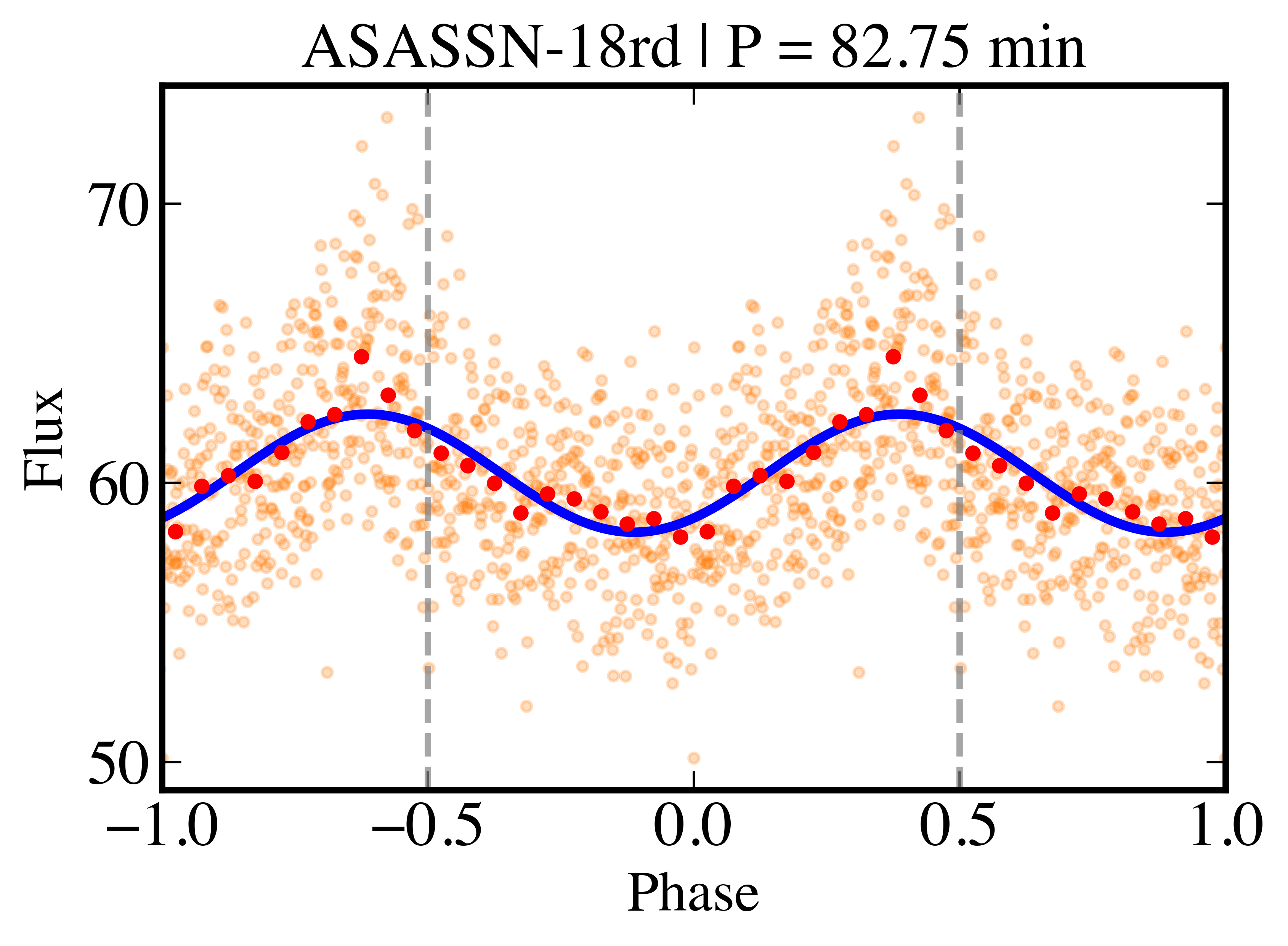}
            \caption{}
        \end{subfigure} &
        \begin{subfigure}{0.32\textwidth}
            \centering
            \includegraphics[width=\linewidth]{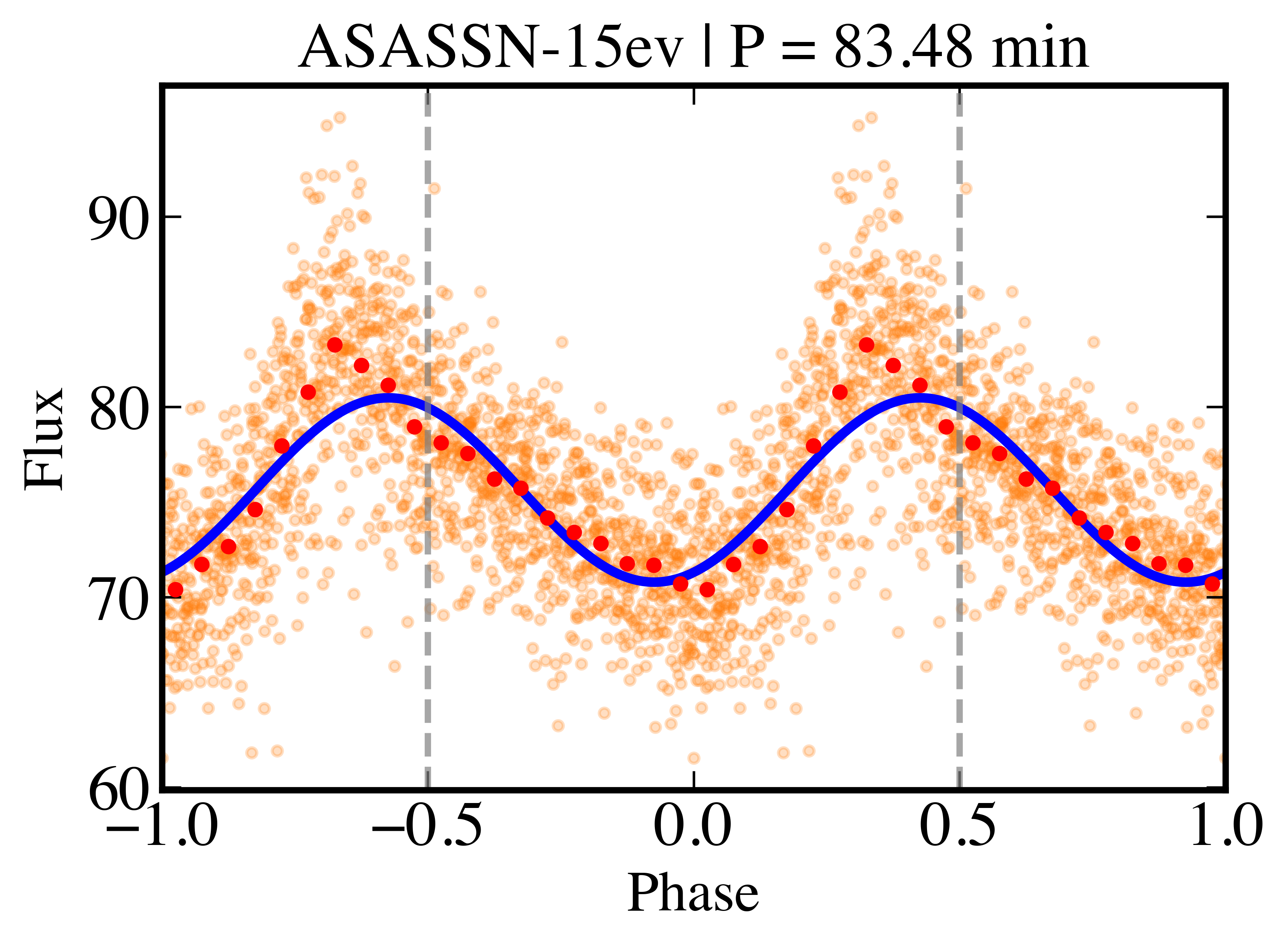}
            \caption{} 
        \end{subfigure} &
        \multicolumn{1}{c}{} \\
        
    \end{tabular}
    
    \caption{Phase-folded TESS light curves (2-min cadence). Orange-themed panels (a), (b), (j), and (k) show systems folded on the superhump period estimated from the SO plateau phase. Blue-themed panels (c)–(i) show systems folded on the orbital period. The solid blue lines represent the best-fitting LS sinusoids, and red circles indicate data averages in bins with width of 0.05 phase. For panels (c), (d), (g) and (i-k), only a subset of data points is plotted to reduce crowding and make the modulation clearer}
    
    \label{fig:phase_plots2}
\end{figure*}

\pagebreak

%\FloatBarrier
\section{Red-Noise Diagnostic} \label{rednoisediagnostic} 
In this appendix we present the RedNoiseFALs periodograms and power law FALs for all 11 targets. The method is described in Section~\ref{sec:rednoise}

\begin{figure*}[p]
    \centering
    \includegraphics[
        width=\textwidth,
        height=0.90\textheight,
        keepaspectratio
    ]{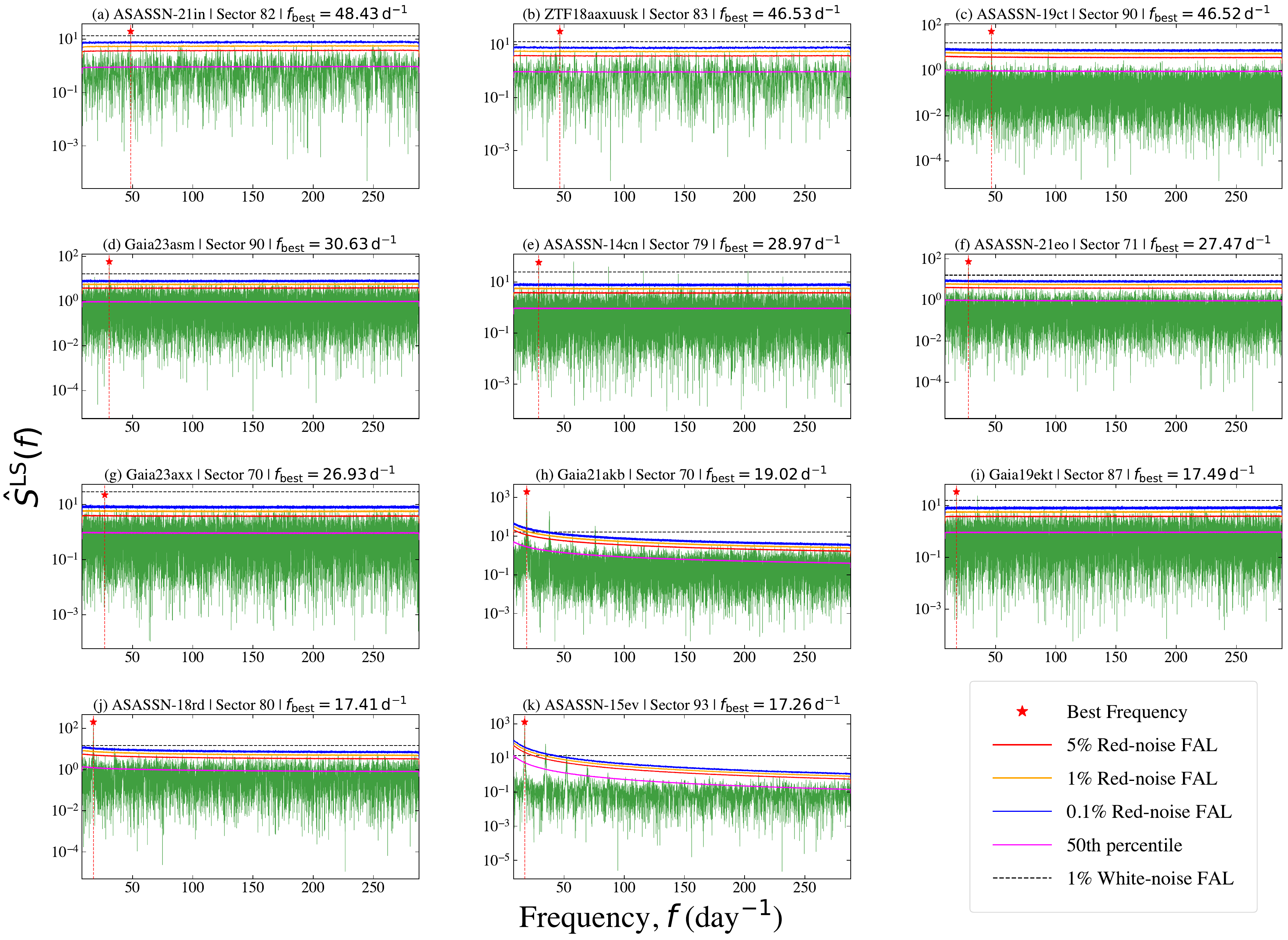}
    \caption{ $\hat{S}^{LS}(f)$ corresponds to the Lomb-Scargle periodogram in power spectral density normalization, as computed using \textit{RedNoiseFALs}. The red stars and red vertical dashed lines indicate the adopted best frequency ($f_{\rm best}$). The colour lines are the 5\% (red), 1\% (orange), and 0.1\% (blue) power law FALs. The fuchsia line shows the 50th percentile of the simulated continuum, while the black dashed line is the 1\%  white noise FAL. Horizontal FAL curves indicate a white continuum, while the slope curves in panels (h), (j), and (k) indicate the presence of red-noise.} 
    \label{fig:rednoise}
\end{figure*}

%\FloatBarrier
\section{Long-term light curves} \label{longterm}

We present the long term ground based light curves of the 4 systems for which we estimate SOs/NOs recurrence times (AM~CVns ASASSN-21in and ZTF18aaxuusk, and CVs ASASSN-18rd and ASASSN-15ev), as well as the six systems (AM CVn candidates Gaia23asm and Gaia23axx, AM CVns ASASSN-14cn and ASASSN-21eo, and CVs Gaia21akb and Gaia19ekt) detected in quiescence with TESS.

\begin{figure*}[!h]
    \centering
    \includegraphics[width=0.99\linewidth]{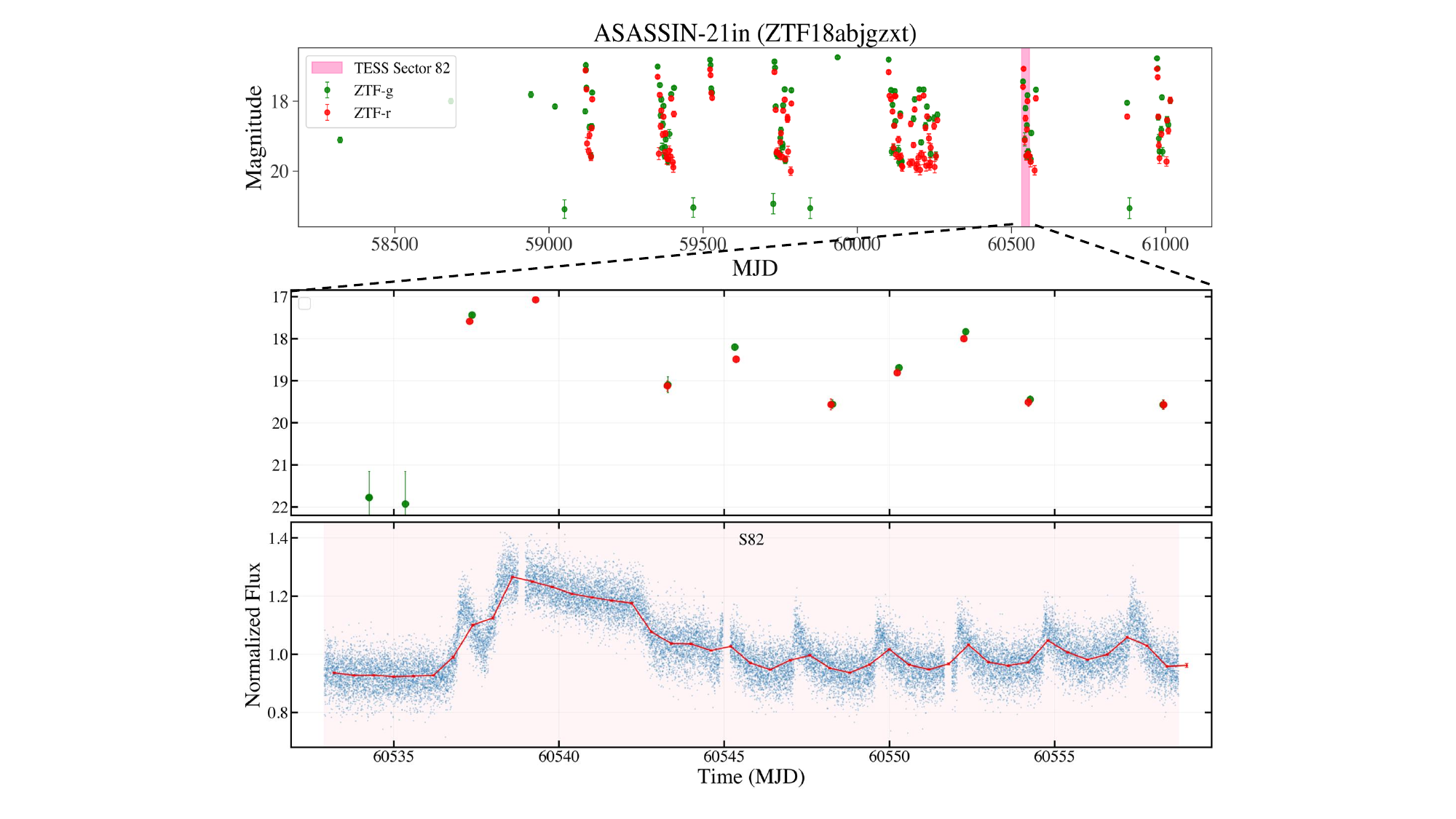} 
    \includegraphics[width=0.99\linewidth]{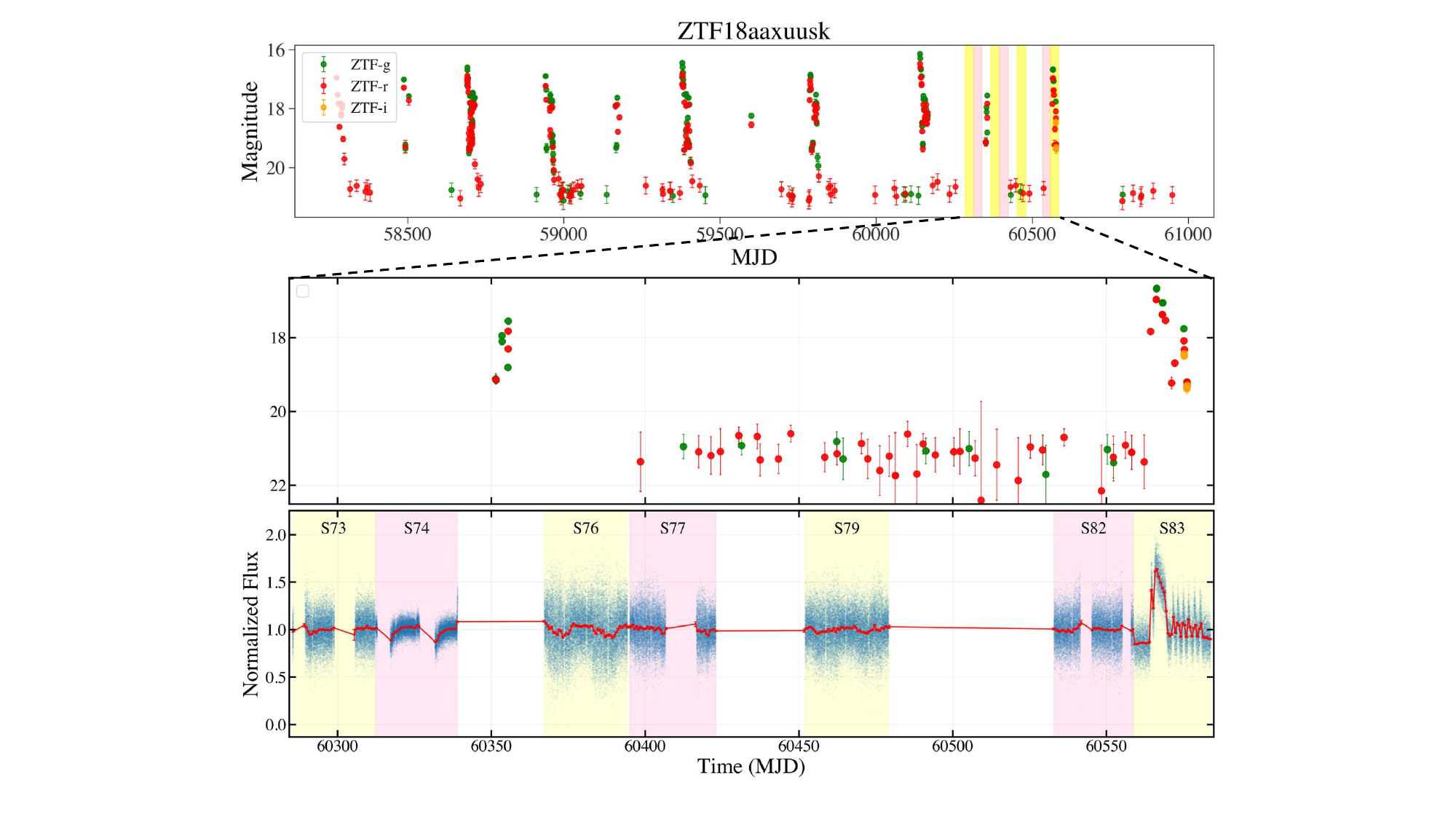}
    
    \caption{Long-term and TESS light curves of ASASSN-21in and ZTF18aaxuusk. For each system, the top panel shows the full ZTF long-term light curve, with the TESS sector coverage indicated by the shaded region. The middle panel shows a zoomed view of the superoutburst detected in the ZTF data. The bottom panel shows the corresponding TESS light curve, with individual sectors highlighted. The red line represents the binned average of the data using a phase-bin width of 0.05.}
    
    \label{fig:Longtermspart1}
\end{figure*}

\begin{figure*}[!h]
    \centering
    \includegraphics[width=0.99\linewidth]{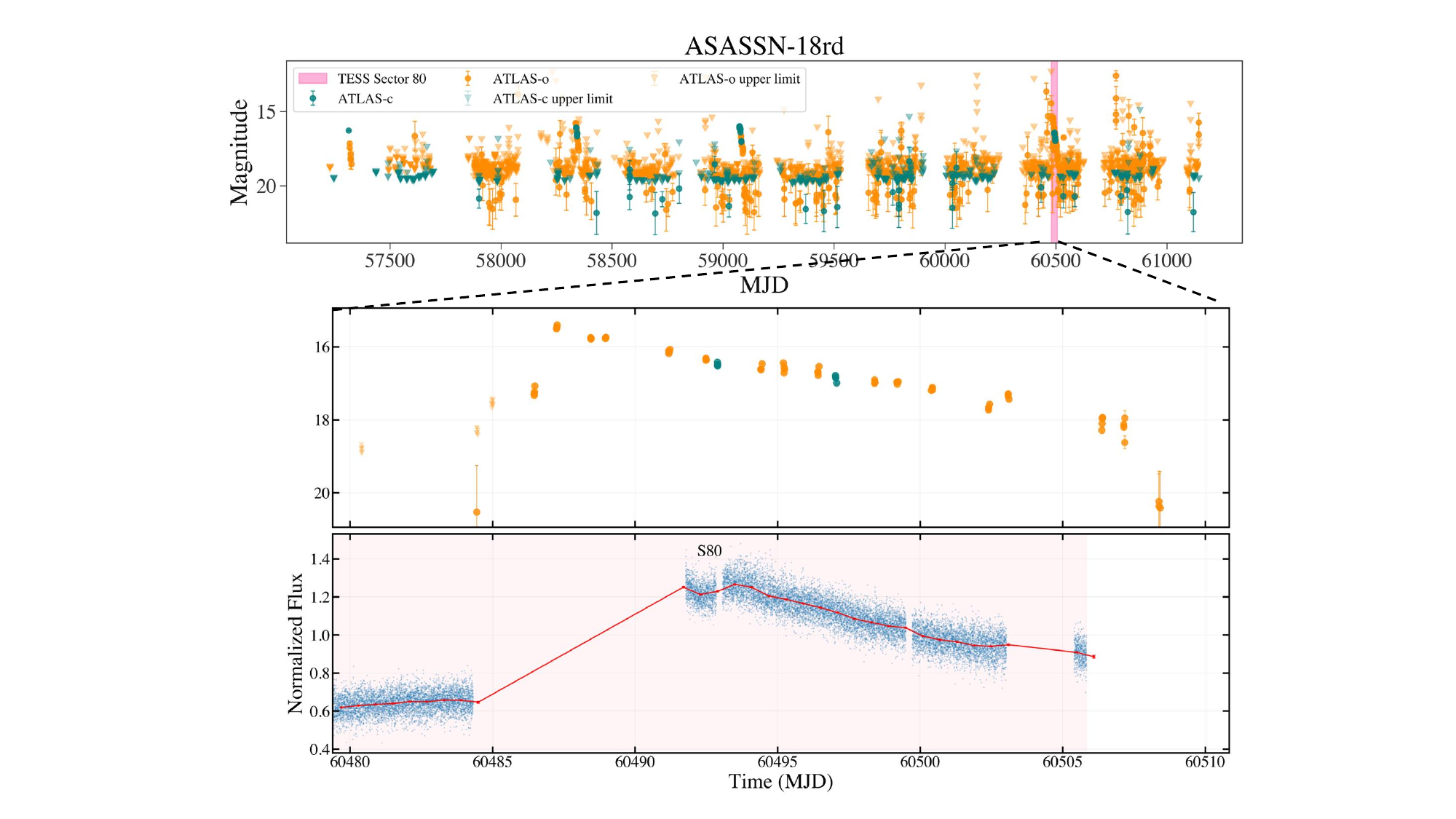}
    \includegraphics[width=0.99\linewidth]{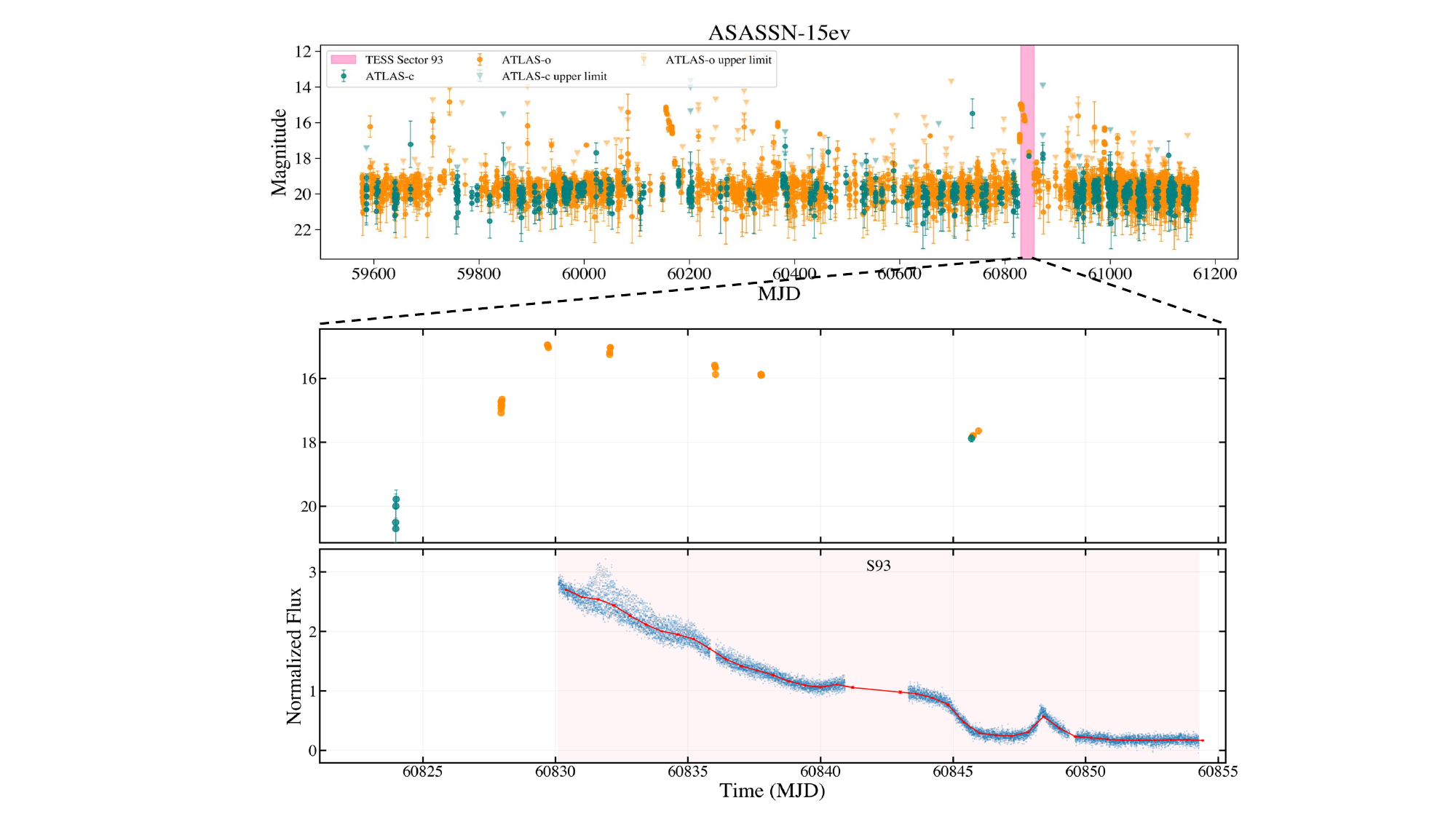} 

    \caption{Long-term and TESS light curves of ASASSN-18rd and ASASSN-15ev. For each system, the top panel shows the full ATLAS long-term light curve, with the TESS sector coverage indicated by the shaded region. The middle panels show a zoomed view of the superoutburst detected in the ATLAS data. The bottom panels show the corresponding TESS light curve, with individual sectors highlighted. The red line represents the binned average of the data using a phase-bin width of 0.05.}    
    \label{fig:Longtermsart2}
    
\end{figure*} 

\begin{figure*}[!h]
    \centering
    \includegraphics[width=0.90\linewidth]{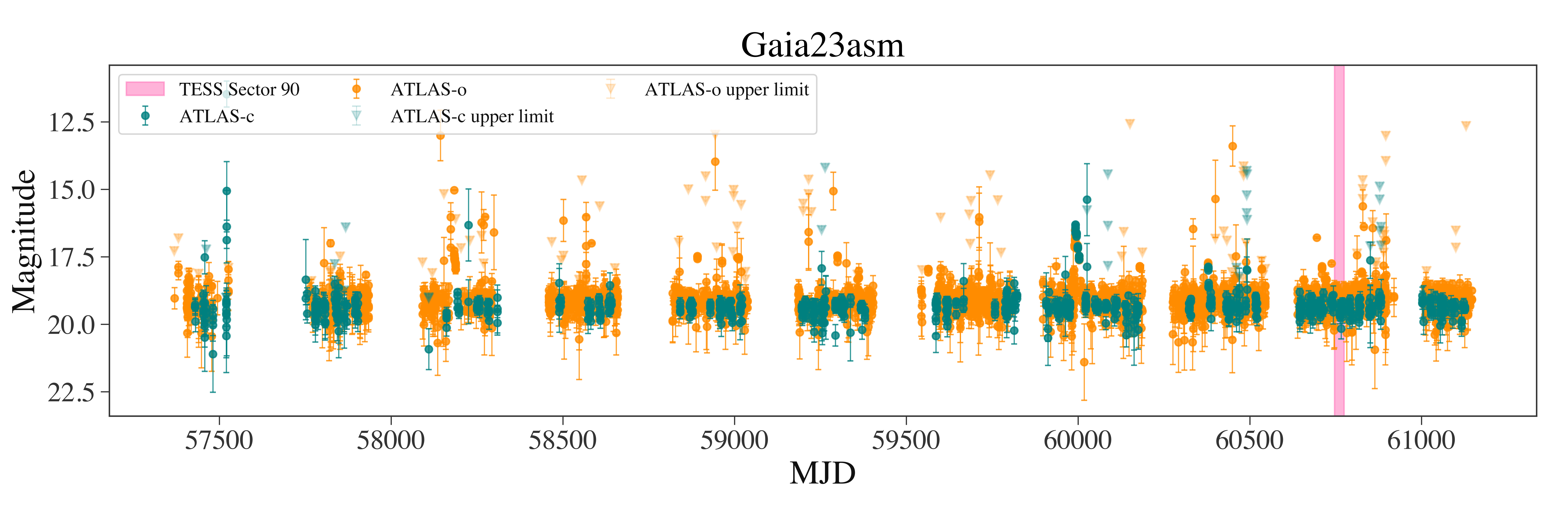}
    \includegraphics[width=0.90\linewidth]{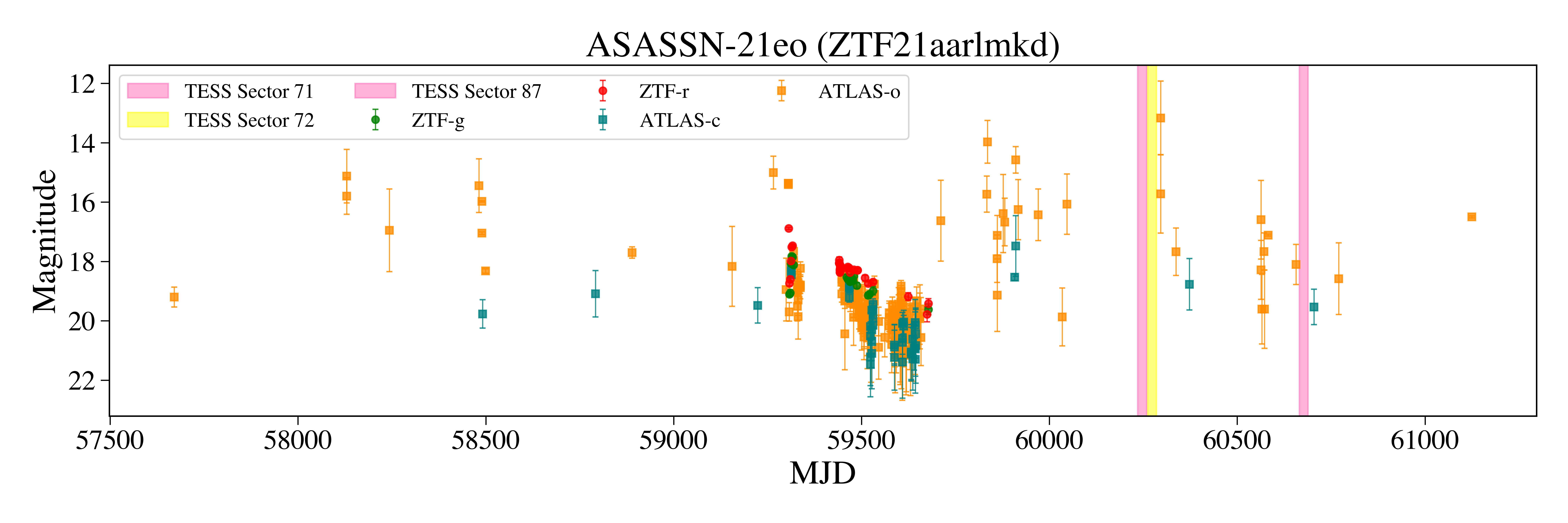}
    \includegraphics[width=0.90\linewidth]{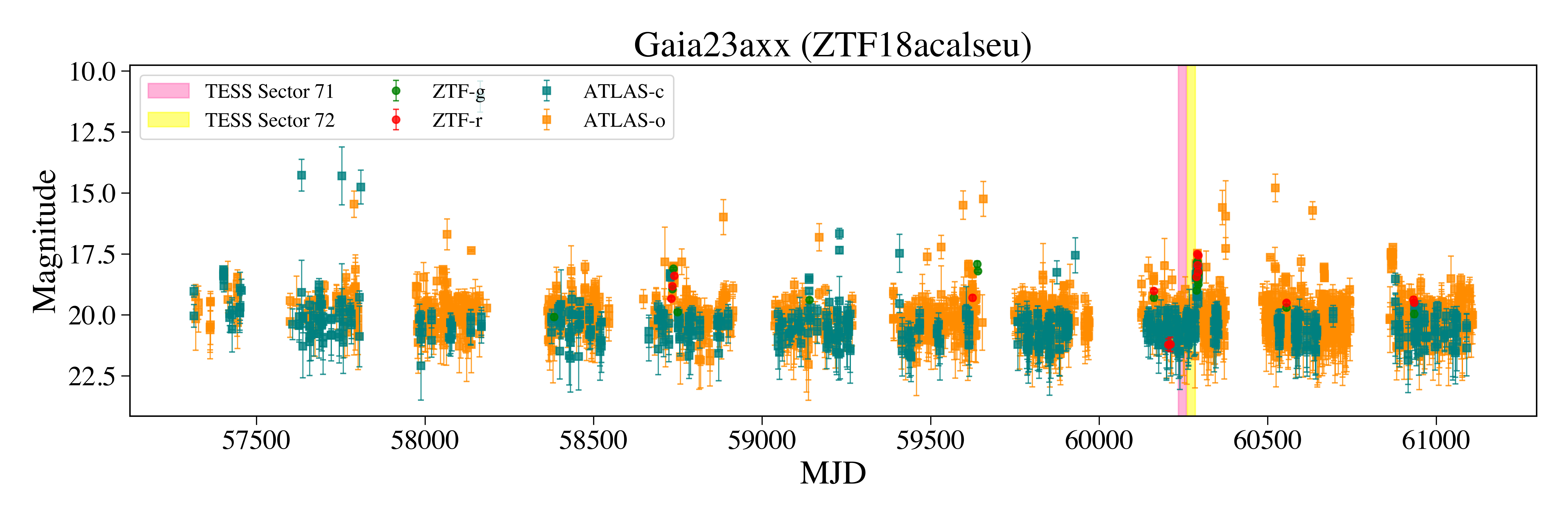}

    \caption{Long-term light curves based on ground based photometric observations. The highlighted regions indicate the TESS Sectors during which the target was observed in this survey.}    
    \label{fig:Longtermsart3}
    
\end{figure*} 

\end{document}